\documentclass[lettersize,journal]{IEEEtran}
\usepackage{amsmath,amsfonts}
\usepackage{algorithmic}
\usepackage{algorithm}
\usepackage{array}
\usepackage[caption=false,font=normalsize,labelfont=sf,textfont=sf]{subfig}
\usepackage{caption}
\usepackage{bm}
\usepackage{bbm}
\usepackage{amsmath,amssymb,amsfonts}
\usepackage{amsthm}
\newtheorem{theorem}{Theorem}
\usepackage{subfig}
\usepackage{multirow}
\usepackage{textcomp}
\usepackage{stfloats}
\usepackage{url}
\usepackage{verbatim}
\usepackage{graphicx}
\usepackage{cite}
\usepackage{authblk}
\usepackage{enumitem}
\newcommand{\myparatight}[1]{\smallskip\noindent{\bf {#1}:}~}
\newcommand{\argmax}{\operatornamewithlimits{argmax}}
\newcommand{\argmin}{\operatornamewithlimits{argmin}}
\newcommand{\certifiedradius}{certified security level}

\newcommand\blfootnote[1]{%
  \begingroup
  \renewcommand\thefootnote{}\footnote{#1}%
  \addtocounter{footnote}{-1}%
  \endgroup
}

\usepackage{xcolor}
\newcommand{\xc}[1]{{#1}}

\begin{document}

\title{FLCert: Provably Secure Federated Learning against Poisoning Attacks}

\author{Xiaoyu Cao, Zaixi Zhang, Jinyuan Jia, and Neil Zhenqiang Gong,~\IEEEmembership{Member,~IEEE}
\thanks{Xiaoyu Cao is with Meta Platforms, 1 Hacker Way, Menlo Park, CA, US. (e-mail: xiaoyucao@fb.com).\par
Zaixi Zhang is with School of Computer Science and Technology, University of Science and Technology of China, No.96, JinZhai Road Baohe District, Hefei, Anhui, China. (e-mail: zaixi@mail.ustc.edu).\par
Jinyuan Jia is with the Department of Computer Science, University of Illinois Urbana-Champaign, 201 N Goodwin Ave, Urbana, IL, US. (e-mail: jinyuan@illinois.edu) \par
Neil Zhenqiang Gong is with the Department of Electrical and Computer Engineering, Duke University, 413 Wilkinson Building, Durham, NC, US. (e-mail: neil.gong@duke.edu)}}

\markboth{Transactions on Information Forensics and Security}%
{Cao \MakeLowercase{\textit{et al.}}: FLCert: Provably Secure Federated Learning against Poisoning Attacks}

\maketitle

\begin{abstract}
\blfootnote{*This work is an extension of \cite{cao2020provably}.}
Due to its distributed nature, federated learning is vulnerable to poisoning attacks, in which  malicious clients  poison the training process via manipulating their local training data and/or local model updates sent to the cloud server, such that the poisoned global model misclassifies many indiscriminate test inputs or attacker-chosen ones. Existing defenses mainly leverage Byzantine-robust federated learning methods or detect malicious clients. However, these defenses do not have provable security guarantees against poisoning attacks and may be vulnerable to more advanced attacks. In this work, we aim to bridge the gap by proposing \emph{FLCert}, an ensemble federated learning framework, that is provably secure against poisoning attacks with a bounded number of malicious clients. Our key idea is to divide the clients into groups, learn a global model for each group of clients using any existing federated learning method, and take a majority vote among the global models to classify a test input. Specifically, we consider two methods to group the clients and propose two variants of FLCert correspondingly, i.e., \emph{FLCert-P} that randomly samples clients in each group, and \emph{FLCert-D} that divides clients to disjoint groups deterministically. Our extensive experiments on multiple datasets show that the label predicted by our FLCert for a test input is provably unaffected by a bounded number of malicious clients, no matter what poisoning attacks they use.
\end{abstract}


\section{Introduction}

Federated learning (FL) \cite{Konen16,McMahan17} is an emerging machine learning paradigm, which enables clients (e.g., smartphones, IoT devices, and organizations) to collaboratively learn a model without sharing their local training data with a cloud server. Due to its promise for protecting privacy of the clients' local training data and the emerging privacy regulations such as General Data Protection Regulation (GDPR), FL has been deployed by industry. For instance, Google has deployed  FL for next-word prediction on Android Gboard.  Existing FL methods mainly follow a \emph{single-global-model} paradigm.  
Specifically, a cloud server maintains a \emph{global model} and each client maintains a \emph{local model}.  
The global model is trained via multiple iterations of communications between the clients and server. 
In each iteration, three steps are performed: 1) the server sends the current global model to the clients; 2) the clients update their local models based on the global model and their local training data, and send the model updates
to the server; and 3) the server aggregates the model updates and uses them to update the global model. 
The learnt global model is then used to predict labels of test inputs. 

However, such single-global-model paradigm is vulnerable to poisoning attacks. In particular, an attacker can inject fake clients to FL or compromise genuine clients, where we call the fake/compromised clients \emph{malicious clients}. For instance, an attacker can use a powerful computer to simulate many fake smartphones. Such malicious clients can corrupt the global model via carefully tampering their local training data or model updates sent to the server. As a result, the corrupted global model has a low accuracy for the normal test inputs \cite{fang2019local,cao2022mpaf} (known as \emph{untargeted poisoning attacks}) or certain attacker-chosen test inputs~\cite{bagdasaryan2020backdoor,bhagoji2019analyzing} (known as \emph{targeted poisoning attacks}). For instance, in an untargeted poisoning attack, the malicious clients can deviate the global model towards the opposite of the direction along which it would be updated without attacks by manipulating their local model updates \cite{fang2019local}. In a targeted poisoning attack, when learning an image classifier, the malicious clients can re-label the cars with certain strips as birds in their local training data and scale up their model updates sent to the server, such that the global model incorrectly predicts a car with the strips as bird \cite{bagdasaryan2020backdoor}.

Various Byzantine-robust FL methods have been proposed to defend against poisoning attacks from malicious clients.  \cite{Blanchard17,Yin18,cao2020fltrust}. The main idea of these methods  is to mitigate the impact of statistical outliers among the clients' model updates. They can bound the difference between the global model parameters learnt without malicious clients and the global model parameters learnt when some clients become malicious. However, these methods cannot provably guarantee that the label predicted by the global model for a test input is not affected by malicious clients. Indeed, studies showed that malicious clients can still substantially degrade the test accuracy of a global model learnt by a Byzantine-robust method via carefully tampering their model updates sent to the server~\cite{bhagoji2019analyzing,fang2019local,cao2022mpaf}.

\myparatight{Our work} In this work, we propose \emph{FLCert}, the first FL framework that is provably secure against poisoning attacks. Specifically, given $n$ clients, we define $N$ groups, each containing a subset of the clients. In particular, we design two methods to group the clients, which corresponds to two variants of FLCert, i.e., \emph{FLCert-P} and \emph{FLCert-D}, where P and D stand for \emph{probabilistic} and \emph{deterministic}, respectively. In FLCert-P,  each of the $N$ groups consists of $k$ clients sampled from the $n$ clients uniformly at random. Note that   there are a total of ${n \choose k}$ possible groups and thus $N$ could be as large as ${n \choose k}$ in FLCert-P. 
In FLCert-D, we divide the $n$ clients into $N$ disjoint groups deterministically.  

For each group, we learn a global model using an arbitrary FL algorithm (called \emph{base FL algorithm}) with the clients in the group. In total, we train $N$ global models.  Given a test input $\bm{x}$, we use each of the $N$ global models to predict its label.  We denote $n_i$ as the number of global models that predict label $i$ for $\bm{x}$ and define $p_i=\frac{n_i}{n}$, where $i=1,2,\cdots,L$. We call $n_i$ \emph{label frequency} and $p_i$ \emph{label probability}. Our \emph{ensemble global model} predicts the label with the largest label frequency/probability for $\bm{x}$. In other words, our ensemble global model takes a majority vote among the $N$ global models to predict label for $\bm{x}$.
Since each global model is learnt using a subset of the clients, a majority of the global models are learnt using benign clients when most clients are benign. Therefore, the majority-vote label among the $N$ global models for a test input is unaffected by a bounded number of malicious clients no matter what poisoning attacks they use.

\myparatight{Theory} Our first major theoretical result is that FLCert provably predicts the same label for a test input $\bm{x}$ when the number of malicious clients is no larger than a threshold, which we call \emph{certified security level}.  Our second major theoretical result is that we prove our derived certified security level is tight, i.e., when no assumptions are made on the base FL algorithm, it is impossible to derive a certified security level that is larger than ours. Note that the certified security level may be different for different test inputs.

\myparatight{Algorithm} Computing our certified security level for $\bm{x}$ requires its largest and second largest label frequencies/probabilities. For FLCert-P, when ${n \choose k}$ is small (e.g., the $n$ clients are dozens of organizations \cite{kairouz2019advances} and $k$ is small), we can compute the largest and second largest label probabilities exactly via training  $N={n \choose k}$ global models. 
However, it is challenging to compute them exactly when ${n \choose k}$ is large. To address the computational challenge, we develop a randomized algorithm to estimate them with probabilistic guarantees via training $N\ll{n \choose k}$ global models. Due to such randomness, FLCert-P achieves probabilistic security guarantee, i.e., FLCert-P outputs an incorrect certified security level for a test input with some probability that can be set to be arbitrarily small.  For FLCert-D, we train $N$ global models and obtain the label frequencies deterministically,  making the security guarantee of FLCert-D deterministic.

\myparatight{Evaluation} We empirically evaluate our method on five datasets from different domains, including three image classification datasets (MNIST-0.1 \cite{lecun2010mnist}, MNIST-0.5 \cite{lecun2010mnist}, and CIFAR-10 \cite{krizhevsky2009learning}), a human activity recognition dataset (HAR) \cite{anguita2013public}, and a next-word prediction dataset (Reddit) \cite{bagdasaryan2020backdoor}.
We distribute the training examples in MNIST and CIFAR to clients to simulate FL scenarios, while the HAR and Reddit dataset represent  real-world FL scenarios, where each user is a client. We also evaluate five different base FL algorithm, i.e., FedAvg~\cite{McMahan17}, Krum~\cite{Blanchard17}, Trimmed-mean~\cite{Yin18}, Median~\cite{Yin18}, and FLTrust~\cite{cao2020fltrust}. 
Moreover, we use \emph{certified accuracy} as our evaluation metric, which is a lower bound of the test accuracy that a method can provably achieve no matter what poisoning attacks the malicious clients use. For instance, our FLCert-D with FedAvg and $N=500$ can achieve a certified accuracy of 81\% on MNIST when evenly distributing the training examples among 1,000 clients and 100 of them are malicious.

In summary, our key contributions are as follows:
\begin{itemize}
\item We propose FLCert, an FL framework with provable security guarantees against poisoning attacks. Specifically, FLCert-P provides probabilistic security guarantee while FLCert-D provides deterministic guarantee. Moreover, we prove that our derived certified security level is tight.
\item  We propose a randomized algorithm to compute our certified security level for FLCert-P.
\item  We evaluate FLCert on multiple datasets from different domains and multiple base FL algorithms. Our results show that FLCert is secure against poisoning attacks both theoretically and empirically.
\end{itemize}

All proofs appear in the Appendix. 

\section{Related Work}

\subsection{Federated Learning (FL)}
Suppose we have $n$ clients $\mathbf{C}=\{C_1, C_2,\cdots, C_n\}$ and a server, where each client $C_i$ ($i=1, 2, \cdots, n$) holds a local training dataset $D_i$. Each client also has a unique user ID assigned when the client registers in the FL system. 
These clients aim to use their local training datasets to collaboratively learn a model that is shared among all  clients, with the help of the  server. We call such shared model  \emph{global model}. 
For simplicity, we use $\bm{w}$ to denote the model parameters of the global model. The parameters $\bm{w}$ are often learnt by solving the following optimization problem
    $\bm{w} = \argmin_{\bm{w}^{\prime}} \sum_{i=1}^n \ell(D_i;\bm{w}^{\prime})$,
where $\ell$ is a loss function, and $\ell(D_i;\bm{w}^{\prime})$ is the empirical loss on the local training dataset $D_i$ of the $i$th client.

The clients and the server collaborate to solve the optimization problem iteratively. Specifically, in each iteration of the training process, the following three steps are performed: 1) the server broadcasts the current global model to (a subset of) clients; 2) the clients initialize their local models as the received global model, train the local models using stochastic gradient descent, and send the local model updates back to the server; 3) the server aggregates the local model updates and updates the global model. 

\subsection{Poisoning Attacks to FL}

Recent works \cite{bagdasaryan2020backdoor,bhagoji2019analyzing,fang2019local,cao2022mpaf,tolpegin2020data,shejwalkarmanipulating}  showed that FL is vulnerable to poisoning attacks. Based on the attacker's goal, poisoning attacks to FL can be grouped into two categories: untargeted poisoning attacks \cite{fang2019local,cao2022mpaf,tolpegin2020data,shejwalkarmanipulating} and  targeted poisoning attacks \cite{bagdasaryan2020backdoor,bhagoji2019analyzing}.

\myparatight{Untargeted poisoning attacks} The goal of untargeted poisoning attacks is to decrease the test accuracy of the learnt global model. The attacks aim to increase the indiscriminate test error rate of the global model as much as possible, which can be considered as Denial-of-Service (DoS) attacks. Depending on how the attack is performed, untargeted poisoning attacks have two variants, i.e., \emph{data poisoning attacks} \cite{tolpegin2020data} and \emph{local model poisoning attacks} \cite{fang2019local,cao2022mpaf}. Data poisoning attacks tamper with the local training data on the malicious clients while assuming the computation process maintains integrity. They  inject fake training data points, delete existing training data points, and/or modify existing training data points. Local model poisoning attacks tamper with the computation process on the malicious clients, i.e., they directly tamper with the malicious clients' local models or model updates sent to the server. Note that in FL, any data poisoning attack can be implemented by a local model poisoning attack. This is because on each malicious client, we can always treat the local model trained using the tampered local training dataset as the tampered local model.

\myparatight{Targeted poisoning attacks} Targeted poisoning attacks aim to force the global model to predict target labels for target test inputs, while its performance on non-target test inputs is unaffected. The target test inputs in a targeted poisoning attack can be  a specific test input \cite{bhagoji2019analyzing}, some test inputs with certain properties \cite{bagdasaryan2020backdoor}, or test inputs with a particular trigger embeded \cite{bagdasaryan2020backdoor}. A targeted poisoning attack is also known as a backdoor attack if its target test inputs are trigger-embedded test inputs. A backdoor attack aims to corrupt the global model such that it predicts the attacker-chosen target label for any test input embedded with the predefined trigger. 

\subsection{Defenses against Poisoning Attacks to FL}

Many defenses \cite{chen2021pois,wen2021great,jia2020certified,steinhardt2017certified,liu2018fine,jia2021intrinsic,wang2020certifying,jia2022certified,wang2021certified,levine2020deep} have been proposed against data poisoning attacks in centralized learning scenarios. However, they are insufficient to defend against poisoning attacks to FL. In particular, the empirical defenses \cite{chen2021pois,wen2021great,liu2018fine} require knowledge about the training dataset, which is usually not available to the FL server for privacy concerns. Moreover, 
the provably secure defenses \cite{wang2020certifying,jia2020certified,steinhardt2017certified,jia2021intrinsic,jia2022certified,wang2021certified,levine2020deep} can guarantee that the label predicted for a test input is unaffected by a bounded number of poisoned training
examples. However, in FL, a single malicious client can poison an
arbitrary number of its local training examples, breaking their
assumption.

Byzantine-robust FL methods \cite{Blanchard17,Yin18,cao2020fltrust} leverage Byzantine-robust aggregation rules to resist poisoning attacks. They share the idea of alleviating the impact of statistical outliers caused by poisoning attacks when aggregating the local model updates. For instance, Krum \cite{Blanchard17} selects a single model update that has the smallest square-Euclidean-distance score as the new global model update; Trimmed-mean \cite{Yin18} computes the coordinate-wise mean of the model update parameters after trim and updates the global model using corresponding mean values; Median \cite{Yin18} updates the global model by computing the coordinate-wise median of the local model updates; FLTrust \cite{cao2020fltrust} leverages a small clean dataset to compute a server update and uses it as a baseline to bootstrap trust to local model updates. These methods all suffer from a key limitation: they cannot provide provable security guarantees, i.e., they cannot ensure that the predicted labels of the global model for testing inputs remain unchanged when there exists a poisoning attack.

Another type of defenses  focused on detecting malicious clients and removing their local models before the aggregation \cite{shen2016auror,fang2019local,zhang2022fldetector}. Shen et al. \cite{shen2016auror} proposed to perform clustering on local models to detect  malicious clients. Fang et al. \cite{fang2019local} proposed two defenses that use a validation dataset to reject potentially malicious local models based on their impact on the error rate or the loss value evaluated on the validation dataset. Zhang et al.~\cite{zhang2022fldetector} proposed to detect malicious clients via checking their model-updates consistency.  These defenses showed some empirical effectiveness in detecting malicious clients. However, they cannot provide provable security guarantees, either. 

Wang et al.~\cite{wang2020certifying} proposed a certified defense against backdoor attacks. A recent work \cite{xie2021crfl} extended this method to FL and called it CRFL. CRFL aims to certify robustness against a particular backdoor attack~\cite{bagdasaryan2020backdoor}, where all malicious clients train their local models using backdoored local training datasets, and scale their model updates before sending them to the server simultaneously in one iteration of FL. They showed that the accuracy of the learned global model under such specific attack could be certified if \emph{the magnitude of the change in malicious clients' local training data is bounded}. However, a malicious client can arbitrarily change its local training data. Therefore,  a single malicious client can break their defense by using poisoning attacks other than the considered particular backdoor attack. On the contrary, our FLCert is provably secure \emph{no matter what attacks the malicious clients perform}.

\begin{figure*}[!t]
    \center
    \captionsetup{justification=centering}
    \includegraphics[width=0.99\textwidth]{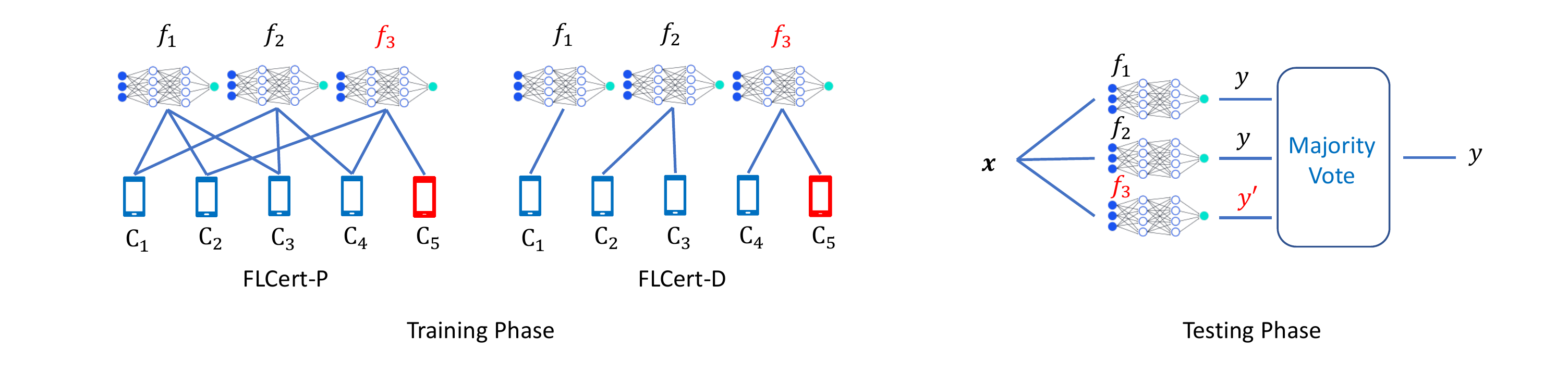}
    \caption{Illustration of FLCert.}
    \label{fig:illustration}
\end{figure*}

\section{Proposed FLCert}
\label{sec:method}

\subsection{Overview}

Figure~\ref{fig:illustration} illustrates FLCert, where there are $n=5$ clients and $N=3$ groups. In FLCert-P, each group contains $k=3$ clients randomly sampled from the $n$ clients, while in FLCert-D, the clients are divided into $N$ disjoint groups. We denote the $N$ groups as $\mathbf{G}_1,\mathbf{G}_2,\cdots,\mathbf{G}_N$.   
We learn a global model for each group of clients using a determinized base FL algorithm. We determinize the base FL algorithm such that we can derive the provable security of FLCert. Since we have $N$ groups, we learn $N$ global models in total. We ensemble the $N$ global models to predict labels for test inputs. Specifically, we take a majority vote among the $N$ global models to predict the label of a test input.

\subsection{Grouping the Clients}
\label{sec:grouping}
In FLCert, the clients are assigned to $N$ groups. 
Considering the provable security and communication/computation overhead, the grouping should satisfy two constraints: 1) a malicious client should influence a small number of groups, which enables FLCert to be secure against more malicious clients, and 2) a client should belong to a small number of groups, which reduces the communication and computation overhead for clients. 
To satisfy the two constraints, we propose two ways to divide the clients, which corresponds to two variants of FLCert, i.e., FLCert-P and FLCert-D.  
Specifically, in FLCert-P, we randomly sample $k$ clients as a group. In FLCert-D, we divide the $n$ clients into $N$ \emph{disjoint} groups using hash values of their user IDs. 

\subsubsection{FLCert-P}
In FLCert-P, the server samples $k$ clients uniformly at random as a group, which results in ${n \choose k}$ possible groups in total. When ${n \choose k}$ is small, we can obtain all possible groups and train one global model for each group. In this case, each client belongs to ${{n-1} \choose {k-1}}$ groups. However, when ${n \choose k}$ is large, we may not be able to train all global models for every possible group. Therefore, in practice, we sample $N\ll{n \choose k}$ groups instead, and each client is expected to belong to a small number (i.e., $\frac{kN}{n}$) of groups. A malicious client can only affect the groups it belongs to. 

\subsubsection{FLCert-D}
In FLCert-D, the server assigns each client to a group by hashing the client's user ID. Therefore, each client belongs to only one group and a malicious client can only influence the group it belongs to. We use clients' user IDs because they cannot be changed after registration, which means that malicious clients cannot change which groups they belong to. 
Formally, we use a hash function with output range $[1,N]$ to compute the hash values of the clients' user IDs; and the clients with hash value $i$ form the $i$th group, where $i=1,2,\cdots,N$.

\subsection{Ensemble Global Model}

After the server assigns the clients into $N$ groups, each group learns a global model using a determinized base FL algorithm $f$. 
In particular, we  determinize a base FL algorithm via fixing the seed of the pseudo-random number generator used by the algorithm $f$. 
We denote by $f_g(\mathbf{G}_g)$ the global model for the $g$th group.  Note that FLCert allows different groups to use different determinized base FL algorithms, e.g., different groups may use the same base FL algorithm with different fixed seeds for the pseudo-random number generator, and different groups may use different base FL algorithms. 

Since there are $N$ groups, we have $N$  global models $f_1(\mathbf{G}_1), f_2(\mathbf{G}_2), \cdots, f_N(\mathbf{G}_N)$ in total. Our FLCert ensembles the $N$ global models to predict labels for test inputs. Specifically, given a test input $\bm{x}$, we first use each global model to predict its label, i.e., $f_g(\mathbf{G}_g, \bm{x})$ is the label predicted by the $g$th global model. 
Then, we compute \emph{label frequency} $n_j(\bm{x})$ for each label $j$, which is the number of global models that predict  $j$ for $\bm{x}$.  Formally, $n_j(\bm{x})$ is defined as follows:
\begin{align}
    n_j(\bm{x}) = \sum_{g\in[1,N]}\mathbbm{1}_{f_g(\mathbf{G}_g,\bm{x})=j}, 
\end{align}
where $\mathbbm{1}$ is the indicator function and $\mathbbm{1}_{f_g(\mathbf{G}_g,\bm{x})=j}=1$ if  $f_g(\mathbf{G}_g,\bm{x})=j$, otherwise $\mathbbm{1}_{f_g(\mathbf{G}_g,\bm{x})=j}=0$. The sum of the label frequencies is $N$. Moreover, we define the \emph{label probability} of label $j$ as $p_j(\bm{x}) = \frac{n_j(\bm{x})}{N}$. Our FLCert takes a majority vote among the $N$ global models to predict the label for $\bm{x}$, i.e., FLCert predicts the label with the largest label frequency/probability for $\bm{x}$. For convenience, we use label probability for FLCert-P and label frequency for FLCert-D in the rest of this paper. We denote our ensemble learning based FLCert algorithm as $F$. The ensemble of the $N$ global models is called \emph{ensemble global model}. Moreover, we denote by $F(\mathbf{C}, \bm{x})$ the label predicted for $\bm{x}$ by our ensemble global model learnt by $F$ on the clients $\mathbf{C}$. Formally, we have:
\begin{align}
\label{eq:define_F}
    F(\mathbf{C}, \bm{x}) = \argmax_{j} \;n_j(\bm{x}) = \argmax_{j} \;p_j(\bm{x}).
\end{align}
When there exist ties, i.e., multiple labels have the same largest label frequency/probability, we use different tie-breaking strategies for FLCert-P and FLCert-D. For FLCert-P, we randomly select a label with the same largest label probability. Such random tie-breaking strategy works for FLCert-P because FLCert-P has probabilistic security guarantees anyway. However, such random tie-breaking strategy invalidates the deterministic  security guarantees of FLCert-D due to the randomness. \xc{While any deterministic tie-breaking strategy can address this challenge, we adopt one that selects the label with the smallest class index as the predicted label in FLCert-D}. For instance, when $n_1(\bm{x})=n_2(\bm{x})>n_j(\bm{x}), \forall j\neq 1 \land j\neq 2$, we have $F(\mathbf{C}, \bm{x})=1$, where $\land$ means logical AND.

\section{Security Analysis}
\label{sec:securityanalysis}

We prove that the label predicted by FLCert for a test input is unaffected by a bounded number of malicious clients no matter what poisoning attacks they use.

\subsection{Certified Security Level} 
Recall that $\mathbf{C}$ is the set of $n$ clean clients. For convenience, we denote by $\mathbf{C}'$ the set of clients including the malicious ones. 
We define the \emph{{\certifiedradius}} $m^*$ of a test input $\bm{x}$ as the maximum number of malicious clients that FLCert can tolerate without predicting a different label for $\bm{x}$. 
Formally, $m^*$ is the largest integer $m$ that satisfies the following:
\begin{align}
    F(\mathbf{C}', \bm{x}) = F(\mathbf{C}, \bm{x}), \forall \mathbf{C}', \xc{|\mathbf{C}' - \mathbf{C}|}\leq m,
\end{align}
where $\xc{|\mathbf{C}' - \mathbf{C}|}$ is the number of malicious clients in $\mathbf{C}'$, compared to $\mathbf{C}$. Note that the {\certifiedradius} $m^*$ may be different for different test inputs. 

\subsection{Deriving Certified Security Level} Next, we derive the {\certifiedradius} $m^*$ for a test input $\bm{x}$. 
Suppose that when there are no malicious clients, FLCert predicts label $y$ for the test input $\bm{x}$, i.e.,  $y=F(\mathbf{C},\bm{x})=\argmax_{j} n_j(\bm{x})$ and $y$ has the largest label frequency. Moreover, we assume $z=\argmax_{j\neq y} n_j(\bm{x})$ is the label with the second largest label frequency. We denote by $p_y$ and $p_z$ respectively their label probabilities. Moreover, we denote by $n_y'$ and $n_z'$ respectively the label frequency, and $p_y'$ and $p_z'$ respectively the label probabilities for $y$ and $z$ in the ensemble global model when there are malicious clients.    

\subsubsection{FLCert-P}  
When ${n \choose k}$ is small (e.g., several hundred), we can create all possible groups and train $N={n \choose k}$ global models. 
Suppose $m$ clients become malicious. Then, $1 - \frac{{n-m \choose k}}{{n \choose k}}$ fraction of groups include at least one malicious client. In the worst-case scenario, for each global model learnt using a group including at least one malicious client, its predicted label for $\bm{x}$ changes from $y$ to $z$. 
Therefore, in the worst-case scenario, the $m$ malicious clients decrease the largest label probability $p_y$ by $1 - \frac{{n-m \choose k}}{{n \choose k}}$ and increase the second largest label probability $p_z$ by $1 - \frac{{n-m \choose k}}{{n \choose k}}$, i.e., we have $p_y'=p_y-(1 - \frac{{n-m \choose k}}{{n \choose k}})$ and $p_z'=p_z + (1 - \frac{{n-m \choose k}}{{n \choose k}})$. Our ensemble global model still predicts label $y$ for $\bm{x}$, i.e., $F(\mathbf{C}',\bm{x})=F(\mathbf{C},\bm{x})=y$, once $m$ satisfies the following inequality:
\begin{align}
\label{inequalitycondition}
p_y' > p_z' 
\Longleftrightarrow p_y-p_z > 2 - 2\frac{{n-m \choose k}}{{n \choose k}}.
\end{align}
In other words, the largest integer $m$ that satisfies the inequality (\ref{inequalitycondition}) is our certified security level $m^*$ for the test input $\bm{x}$. The inequality (\ref{inequalitycondition}) shows that our certified security level is related to the gap $p_y-p_z$ between the largest and second largest label probabilities in the ensemble global model trained on the clients $\mathbf{C}$. 
For instance, when a test input has a larger  gap $p_y-p_z$, the inequality (\ref{inequalitycondition}) may be satisfied by a larger $m$, which means that our ensemble global model may have a larger certified security level for the test input.

However, when ${n\choose k}$ is large,  it is computationally challenging to compute the exact label probabilities via training ${n\choose k}$ global models. For instance, when $n=100$ and $k=10$, there are already $1.73\times 10^{13}$ global models, training all of which is computationally intractable in practice. Therefore, we also derive certified security level using a lower bound $\underline{p_y}$ of $p_y$ (i.e., $\underline{p_y}\leq p_y$) and an upper bound $\overline{p}_z$  of $p_z$ (i.e., $\overline{p}_z\geq p_z$). We use a lower bound $\underline{p_y}$ of $p_y$ and an upper bound $\overline{p}_z$  of $p_z$ because our certified security level is related to the gap $p_y-p_z$ and we aim to estimate a lower bound of the gap. 

The lower bound $\underline{p_y}$ and upper bound $\overline{p}_z$ may be estimated by different methods. We propose a Monte Carlo algorithm to estimate a lower bound $\underline{p_y}$ and an upper bound $\overline{p}_z$  via only training $N$ of the ${n\choose k}$ global models. Specifically, we sample $N$ 
groups, each of which includes $k$ clients sampled from the $n$ clients uniformly at random, and we use them to train $N$ global models $f_1(\mathbf{G}_1), f_2(\mathbf{G}_2), \cdots, f_N(\mathbf{G}_N)$. We use the $N$ global models to predict labels for $\bm{x}$ and count the frequency of each label. We treat the label with the largest frequency as the predicted label $y$. Recall that, based on the definition of label probability, a global model trained on a random group with $k$ clients  predicts label $y$ for  $\bm{x}$ with the label probability $p_{y}$. Therefore, the frequency $N_{y}$ of the label $y$ among the $N$ global models follows a binomial distribution $B(N, p_{y})$ with parameters $N$ and $p_{y}$. Thus, given $N_{y}$ and $N$, we can use the standard one-sided Clopper-Pearson method~\cite{clopper1934use} to estimate a lower bound $\underline{p_{y}}$ of ${p_{y}}$ with a confidence level $1-\alpha$. Specifically, we have $\underline{p_{y}}=\mathcal{B}\left({\alpha};N_{y},N-N_{y}+1\right)$, where $\mathcal{B}(q; v,w)$ is the $q$th quantile from a beta distribution with shape parameters $v$ and $w$.
Moreover, we can estimate $\overline{p}_{y}=1-\underline{p_{y}}\geq 1-p_{y}\geq {p}_{z}$ as an upper bound of ${p}_{y}$.  

\begin{figure}[!t]
    \center
   {\includegraphics[width=0.4\textwidth]{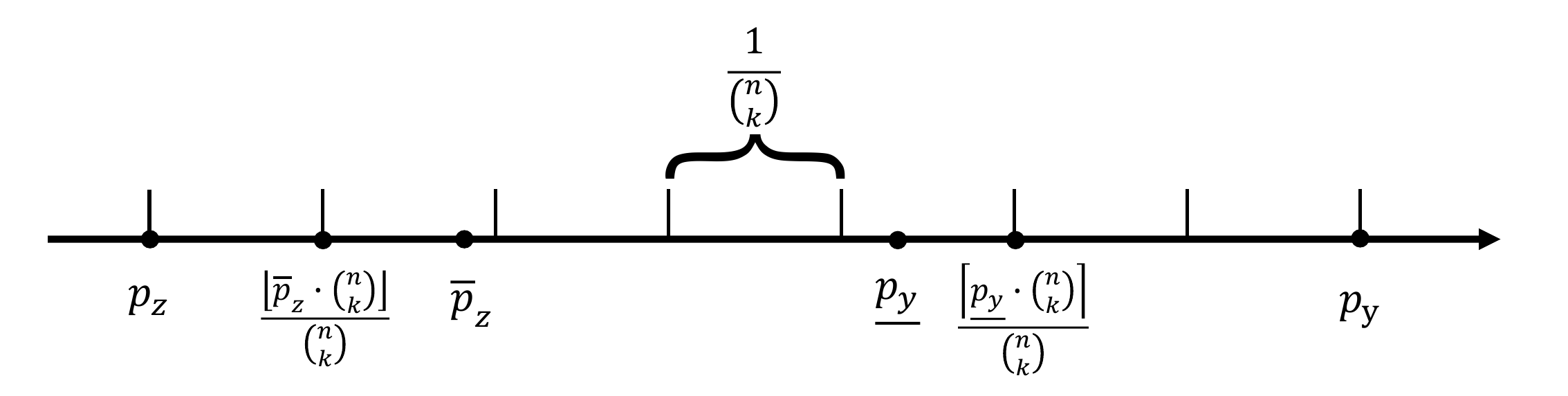}}  
   \vspace{-3mm}
    \caption{An example to illustrate the relationships between $p_y, \underline{p_y}$, and  $\frac{\left\lceil\underline{p_y} \cdot {{n \choose k}}\right\rceil}{{n \choose k}}$ as well as  $p_z, \overline{p}_z$, and $\frac{\left\lfloor\overline{p}_z \cdot {n \choose k}\right\rfloor}{{n\choose k}}$.}
    \label{fig:remark}
\vspace{-3mm}
\end{figure}

Next, we derive our certified security level based on the probability bounds $\underline{p_y}$ and $\overline{p}_z$. 
One way  is to replace $p_y$ and $p_z$ in inequality (\ref{inequalitycondition}) as $\underline{p_y}$ and $\overline{p}_z$, respectively. Formally, we have the following inequality: 
\begin{align}
\label{inequalitycondition1}
 \underline{p_y}-\overline{p}_z > 2 - 2\frac{{n-m \choose k}}{{n \choose k}}.
\end{align}
If an $m$ satisfies inequality (\ref{inequalitycondition1}), then the $m$ also satisfies inequality (\ref{inequalitycondition}), because  $\underline{p_y}-\overline{p}_z \leq p_y-p_z$. 
Therefore, we can find the largest integer $m$ that satisfies the inequality (\ref{inequalitycondition1}) as the certified security level $m^*$. However, we found that the certified security level $m^*$ derived based on inequality (\ref{inequalitycondition1}) is not tight, i.e., our ensemble global model may still predict label $y$ for $\bm{x}$ even if the number of malicious clients is larger than $m^*$ derived based on inequality (\ref{inequalitycondition1}). The key reason is that the label probabilities are integer multiplications of $\frac{1}{{n \choose k}}$. Therefore, we normalize $\underline{p_y}$ and $\overline{p}_z$ as integer multiplications of $\frac{1}{{n \choose k}}$ to derive a tight certified security level. Specifically, we derive the certified security level as the largest integer $m$ that satisfies the following inequality (formally described in Theorem~\ref{certified_radius_p}):
\begin{align}
\label{eq:certified_condition1}
\frac{\left\lceil\underline{p_y} \cdot {n \choose k}\right\rceil}{{n \choose k}} - \frac{\left\lfloor\overline{p}_z \cdot {n \choose k}\right\rfloor}{{n\choose k}}> 2 - 2\cdot \frac{{n-m \choose k}}{{n \choose k}}.
\end{align}
Figure~\ref{fig:remark} illustrates the relationships between $p_y, \underline{p_y}$, and  $\frac{\left\lceil\underline{p_y} \cdot {{n \choose k}}\right\rceil}{{n \choose k}}$ as well as  $p_z, \overline{p}_z$, and $\frac{\left\lfloor\overline{p}_z \cdot {n \choose k}\right\rfloor}{{n\choose k}}$. 
When an $m$ satisfies  inequality (\ref{inequalitycondition1}), the $m$ also satisfies  inequality (\ref{eq:certified_condition1}), because $\underline{p_y}-\overline{p}_z \leq \frac{\left\lceil\underline{p_y} \cdot {{n \choose k}}\right\rceil}{{n \choose k}} - \frac{\left\lfloor\overline{p}_z \cdot {n \choose k}\right\rfloor}{{n\choose k}}$. Therefore, the certified security level derived based on inequality (\ref{inequalitycondition1}) is smaller than or equals the certified security level derived based on inequality (\ref{eq:certified_condition1}). 
Note that when $\underline{p_y}=p_y$ and $\overline{p}_z=p_z$, both (\ref{inequalitycondition1}) and (\ref{eq:certified_condition1}) reduce to (\ref{inequalitycondition}) as the label probabilities are integer multiplications of $\frac{1}{{n \choose k}}$.  
The following  theorem formally summarizes our certified security level.

\begin{algorithm}[tb]
   \caption{Computing Predicted Label and Certified Security Level in FLCert-P}
   \label{alg:certify}
\begin{algorithmic}[1]
   \STATE {\bfseries Input:} $\mathbf{C}$, $f$, $k$, $N$, $\mathcal{D}$, $\alpha$.
   \STATE {\bfseries Output:} Predicted label and certified security level for each test input in $\mathcal{D}$. \\
   \STATE $f_1(\mathbf{G}_1),\cdots,f_N(\mathbf{G}_N) \gets  \textsc{Sample\&Train}(\mathbf{C},f,k,N)$ \\
   \FOR{$\bm{x}_t$ {\bfseries in} $\mathcal{D}$}
   \STATE counts$[i] \gets \sum_{l=1}^{N}\mathbb{I}(f_{l}(\mathbf{G}_l, \bm{x}_t)=i), i\in \{1,2,\cdots,L\} $   
   \STATE /* $\mathbb{I}$ is the indicator function */\\ 
   \STATE $\hat{y}_t\gets$  index of the largest entry in counts \\
   \STATE $\underline{p_{\hat{y}_t}}\gets \mathcal{B}\left(\frac{\alpha}{|\mathcal{D}|};N_{\hat{y}_t},N-N_{\hat{y}_t}+1\right)$ \\
   \STATE $ \overline{p}_{\hat{z}_t} \gets 1-\underline{p_{\hat{y}_t}}$
   \IF{$\underline{p_{\hat{y}_t}} > \overline{p}_{\hat{z}_t}$}
   \STATE $\hat{m}_t^* \gets \textsc{SearchLevel} (\underline{p_{\hat{y}_t}}, \overline{p}_{\hat{z}_t}, k, |\mathbf{C}|)$ \\
   \ELSE
   \STATE $\hat{y}_t \gets \text{ABSTAIN}$, $\hat{m}_t^* \gets \text{ABSTAIN}$
   \ENDIF
   \ENDFOR
  \STATE \textbf{return} $\hat{y}_1,\hat{y}_2,\cdots, \hat{y}_d$ and $\hat{m}_1^*,\hat{m}_2^*,\cdots, \hat{m}_d^*$
\end{algorithmic}
\end{algorithm}

\begin{theorem}
\label{certified_radius_p}
Given $n$ clients $\mathbf{C}$, an arbitrary base federated learning algorithm $f$,  a group size $k$, and a test input $\bm{x}$, we define an ensemble global model $F$ as Equation (\ref{eq:define_F}). $y$ and $z$ are the labels that have the largest and second largest label probabilities for $\bm{x}$ in the ensemble global model. $\underline{p_y}$ is a lower bound of $p_y$ and $\overline{p}_z$ is an upper bound of $p_z$. Formally, $\underline{p_y}$ and $\overline{p}_z$ satisfy the following conditions:
\begin{equation}
\label{eq:prob_condition}
    \max_{i\neq y} p_i = p_z \le \overline{p}_z \le \underline{p_y} \le p_y.
\end{equation} 
Then, $F$ provably predicts $y$ for  $\bm{x}$ when at most $m^*$ clients in $\mathbf{C}$ become malicious, i.e., we have:
{
\begin{align}
      \;F(\mathbf{C'},  \bm{x}) =F(\mathbf{C}, \bm{x}) =y, \forall \mathbf{C}', \xc{|\mathbf{C}' - \mathbf{C}|} \leq m^*,
\end{align}
}
where $m^*$ is the largest integer $m$ ($0 \le  m \le n-k$) that satisfies inequality (\ref{eq:certified_condition1}). 
\end{theorem}

Our Theorem~\ref{certified_radius_p} is applicable to any base federated learning algorithm, any  lower bound $\underline{p_y}$ of $p_y$ and any upper bound $\overline{p}_z$ of $p_z$ that satisfy (\ref{eq:prob_condition}). When the lower bound $\underline{p_y}$ and upper bound $\overline{p}_z$ are estimated more accurately, i.e., $\underline{p_y}$ and $\overline{p}_z$ are respectively closer to $p_y$ and $p_z$, our certified security level may be larger. 
The following theorem shows that our derived certified security level is tight, i.e., when no assumptions on the base federated learning algorithm are made, it is impossible to derive a certified security level that is larger than ours for the given probability bounds $\underline{p_y}$ and $\overline{p}_z$.

\begin{theorem}
\label{tightness_theorem}
Suppose $\underline{p_y} + \overline{p}_z \le 1$. 
For any $\mathbf{C'}$ satisfying $\xc{|\mathbf{C}' - \mathbf{C}|}>m^*$, i.e., at least $m^*+1$ clients are malicious, 
there exists a base federated learning algorithm $f^*$ that satisfies (\ref{eq:prob_condition}) but $F(\mathbf{C'}, \bm{x}) \neq y$ or there exist ties.
\end{theorem}

Given a test set $\mathcal{D}$, Algorithm~\ref{alg:certify} shows our algorithm to compute the predicted labels and certified security levels for all test inputs in $\mathcal{D}$. The function \textsc{Sample\&Train} randomly samples $N$ groups with $k$ clients and trains $N$ global models using the base federated learning algorithm $f$. \xc{We use $1-\underline{p_{\hat{y}_t}}$ as an upper bound for $p_{\hat{z}_t}$ because we want to limit the probability of incorrect bound pairs $(\overline{p}_{\hat{z}_t},\underline{p_{\hat{y}_t}})$ within $\alpha/|\mathcal{D}|$}.
Given the probability bounds $\underline{p_{\hat{y}_t}}$ and $\overline{p}_{\hat{z}_t}$ for a test input $\bm{x}_t$, the function \textsc{SearchLevel} finds the certified security level $\hat{m}_t^*$  via finding the largest integer $m$ that satisfies (\ref{eq:certified_condition1}).  For example, \textsc{SearchLevel} can simply start $m$ from 0 and iteratively increase it by one until finding $\hat{m}_t^*$. 

In Algorithm~\ref{alg:certify}, since we estimate the lower bound $\underline{p_{\hat{y}_t}}$ using the Clopper-Pearson method, there is a probability that the estimated lower bound is incorrect, i.e., $\underline{p_{\hat{y}_t}} > p_{\hat{y}_t}$. When the lower bound is estimated incorrectly for a test input $\bm{x}_t$, the certified security level $\hat{m}_t^*$ output by Algorithm~\ref{alg:certify} for $\bm{x}_t$ may also be incorrect, i.e., there may exist an $\mathbf{C}'$ such that $\xc{|\mathbf{C}' - \mathbf{C}|}\leq \hat{m}_t^*$ but $F(\mathbf{C}',\bm{x}_t)\neq \hat{y}_t$. In other words, our Algorithm~\ref{alg:certify} has probabilistic guarantees for its output certified security levels. 
However, in the following theorem, we prove  
the probability that Algorithm~\ref{alg:certify}  returns an incorrect certified security level for at least one test input is at most $\alpha$. 
\begin{theorem}
\label{probability_of_certify}
The probability that Algorithm~\ref{alg:certify} returns an incorrect certified security level for at least one testing example in $\mathcal{D}$ is bounded by $\alpha$, which is equivalent to: 
{\small
\begin{align}
     \text{Pr}&(\cap_{\bm{x}_t \in \mathcal{D}} (h(\mathbf{C}',\bm{x}_t)=\hat{y}_t, \forall \mathbf{C}', M(\mathbf{C}')\leq \hat{m}_t^*|\hat{y}_t\neq \text{ABSTAIN})) \nonumber\\&\geq 1 -\alpha.
\end{align}
}
\end{theorem}

Note that when the probability bounds are estimated deterministically, e.g., when ${n \choose k}$ is small and the exact label probabilities can be computed  via training $N={n \choose k}$ global models, the certified security level obtained from our Theorem~\ref{certified_radius_p} is also deterministic. \xc{When there are many clients or $N$ is too small, it is possible that not all clients participate in training the $N$ global models. However, our probabilistic guarantee still holds in this case.}

\subsubsection{FLCert-D}
FLCert-D still predicts label $y$ for $\bm{x}$ after attacks if the following condition is satisfied: 
  \begin{align}
     n_y'(\bm{x}) > n_z'(\bm{x}) \text{ or } (n'_y(\bm{x}) = n'_z(\bm{x}) \land y<z),
    \label{eq:csl}
\end{align}
where FLCert-D still predicts label $y$ when $(n'_y(\bm{x}) = n'_z(\bm{x}) \land y<z)$ holds because of our tie-breaking strategy.  Next, we derive a lower bound of $n'_y(\bm{x})$ and an upper bound of $n'_z(\bm{x})$, which depend on $m$, the number of malicious clients in $\mathbf{C}'$. The largest $m$, for which the lower bound of $n'_y(\bm{x})$ and the upper bound of $n'_z(\bm{x})$ satisfy Equation (\ref{eq:csl}), is our {\certifiedradius} $m^*$ for $\bm{x}$. 
 
If a group $\mathbf{G}_g'$ ($g=1,2,\cdots, N$) after attack does not include malicious clients, then we know that $\mathbf{G}_g'$ and $\mathbf{G}_g$ include the same clients, i.e., $\mathbf{G}_g'=\mathbf{G}_g$. Moreover, since our base FL algorithm for each group is determinized, the global models learnt for $\mathbf{G}_g'$ and $\mathbf{G}_g$ predict the same label for $\bm{x}$, i.e., we have $f_g(\mathbf{G}_g', \bm{x})=f_g(\mathbf{G}_g, \bm{x})$. If a group $\mathbf{G}_g'$ includes at least one malicious client, the global model learnt for this group changes its predicted label for $\bm{x}$ from $y$ to $z$ after attack in the worst case. In other words, when a group includes malicious clients, the label frequency $n_y'(\bm{x})$ decreases by 1 and the label frequency $n_z'(\bm{x})$ increases by 1. According to our grouping strategy, $m$ malicious clients influence at most $m$ groups. Therefore, we have the following bounds in the worst case: 
\begin{align}
    n_y'(\bm{x}) \geq n_y(\bm{x}) - m,\  n_z'(\bm{x}) \leq n_z(\bm{x}) + m.
\end{align}
 The condition in Equation (\ref{eq:csl}) is satisfied, i.e., $F(\mathbf{C}', \bm{x}) = F(\mathbf{C}, \bm{x})=y$,  when $m$ is bounded as follows:
\begin{align}
\label{eq:csl_fancy}
m\leq\frac{n_y(\bm{x}) - [n_z(\bm{x}) + \mathbbm{1}_{z < y}]}{2}.
\end{align} 
Therefore, we have our {\certifiedradius} $m^*=\left\lfloor{\frac{n_y(\bm{x}) - [n_z(\bm{x}) + \mathbbm{1}_{z < y}]}{2}}\right\rfloor$ for a test input $\bm{x}$. 
We summarize our provable security  of FLCert in the following theorem.

\begin{theorem}
\label{certified_radius_d}
Given $n$ clients $\mathbf{C}$ that are divided into $N$ groups by hashing their user IDs,  a determinized training algorithm for each group, and a test input $\bm{x}$.  
$y$ and $z$ are the labels that have the largest and second largest label frequencies for $\bm{x}$, where ties are broken by selecting the label with the smallest class index. 
Then, $F$ provably predicts label $y$ for  $\bm{x}$ when  at most $m^*$ clients become malicious, i.e., we have:
\begin{align}
      F(\mathbf{C}', \bm{x}) = F(\mathbf{C}, \bm{x})=y, \forall \mathbf{C}', \xc{|\mathbf{C}' - \mathbf{C}|}\leq m^*,
\end{align}
where $m^*=\left\lfloor{\frac{n_y(\bm{x}) - [n_z(\bm{x}) + \mathbbm{1}_{z < y}]}{2}}\right\rfloor$.
\end{theorem}

\section{Communication/Computation Cost Analysis}
\label{sec:communicationcost}
We compare the communication and computation cost of our FLCert with the conventional single-global-model setting in which the base FL algorithm is used to learn a single global model for all the clients. Note that, for each client, its communication and computation cost is linear to the number of global iterations  the client involves in. 
Therefore, in the following analysis, we focus on the average number of global iterations a client involves in for our FLCert and the single-global-model setting.

Suppose we are given a base FL algorithm $f$. In the single-global-model setting,  $f$ is used to learn a single global model with all the clients. 
Assume $f$ performs $T$ global iterations between the clients and server to learn the global model. 
In each global iteration, the server randomly selects $\beta$ fraction of the clients and sends the current global model to them; the selected clients update their local models and send the updated local models to the server; and the server aggregates the local models as a new global model. $\beta$ is often set to be smaller than 1 to save communication cost per global iteration. Therefore, the communication and computation cost for a client is $O(\beta T)$  on average in the single-global-model setting. 

In FLCert, we use a base FL algorithm to train a global model for each group of clients. 
When learning a global model for a group of clients, the server randomly selects $\beta_e$ fraction of the clients in the group in  each global iteration. 
Assume each global model is learnt via $T_e$ global iterations between the corresponding clients and the server. 
 Each client is involved in $\frac{kN}{n}$ global models on average in FLCert-P, while each client is involved in only one global model in FLCert-D. Therefore, the communication and computation cost for a client is $O(\frac{kN\beta_eT_e}{n})$ for FLCert-P and  $O(\beta_eT_e)$ for FLCert-D. 
 
We note that each global model in our FLCert is learnt to fit local training data on a group of clients, while the global model in the single-global-model setting is learnt to fit local training data on all the clients. Therefore, learning each global model in our FLCert may require fewer global iterations, i.e.,  $T_e< T$. 
Moreover, when setting $kN=n$ and $T_e= \beta T /\beta_e$, our FLCert and the single-global-model setting have the same communication and computation cost for the clients.  
In other words, compared to the single-global-model setting, our FLCert can provide provable security against malicious clients without incurring additional communication and computation cost for the clients. 

\section{Evaluation}
\subsection{Experimental Setup}
\subsubsection{Datasets}
We use multiple datasets from different domains for evaluation, including three image classification datasets (MNIST-0.1, MNIST-0.5, and CIFAR-10), a human activity recognition dataset (HAR), and a next-word prediction dataset (Reddit). By default, we use MNIST-0.5 unless otherwise mentioned.

\xc{\myparatight{MNIST-0.1} %
We follow \cite{fang2019local} to distribute the training examples to the clients. 
In their work, a parameter called degree of Non-iid is proposed to control the distribution of data among clients, where a larger value indicates the data are further from independently and identically distributed (IID).
We set the degree of Non-iid to 0.1 in MNIST-0.1, which indicates IID data among clients.

\myparatight{MNIST-0.5} 
Clients in FL often have non-IID local training data. Therefore, in MNIST-0.5, we set the degree of Non-iid to 0.5 when distributing the training examples to clients to simulate non-IID local training data. }

\myparatight{CIFAR-10} 
Like MNIST-0.5, we set the degree of Non-iid to 0.5 when distributing training examples to clients to simulate  non-IID local training data. 

\myparatight{Human Activity Recognition (HAR)} HAR \cite{anguita2013public} is a real-world dataset consisting of human activity data collected from 30 users. The task is to predict a user's activity among six possible activities (e.g., WALKING, STANDING, and LAYING), given the sensor signal data collected from the user's smartphone. There are in total 10,299 examples in HAR dataset. We randomly select 75\% of each user's examples as the training dataset and use the rest of examples as the test dataset. In HAR, each user is considered as a client naturally.

\myparatight{Reddit} Reddit \cite{bagdasaryan2020backdoor} is a next-word prediction dataset consisting of posts collected from Reddit  in a randomly chosen month (November 2017). 
Given a sequence of words, the task is to predict which word will appear next. For Reddit dataset, each Reddit user is a client. There are 80,000 users in total and each user has  247 posts  on average.

\subsubsection{Evaluated Base FL Algorithms}

FLCert can use any base FL algorithm. To show such generality, we evaluate multiple popular base FL algorithms, including FedAvg~\cite{McMahan17}, Krum \cite{Blanchard17}, Trimmed-mean \cite{Yin18}, Median \cite{Yin18}, and FLTrust \cite{cao2020fltrust}. 
These base FL algorithms essentially use different aggregation rules to aggregate the clients' local model updates and update the global model in each global iteration. For FLTrust, we assume the server has a small clean dataset with 100 training examples randomly sampled from the training dataset, as suggested by the authors \cite{cao2020fltrust}. By default, we use FedAvg in our experiments unless otherwise mentioned.

\subsubsection{Global Model Architectures}
To show the generality of our FLCert, 
we use different neural network architectures as the global models on different datasets. Specifically, we borrow the CNN from \cite{cao2020provably} for MNIST-0.1 and MNIST-0.5. For CIFAR-10, we use the popular ResNet20 \cite{he2016deep} architecture. For HAR, we use a fully connected neural network with two hidden layers, each of which includes 256 neurons and uses ReLU activation. For Reddit, we follow \cite{inan2017tying} to use a 2-layer LSTM as global model.

\subsubsection{Evaluation Metrics} We use \emph{certified accuracy (CA)} to evaluate the provable security of FLCert against poisoning attacks. Given a test dataset and a number of malicious clients $m$, we define the certified accuracy \emph{CA@$m$}  as the fraction of test inputs that 1) are correctly classified by the FL algorithm and 2) have certified security levels no smaller than $m$.  
For untargeted attacks, the test dataset includes the normal test examples. For backdoor attacks, the test dataset includes the test examples embedded with a pattern trigger. CA@$m$ is a lower bound of test accuracy that a method can achieve no matter what poisoning attacks the malicious clients use once there are at most $m$ of them.  

Moreover, for both FLCert and the single-global-model setting, we also use the standard \emph{test accuracy} and \emph{attack success rate} to evaluate the \emph{empirical} performance against an existing untargeted poisoning attack~\cite{fang2019local} and backdoor attack \cite{bagdasaryan2020backdoor}, respectively. Attack success rate is the fraction of trigger-embedded test inputs that are classified as the attacker-chosen target label. We note that CA@0  reduces to the standard test accuracy when there are no attacks in deterministic scenarios, i.e., for FLCert-D and if we can train all ${n \choose k}$ global models for FLCert-P.  However, when ${n \choose k}$ is too large and we  sample $N$ groups of clients in FLCert-P, the empirical test accuracy may be different from CA@0.

\begin{figure*}[!t]
    \center
    \vspace{-1mm}
    \subfloat[MNIST-0.1]{\includegraphics[width=0.195\textwidth]{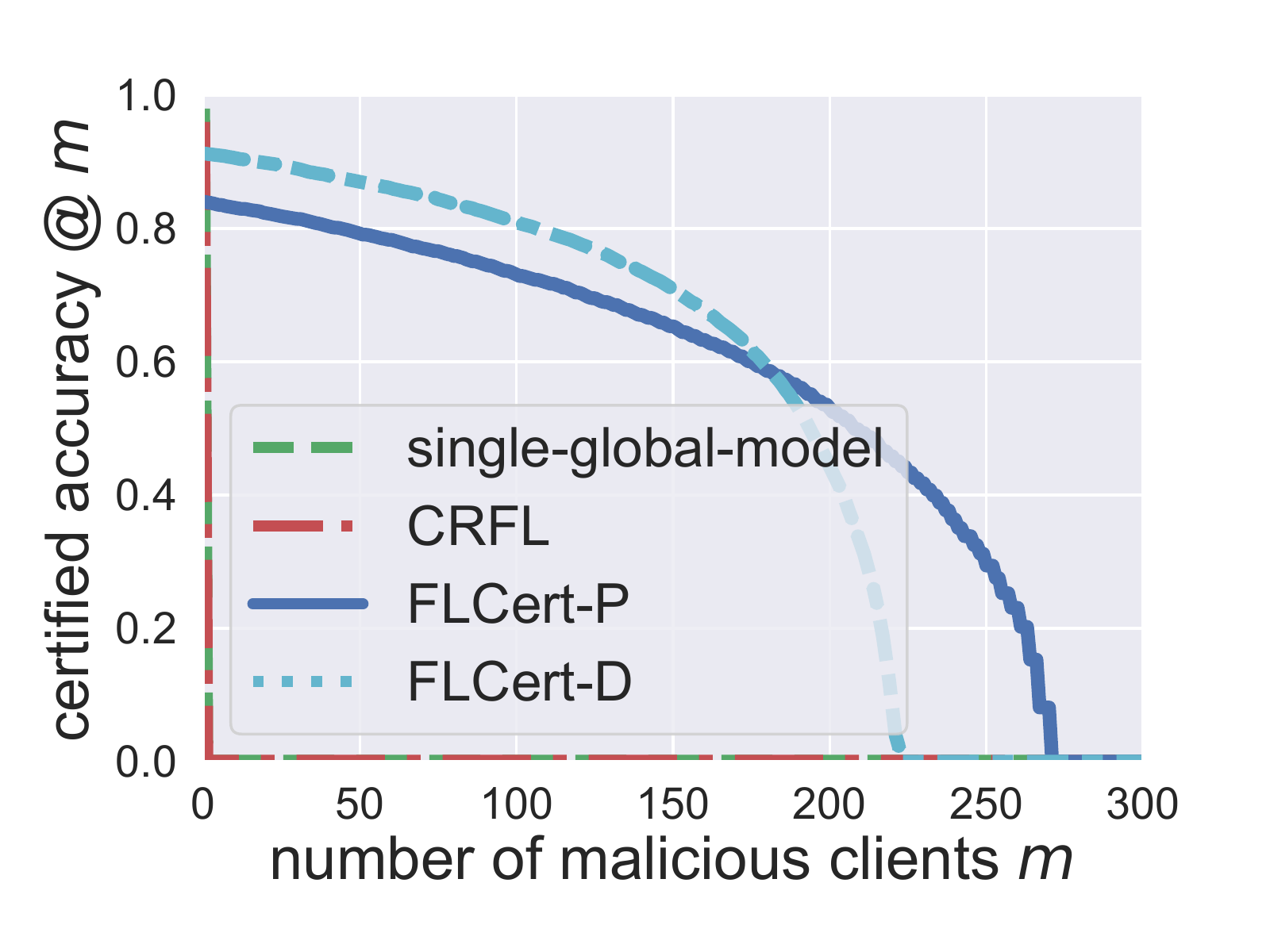}\label{fig:mnist0.1_mean_compare}}
    \subfloat[MNIST-0.5]{\includegraphics[width=0.195\textwidth]{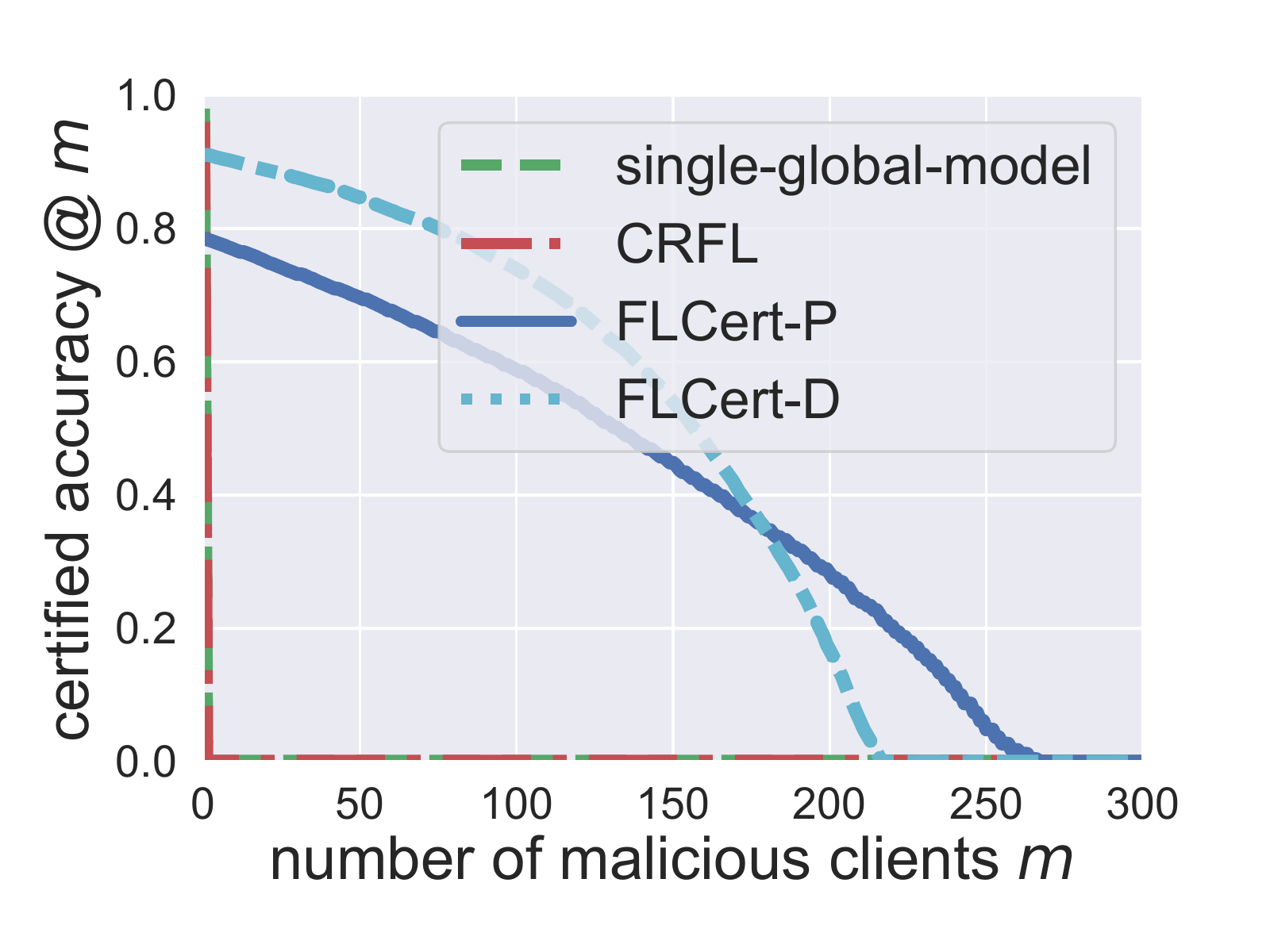}\label{fig:mnist0.5_mean_compare}}
    \subfloat[CIFAR-10]{\includegraphics[width=0.195\textwidth]{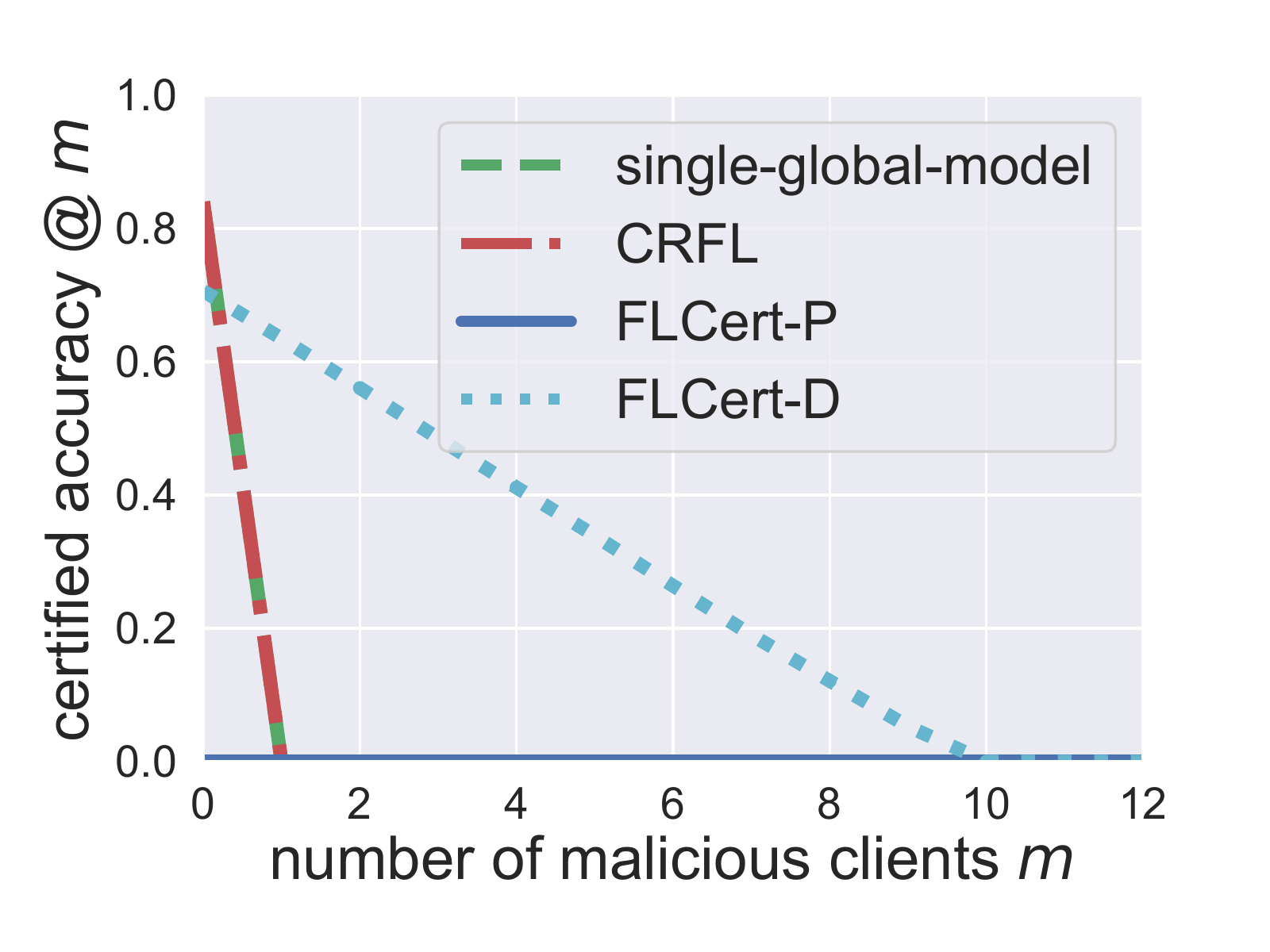}\label{fig:cifar_mean_compare}}
    \subfloat[HAR]{\includegraphics[width=0.195\textwidth]{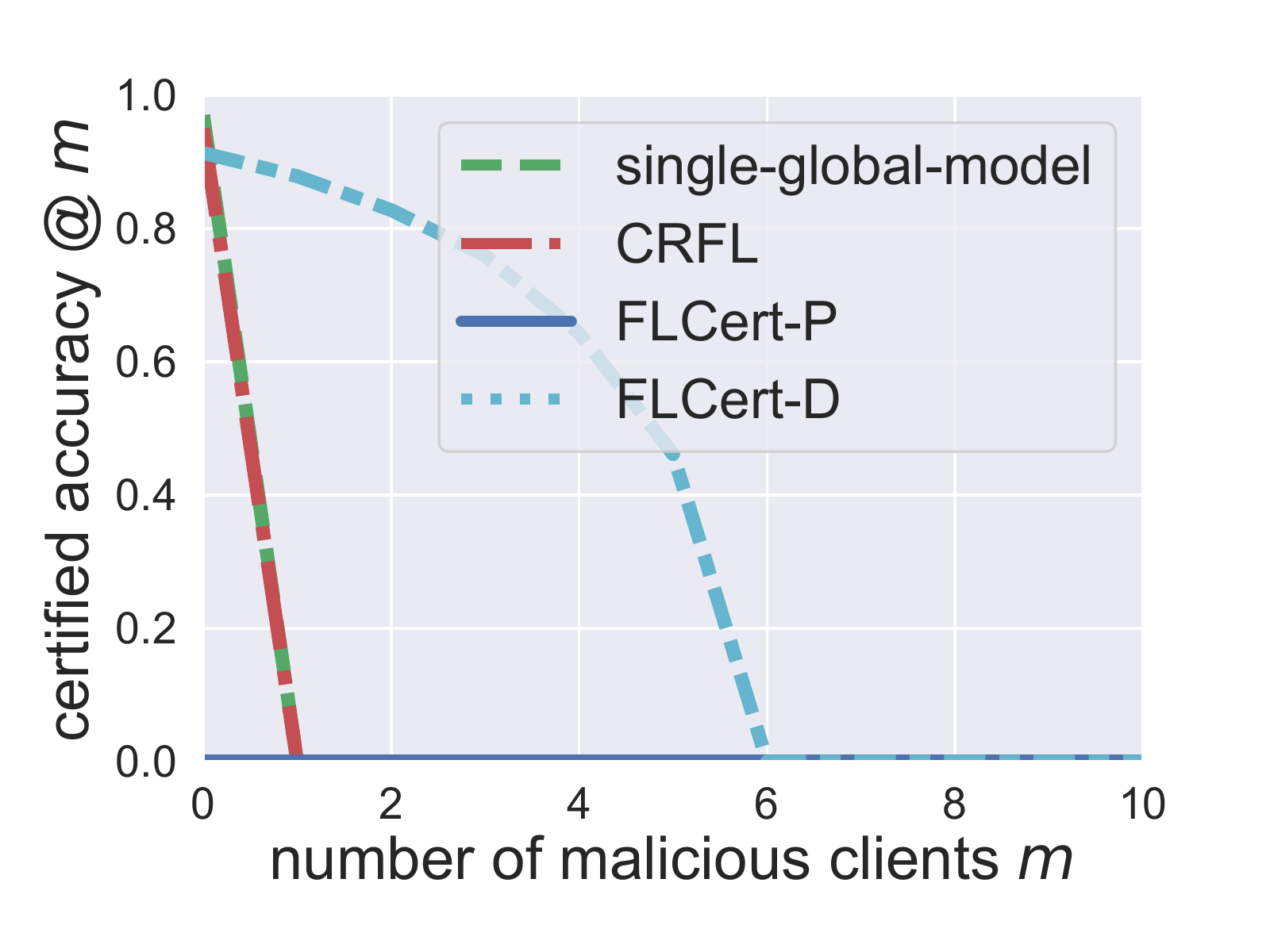}\label{fig:har_mean_compare}}
    \subfloat[Reddit]{\includegraphics[width=0.195\textwidth]{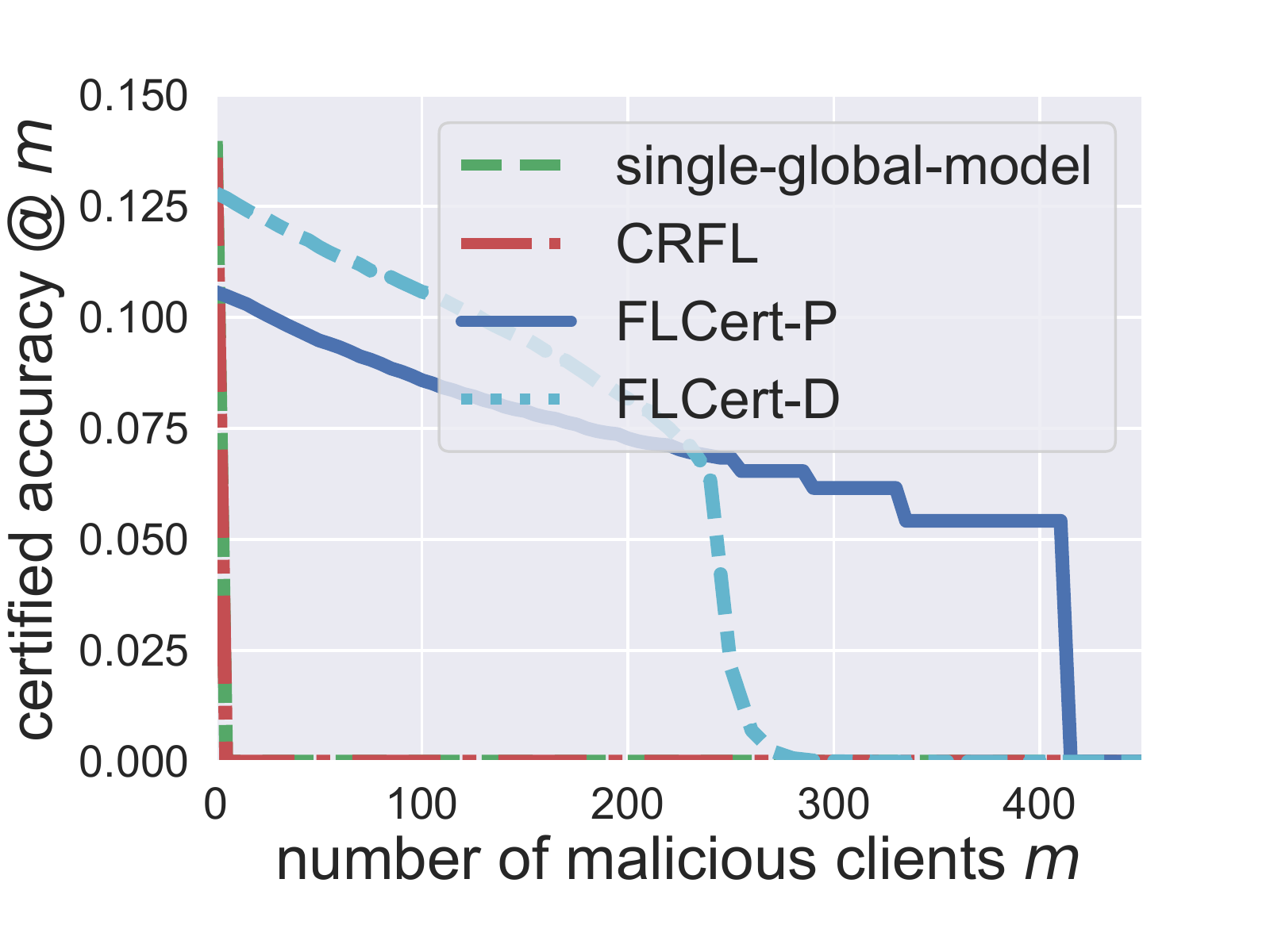}\label{fig:reddit_mean_compare}}
    \caption{Comparing FLCert with single-global-model and CRFL on different datasets when FedAvg is the base FL algorithm. }
    \label{fig:comparison}
 \vspace{-2mm}
\end{figure*}

\subsubsection{\xc{Existing} Poisoning Attacks} To calculate the standard {test accuracy} and {attack success rate}, we need \xc{existing} poisoning attacks. For untargeted attacks, we use the full-knowledge attacks proposed by Fang et al.~\cite{fang2019local}, i.e., Krum attack for Krum, and Trim attack for Trimmed-mean, Median, and FLTrust. For FedAvg, since a single malicious client can arbitrarily manipulate the global model \cite{Yin18}, we use an attack that forces the aggregated model update to be 0. In other words, a global model will learn nothing from the training process if there is any malicious client in its training process. For backdoor attacks, we consider the same trigger as proposed by Gu et al. \cite{Gu17} and set 0 as the target label for MNIST-0.1 and MNIST-0.5. For CIFAR-10, we use the same pixel-pattern trigger and the target label ``bird" in \cite{bagdasaryan2020backdoor}. For HAR, we design our  trigger by setting a feature value to 0 for every 20 features and we choose ``WALKING\_UPSTAIRS" as the target label. For Reddit, we consider the text trigger ``pasta from Astoria is" and the target label ``delicious" as proposed in \cite{bagdasaryan2020backdoor}.

\subsubsection{Parameter Settings} 
We consider different numbers of clients for different datasets to show generality of FLCert. 
Specifically, we assume 1,000 clients for MNIST-0.1 and MNIST-0.5 datasets. For CIFAR-10 dataset, we assume 100 clients.  For HAR and Reddit, each user is considered as a client naturally, which results in 30 and 80,000 clients, respectively.  In FLCert-D, we randomly generate a 64-bit user ID for each client and use the Python built-in \emph{hash()} function to divide clients into $N$ groups based on user IDs for each dataset. In FLCert-P, we choose  $k=\frac{n}{N}$ such that the expected communication/computation cost for each client is the same as FLCert-D; and  we set $\alpha=0.001$, i.e., given a test dataset, FLCert-P outputs an incorrect certified security level for at least one test input in 1 out of 1,000 runs on average. Moreover, we use the same fixed seed for a base FL algorithm when learning the global models for different groups. 

In FLCert, unless otherwise mentioned, we set $\beta_e=1$ (i.e., the server selects all the clients in a group in each global iteration when learning a global model for the group) in all the datasets except Reddit as the groups are small in these datasets. For Reddit, we set $\beta_e=0.025$, i.e., 2.5\% of clients in a group are chosen in each global iteration when learning a global model for the group. A base FL algorithm uses 200, 200, 200, 200, and 1,000 global iterations when learning a global model in either FLCert or the single-global-model setting for MNIST-0.1, MNIST-0.5, CIFAR-10, HAR, and Reddit, respectively.   In each global iteration, each client trains its local model for 5, 5, 40, 5, and 12 local iterations  using  stochastic gradient descent with learning rates  0.001, 0.001, 0.01, 0.001, and 20 as well as  batch sizes  32, 32, 64, 32, and 64 for the five datasets, respectively. We set $N=500, 500, 20, 15$ and 500 for the five datasets, respectively. We  consider different parameters for different datasets because of their different data characteristics.  

\subsubsection{Equipment and Environment} 
\xc{We run our experiments on a server with Intel Xeon Gold 6230R Processor and 10 NVIDIA Quadro RTX 6000 GPUs. We train our models using the MXNet GPU framework. }

\subsection{Experimental Results}
\subsubsection{Certified Accuracy}

\begin{figure}[!t]
    \center
    \subfloat[FLCert-P]{\includegraphics[width=0.24\textwidth]{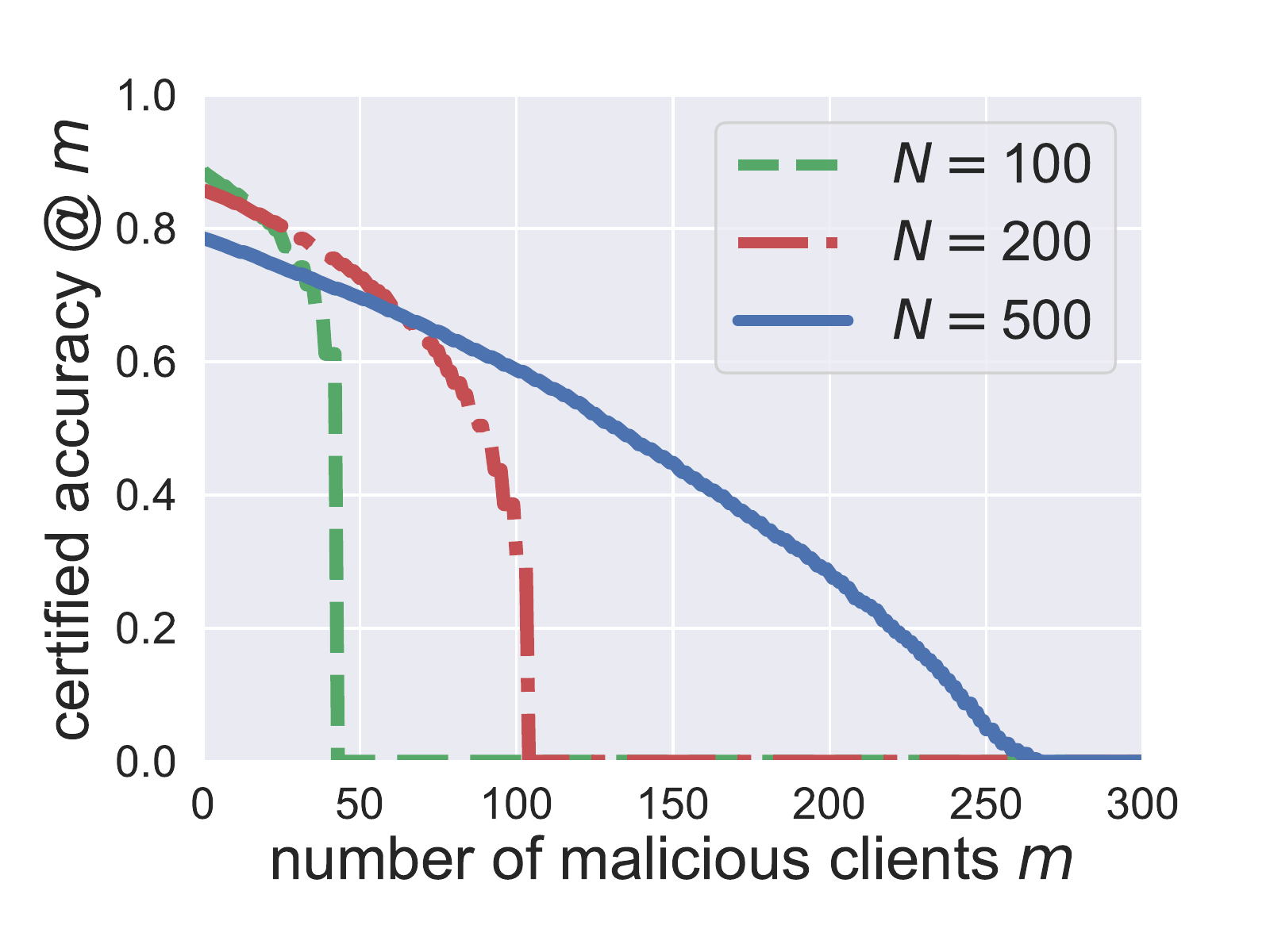}\label{fig:mnist0.5_p_mean_N}}
    \subfloat[FLCert-D]{\includegraphics[width=0.24\textwidth]{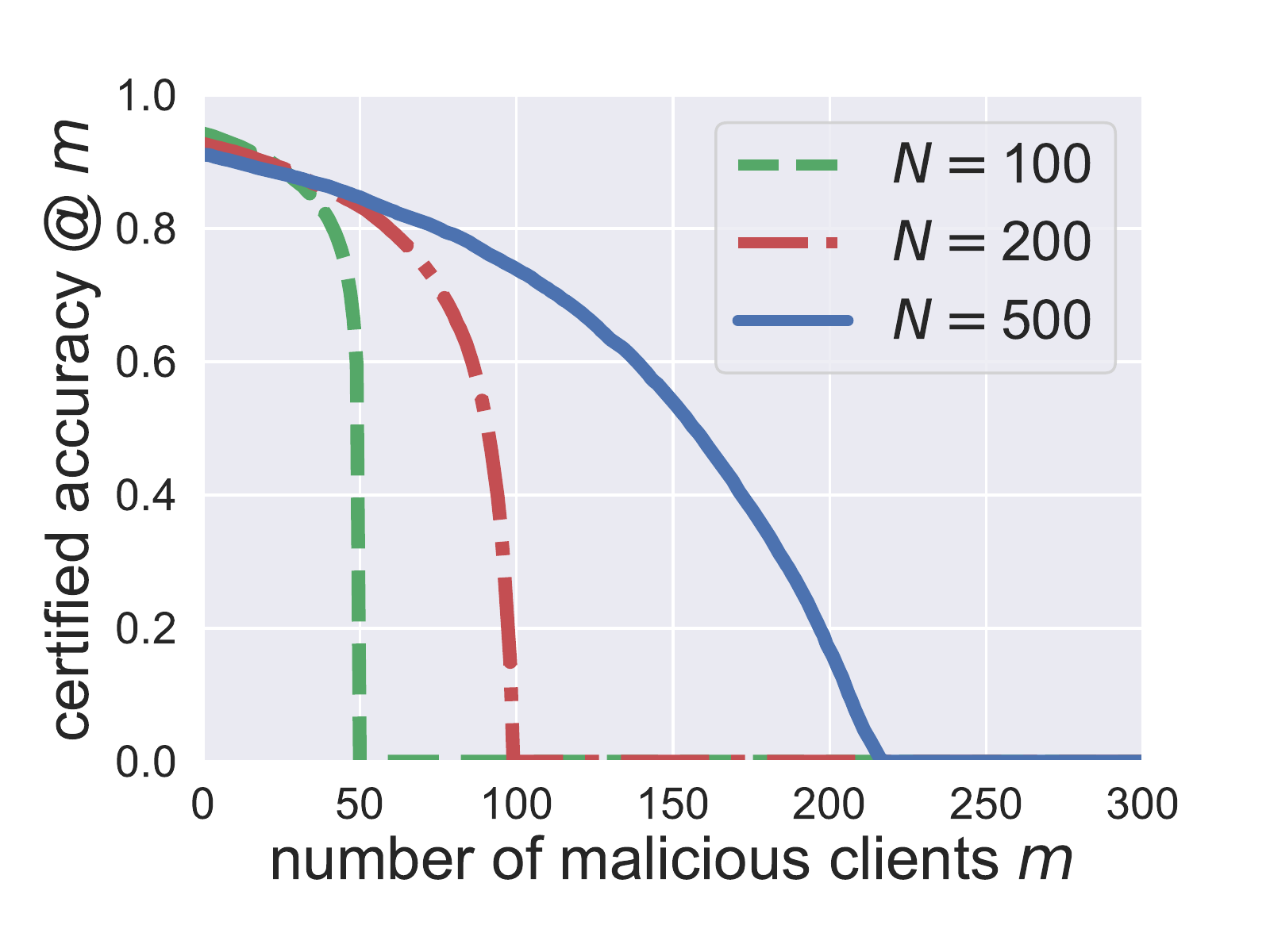}\label{fig:mnist0.5_d_mean_N}}
    \caption{Impact of $N$ on certified accuracy of FLCert.}
    \label{fig:mnist-0.5-N}
\end{figure}

We first show results on certified accuracy. 

\myparatight{Comparing FLCert with single-global-model and CRFL} We compare the two variants of FLCert with two baselines, i.e.,  single-global-model FL and CRFL \cite{xie2021crfl}. For CRFL, we use their default parameter settings, i.e., the clipping threshold $\rho=15$, standard deviation $\sigma=0.01$, and sample size $M=1,000$. Figure \ref{fig:comparison} shows the results on all five datasets, where FedAvg is the base FL algorithm. We observe that when there are no malicious clients (i.e., $m=0$), single-global-model FedAvg and CRFL have larger certified accuracy than FLCert. However, single-global-model FedAvg and CRFL have 0 certified accuracy when just one client is malicious. This is because a single malicious client can arbitrarily manipulate the global model learnt by FedAvg~\cite{Yin18}, reducing the certified accuracy of single-global-model FedAvg to 0, and  its local training data, reducing the certified accuracy of CRFL to 0. 

Moreover, we notice that there is a trade-off between the two variants of FLCert for MNIST-0.1, MNIST-0.5, and Reddit, where $N$ is large. Specifically, when $m$ is small, FLCert-D has higher certified accuracy. This is because in FLCert-P, we estimate the label probabilities, which reduces the gap between the largest and the second largest label probabilities. On the contrary, when $m$ is large, FLCert-P achieves higher certified accuracy, which is because FLCert-P can naturally certify more malicious clients than FLCert-D in the extreme cases if we compare Equation \ref{eq:certified_condition1} with Equation \ref{eq:csl_fancy}. However, FLCert-P cannot certify any malicious client on CIFAR-10 and HAR.  \xc{This is because the number of groups $N$ is small for these two datasets, e.g., $N=15$ for HAR, which results in inaccurate estimation of label probabilities. A possible solution is to design probability estimation methods that are more accurate with a small number of samples.} We also found that if we increase $N$ to 500 while keeping $k=2$, the certified accuracy of FLCert-P is still larger than 0 when up to 8 out of 30 clients are malicious for HAR. However, FLCert-P incurs a much larger  communication/computation overhead than FLCert-D in such scenario.

\begin{figure}[!t]
    \center
    \subfloat[FLcert-P]{\includegraphics[width=0.24\textwidth]{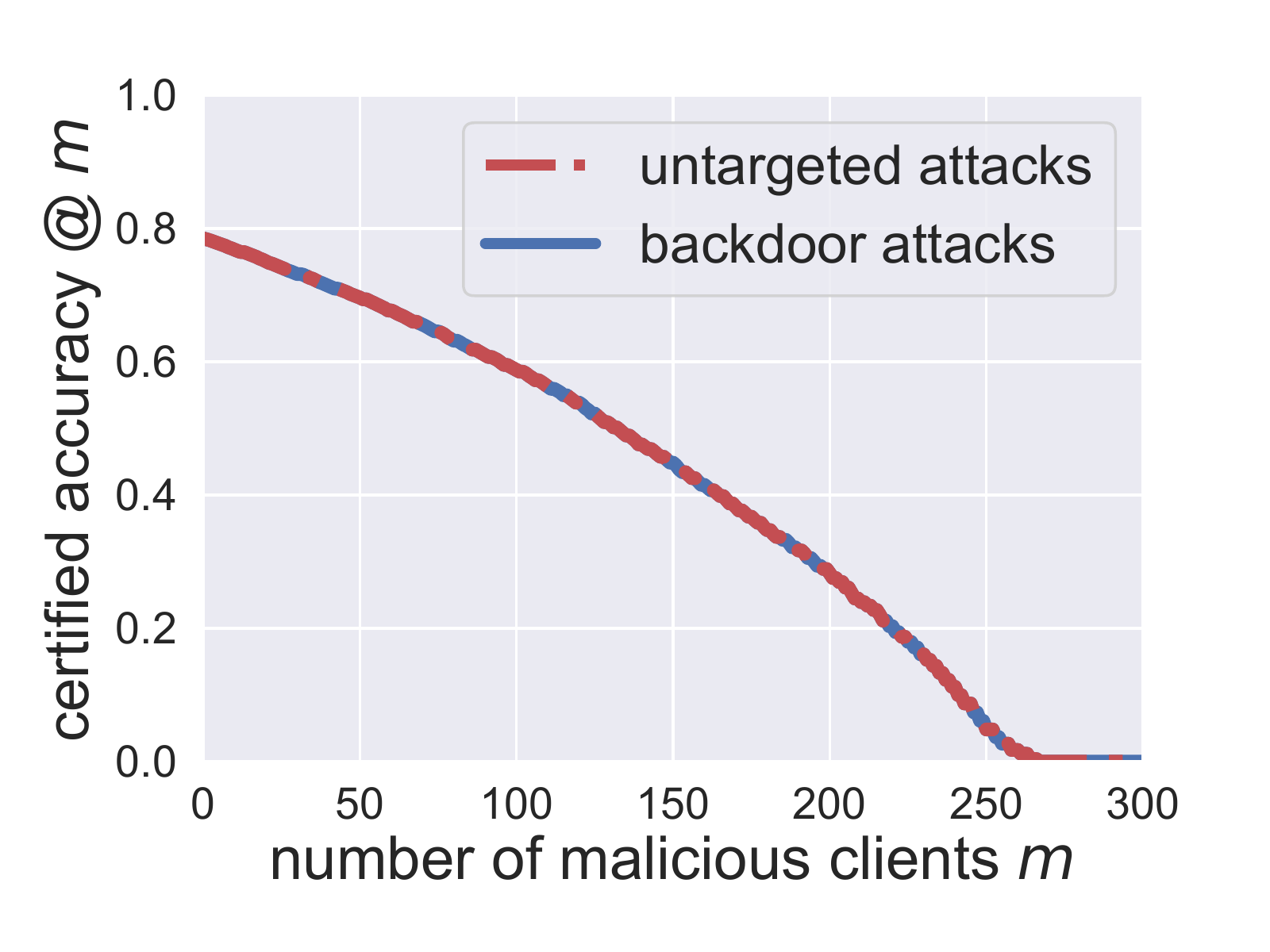}\label{fig:mnist0.5_p_attack}}
    \subfloat[FLCert-D]{\includegraphics[width=0.24\textwidth]{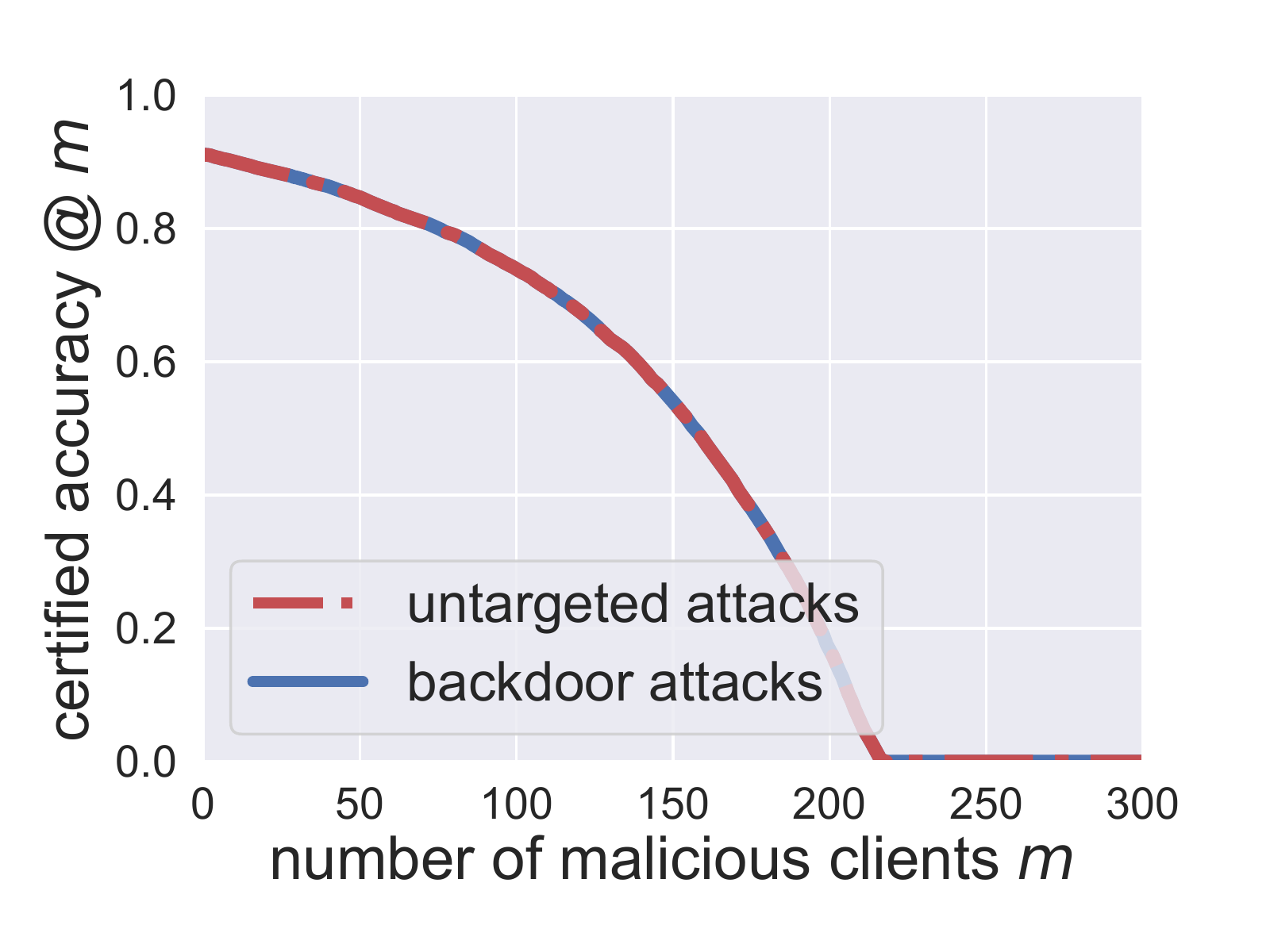}\label{fig:mnist0.5_d_attack}}
    \caption{Certified accuracy of FLCert against untargeted attacks and backdoor attacks.}
    \label{fig:untargeted_backdoor}
\vspace{-4mm}
\end{figure}

\begin{figure}[!t]
    \center
    \subfloat[FLCert-P]{\includegraphics[width=0.25\textwidth]{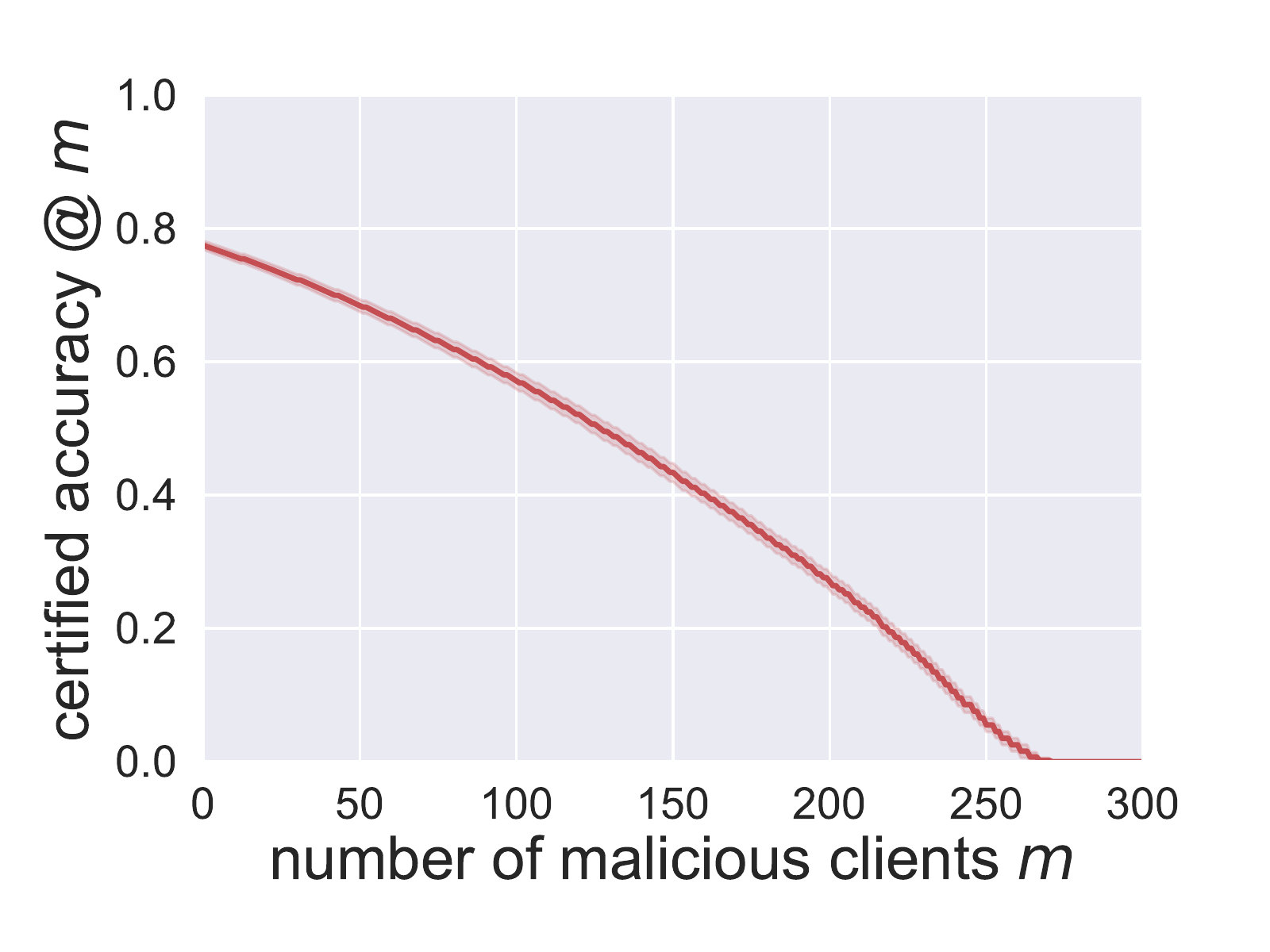}}
    \subfloat[FLCert-D]{\includegraphics[width=0.25\textwidth]{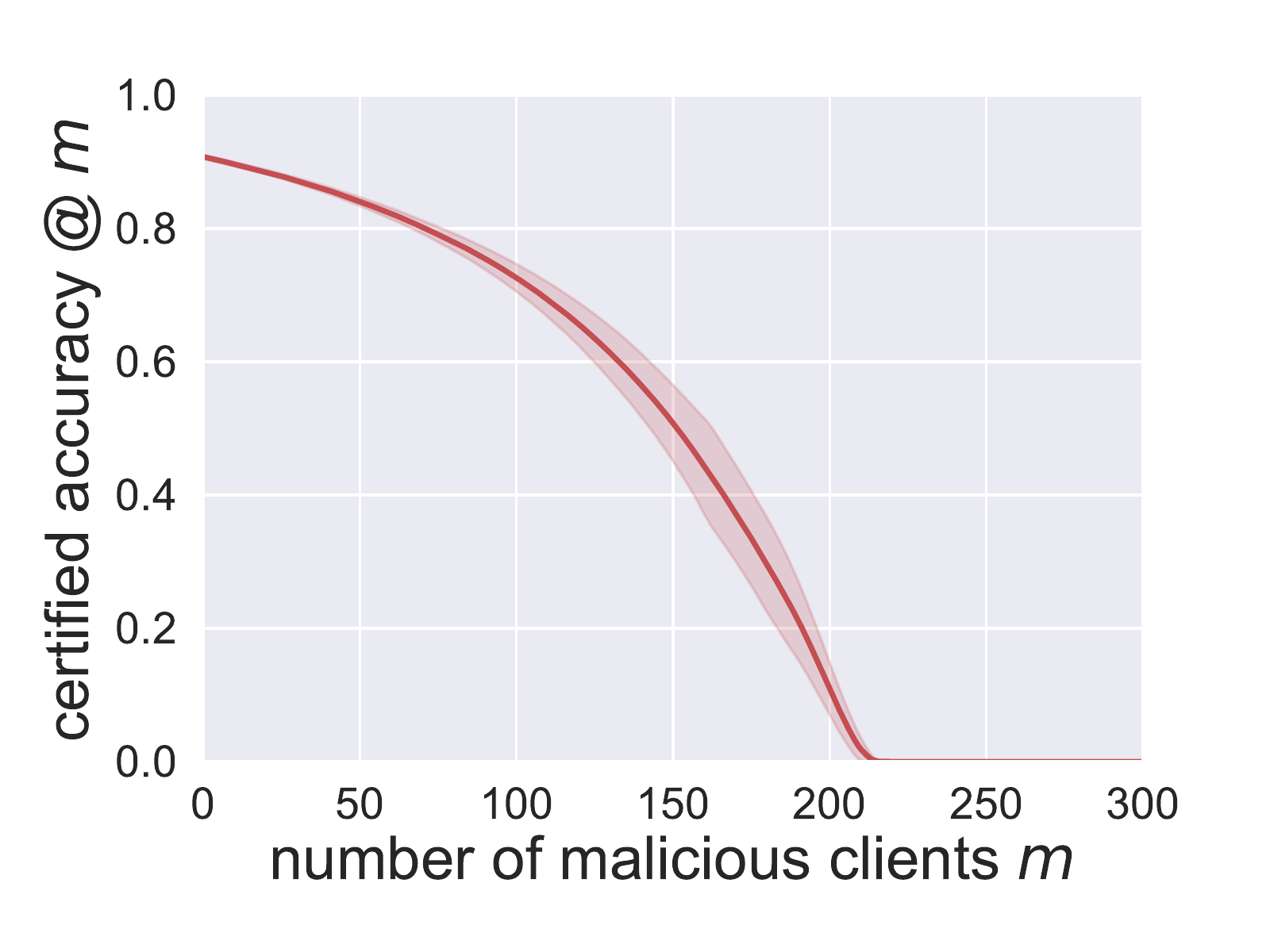}}
    \caption{Impact of seeds in the hash function used to group clients and the base FL algorithm on the certified accuracy of FLCert,  where  dataset is MNIST-0.5,  $N=500$, and FedAvg is  the base FL algorithm. We repeat the experiments for 50 times, each of which uses  a distinct seed for the hash function and a distinct seed for FedAvg. The solid line shows the mean certified accuracy of the 50 trials and the shade represents the standard deviation.}
    \label{fig:randomness}
\vspace{-2mm}
\end{figure}

\myparatight{$N$ achieves a trade-off between accuracy and provable security} We explore the impact of the number of groups $N$ on FLCert. Figure \ref{fig:mnist-0.5-N} shows the results. We observe that $N$ controls a trade-off between  \emph{accuracy under no attacks} (i.e., $m=0$) and provable security. Specifically, FLCert with a larger $N$ has a lower accuracy under no attacks but can tolerate more malicious clients. 
This is because when  $N$ is larger, the average number of clients in each group becomes smaller. Therefore, the  accuracy of each individual global model becomes lower, leading to a lower  accuracy for the ensemble global model under no attacks. Meanwhile, since the number of groups is larger, our FLCert can tolerate more malicious clients. 

\myparatight{Untargeted attacks vs. backdoor attacks} We evaluate the certified accuracy for both untargeted attacks and targeted attacks. Figure \ref{fig:untargeted_backdoor} shows the results. We observe that the certified accuracy of both FLCert-P and FLCert-D against untargeted  attacks is similar to that against backdoor attacks. The reason is that  the trigger in a backdoor attack is often small to be stealthy, and thus a clean  global model's predicted label for a test input is unaffected by the trigger.

\begin{figure*}[!t]
    \center
    \subfloat{\includegraphics[width=0.5\textwidth]{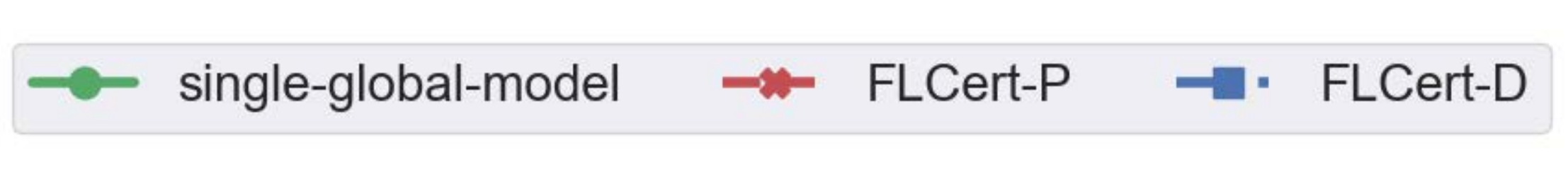}}\\\vspace{-5mm}
    \subfloat{\includegraphics[width=0.19\textwidth]{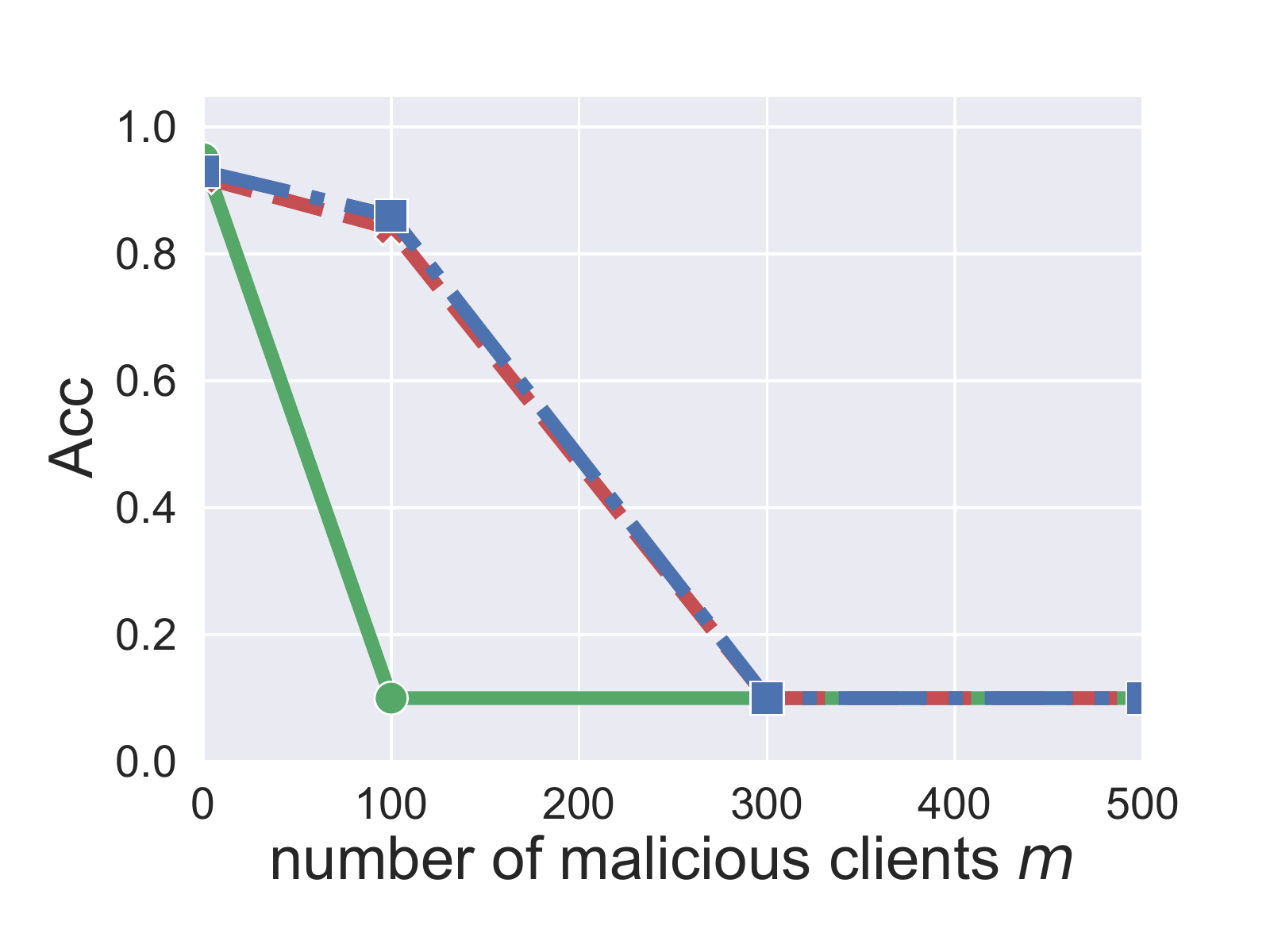}\label{fig:mnist0.5_mean_ter}}
    \subfloat{\includegraphics[width=0.19\textwidth]{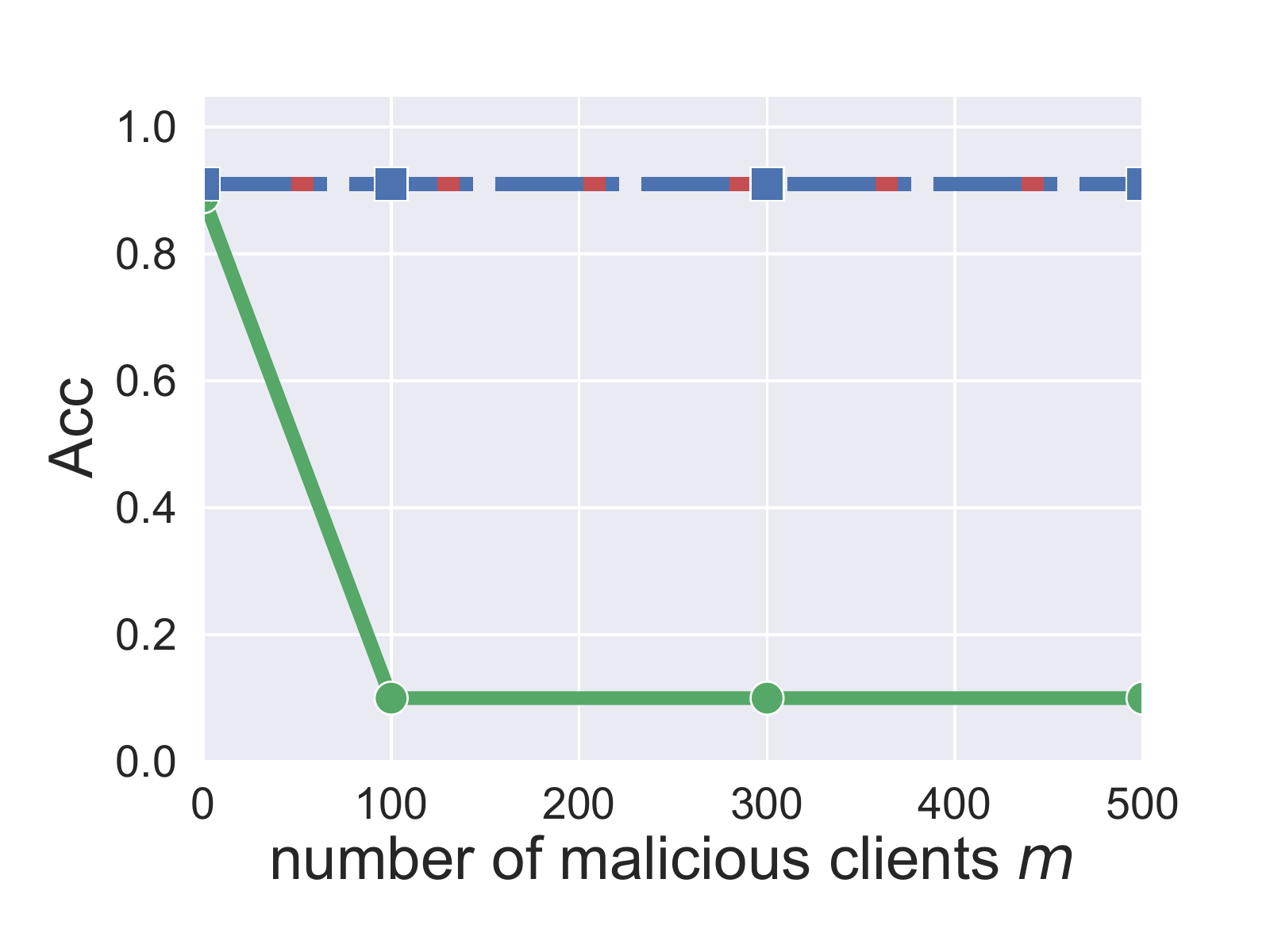}\label{fig:mnist0.5_krum_ter}}
    \subfloat{\includegraphics[width=0.19\textwidth]{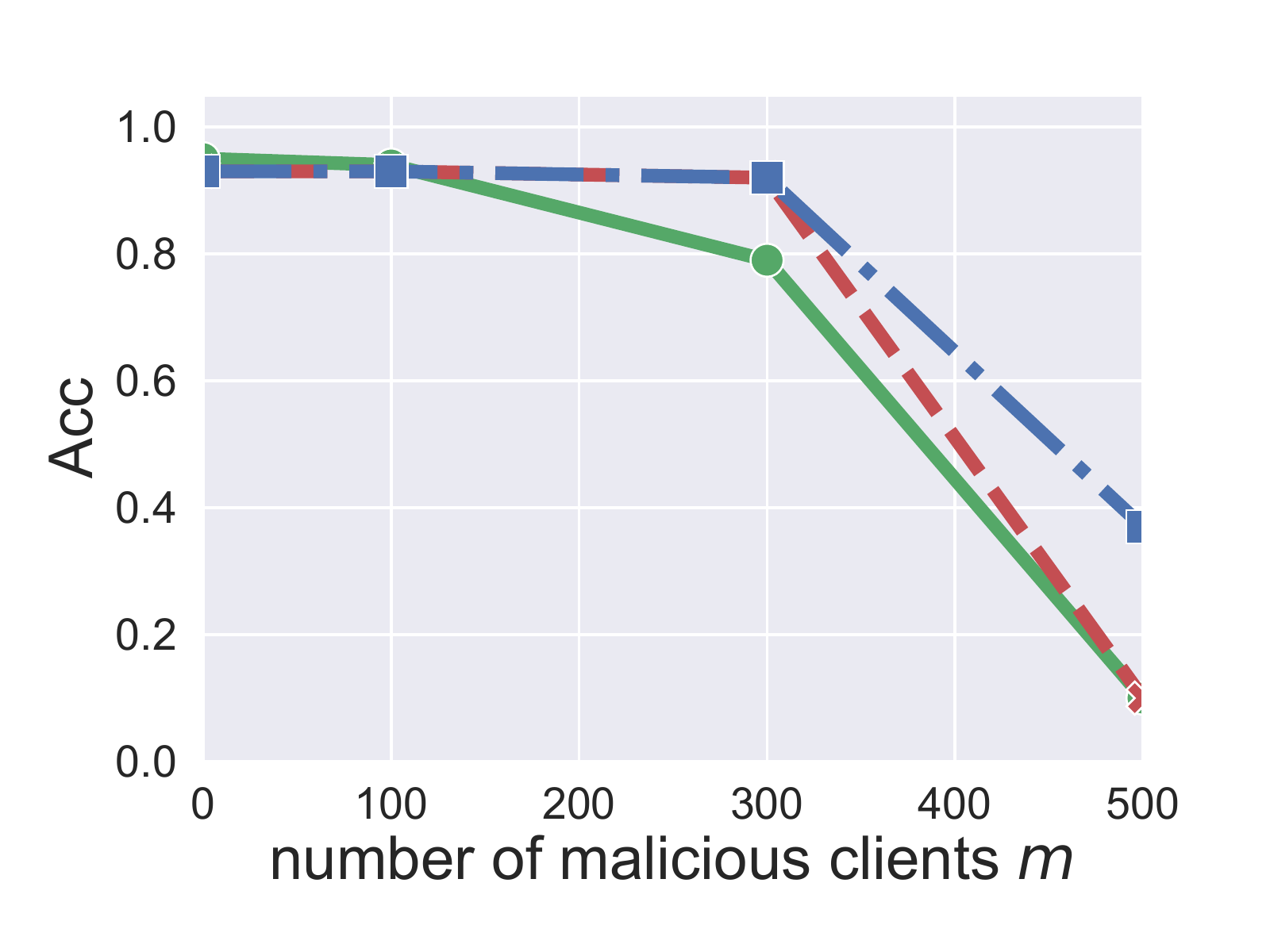}\label{fig:mnist0.5_trim_ter}}
    \subfloat{\includegraphics[width=0.19\textwidth]{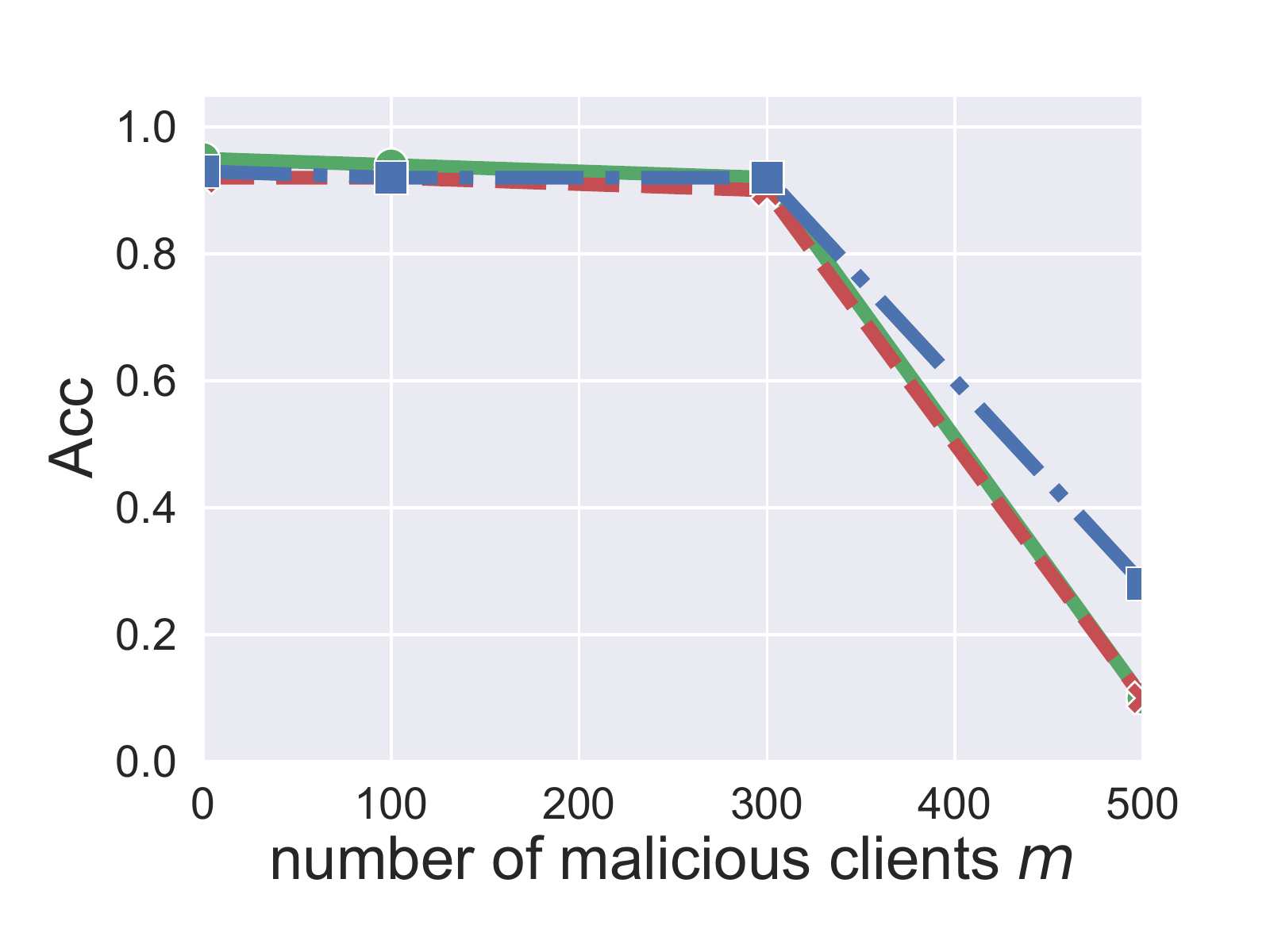}\label{fig:mnist0.5_median_ter}}
    \subfloat{\includegraphics[width=0.19\textwidth]{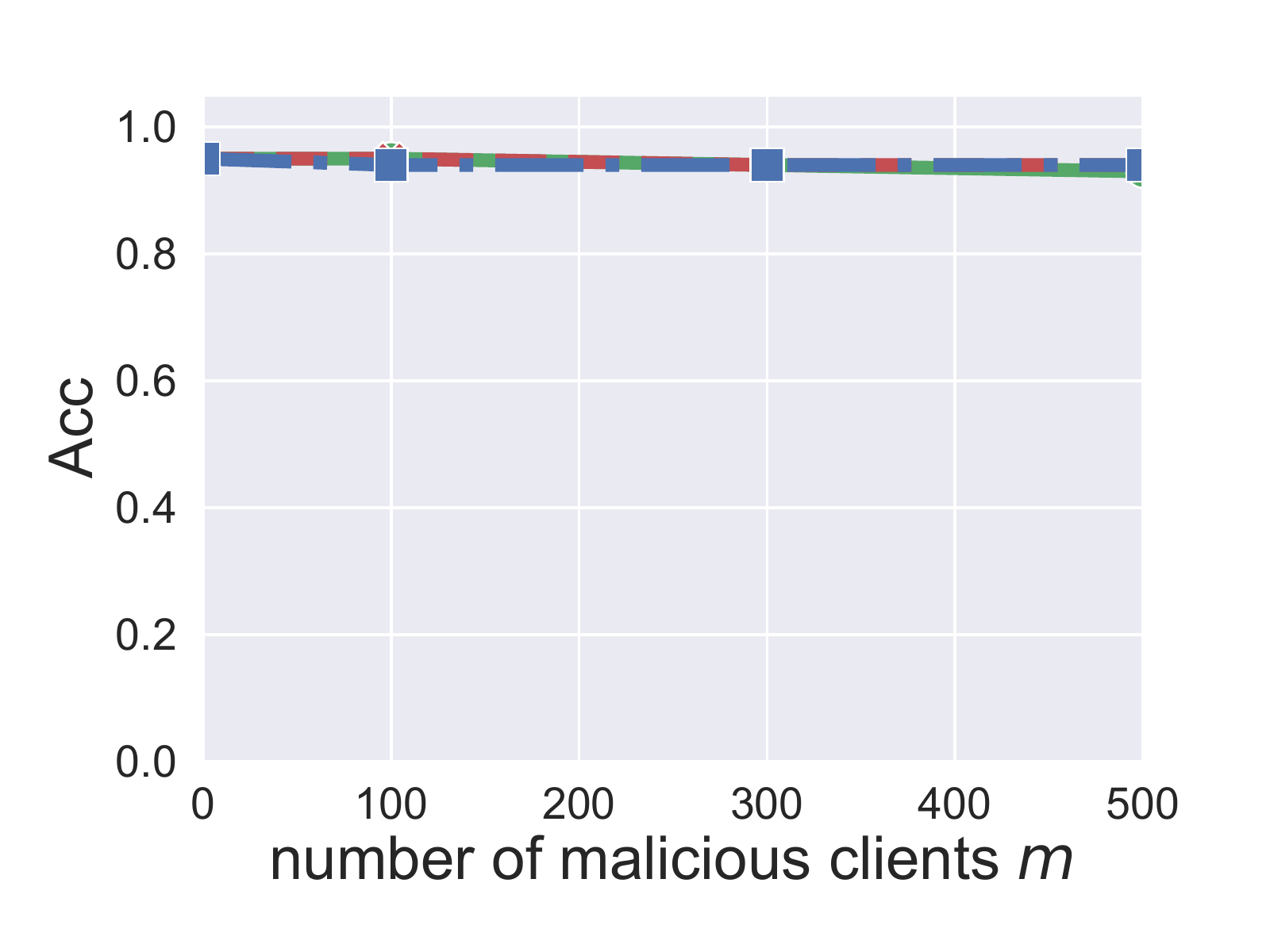}\label{fig:mnist0.5_fltrust_ter}}\\\vspace{-5mm}
    \addtocounter{subfigure}{-6}
    \subfloat[FedAvg]{\includegraphics[width=0.19\textwidth]{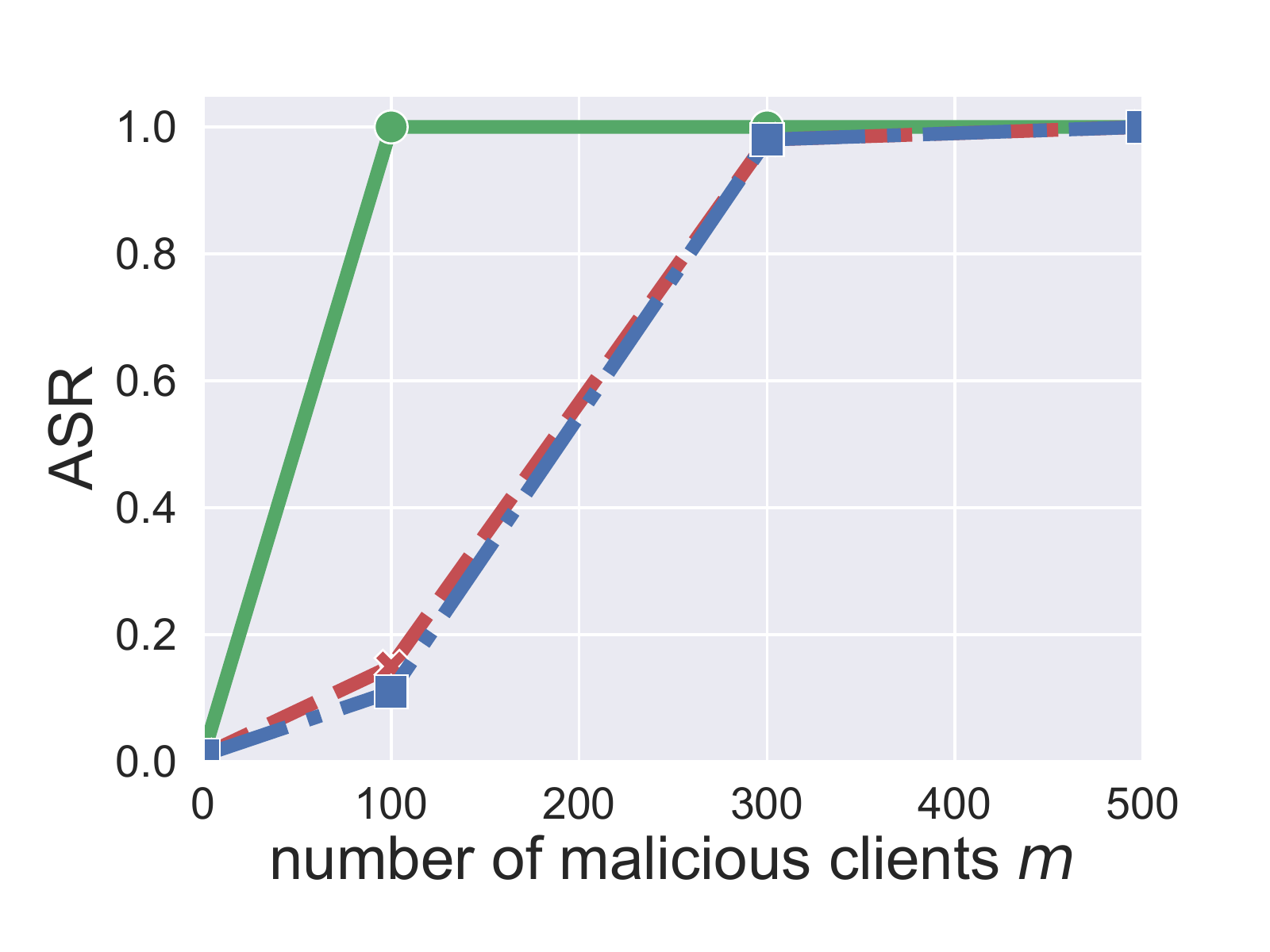}\label{fig:mnist0.5_mean_asr}}
    \subfloat[Krum]{\includegraphics[width=0.19\textwidth]{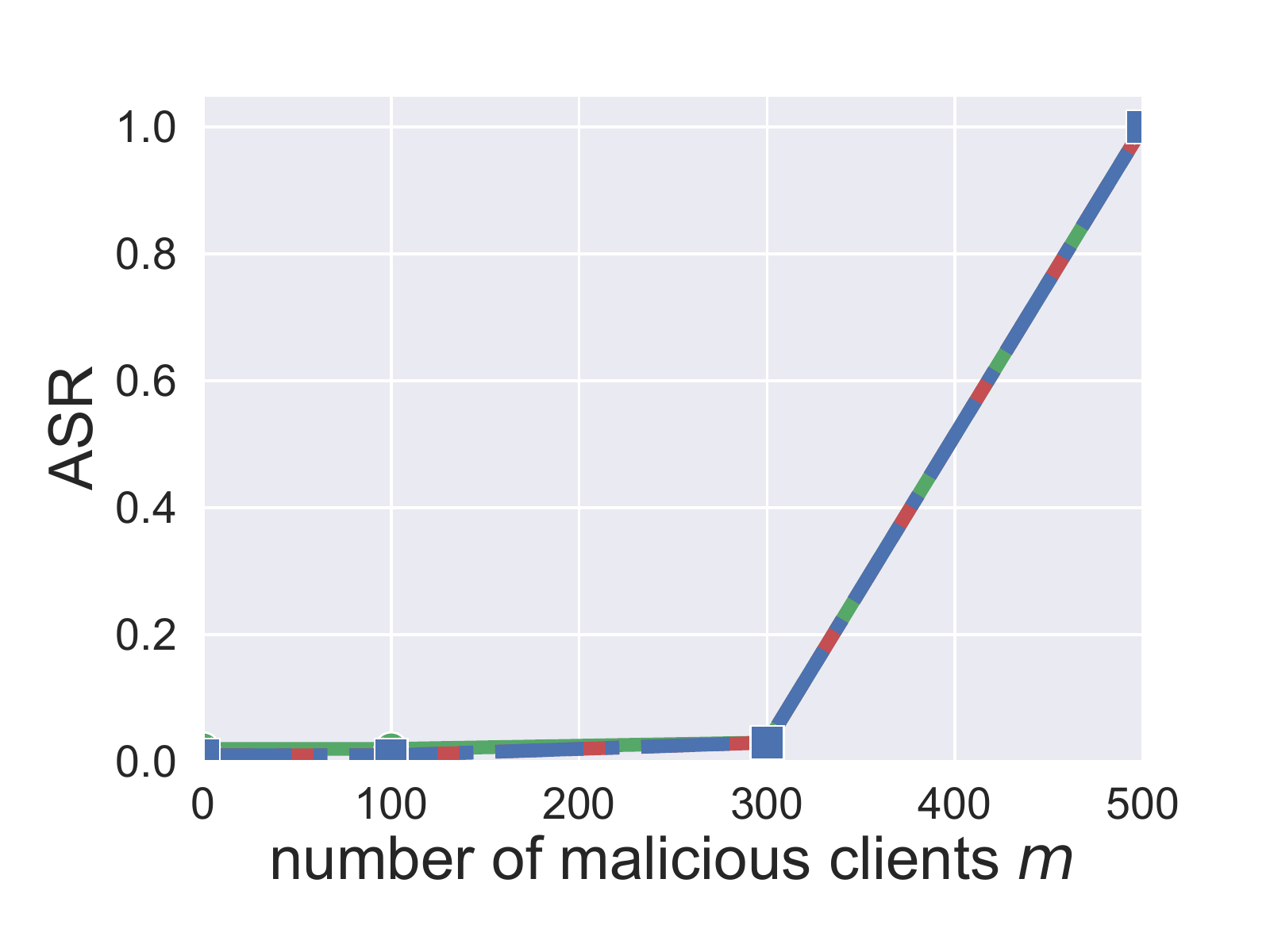}\label{fig:mnist0.5_krum_asr}}
    \subfloat[Trimmed-mean]{\includegraphics[width=0.19\textwidth]{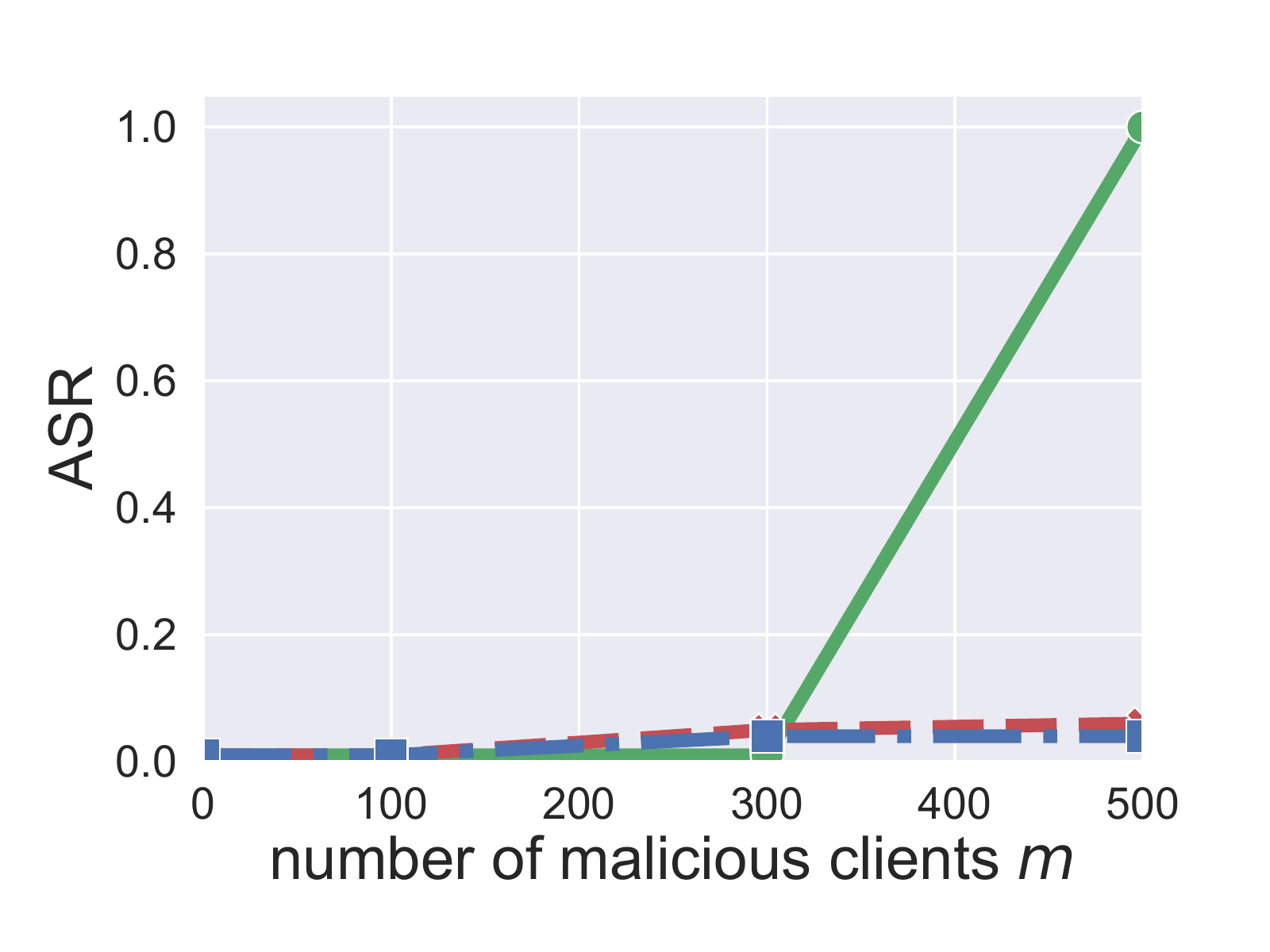}\label{fig:mnist0.5_trim_asr}}
    \subfloat[Median]{\includegraphics[width=0.19\textwidth]{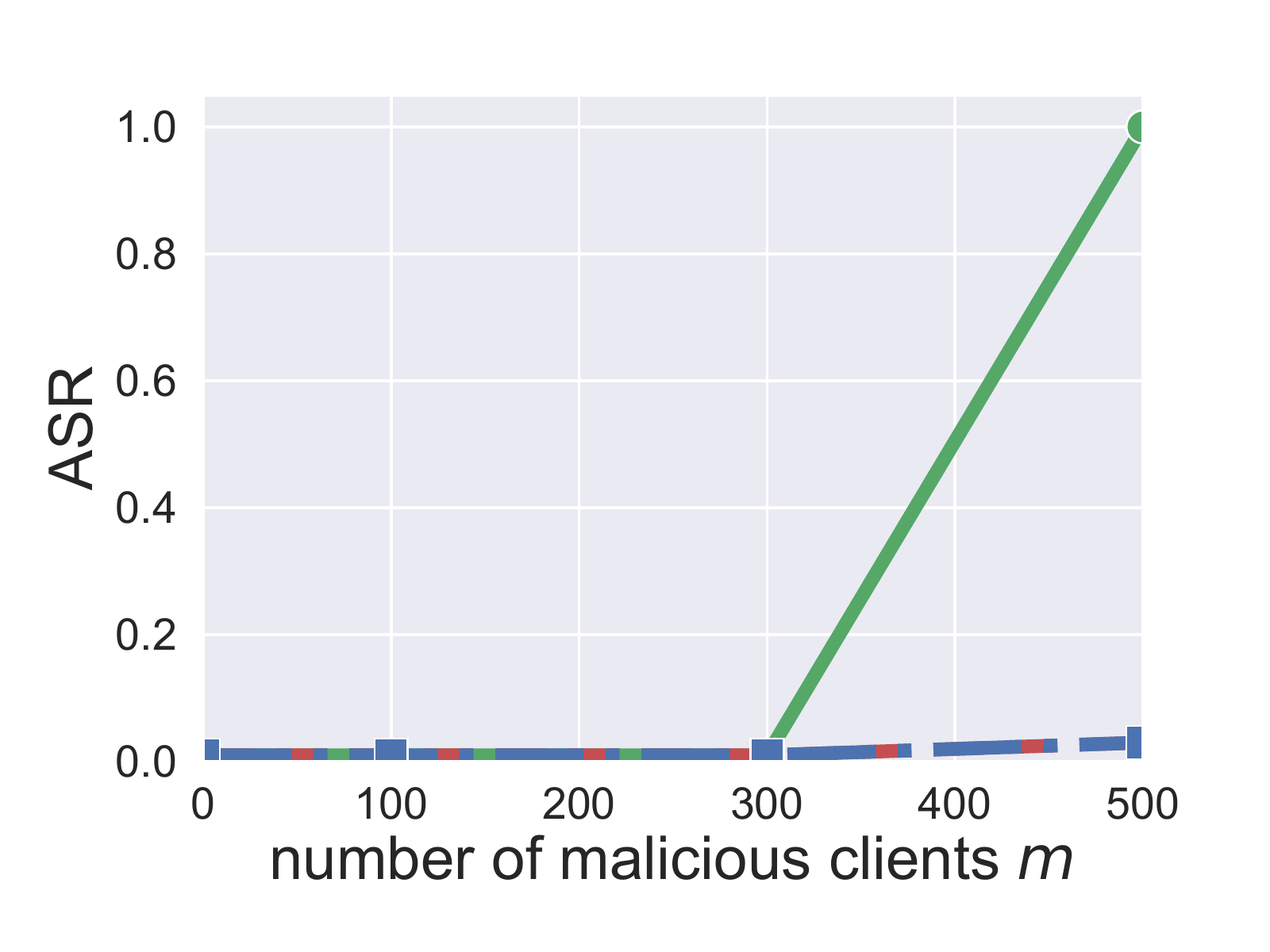}\label{fig:mnist0.5_median_asr}}
    \subfloat[FLTrust]{\includegraphics[width=0.19\textwidth]{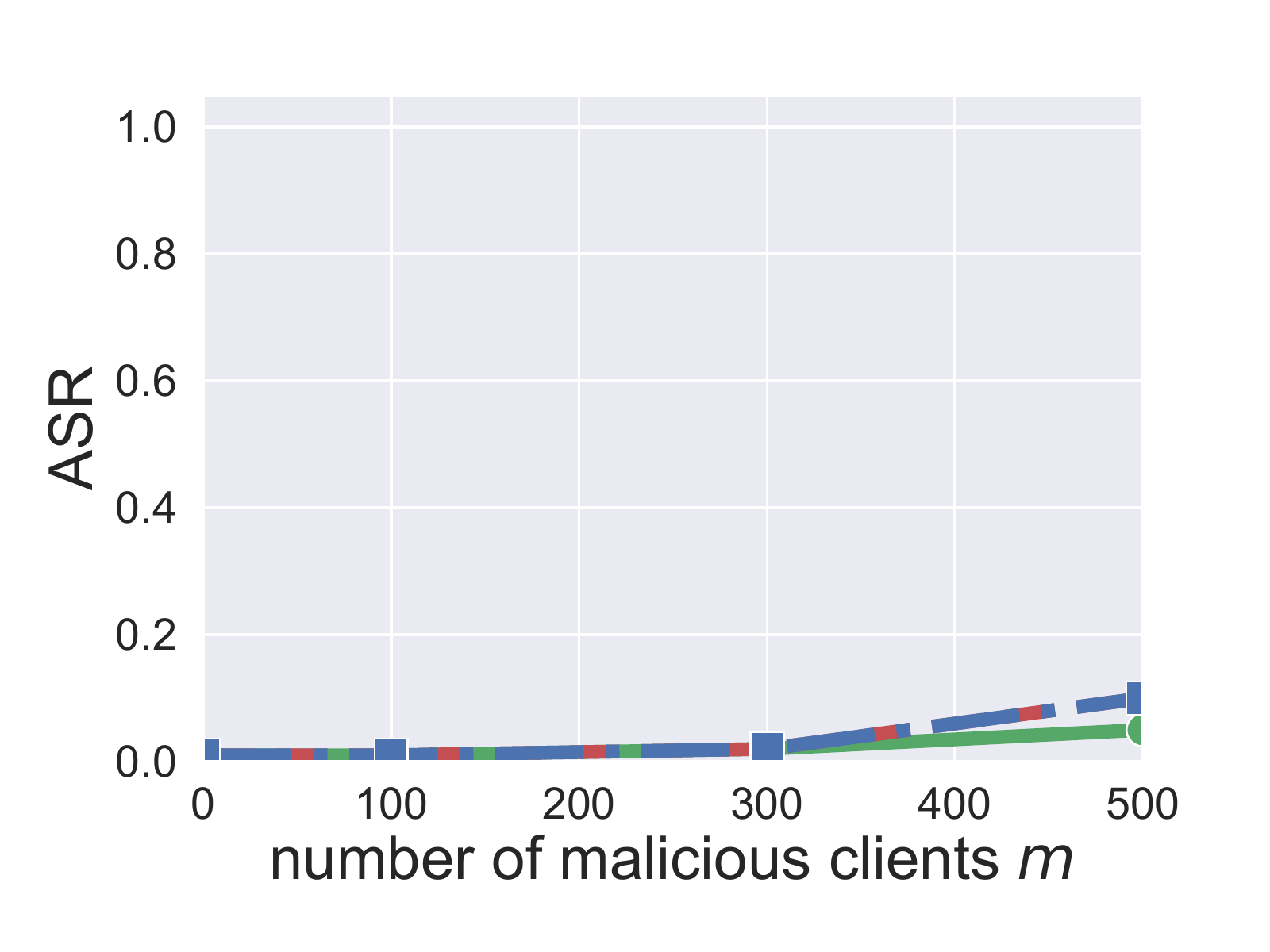}\label{fig:mnist0.5_fltrust_asr}}
    \caption{\xc{Test accuracy (Acc) against untargeted attacks and attack success rate (ASR) against backdoor attacks for the single-global-model settings and FLCert when different base FL algorithms are used for MNIST-0.5 and $N=200$. A higher test accuracy and a lower ASR mean better empirical performance. }}
    \label{fig:empirical}
 \vspace{-3mm}
\end{figure*}

\myparatight{Impact of  seeds in hashing and base FL algorithms} Recall that we use the built-in Python \emph{hash()} function with seeds to divide clients into groups for FLCert-D and we determinize a base FL algorithm via fixing the seed for both FLCert-P and FLCert-D. 
We study the impact of the seeds on the certified accuracy of FLCert. Specifically, we generate 50 pairs of seeds for the hash() function and base FL algorithm. 
Then, we run our FLCert for 50 times on a dataset, each of which uses a distinct pair of seeds. In each run, we use the same seed for different groups. 
Figure \ref{fig:randomness} shows the certified accuracy of FLCert with $N=500$ on MNIST-0.5 when the base FL algorithm is FedAvg. 
We observe that the standard deviation is relatively small compared to the average certified accuracy, which means that our FLCert is insensitive to the seed in the \emph{hash()} function used to group clients and the seed in the base FL algorithm.

\begin{figure}[!t]
    \center
    \subfloat[FLCert-P]{\includegraphics[width=0.24\textwidth]{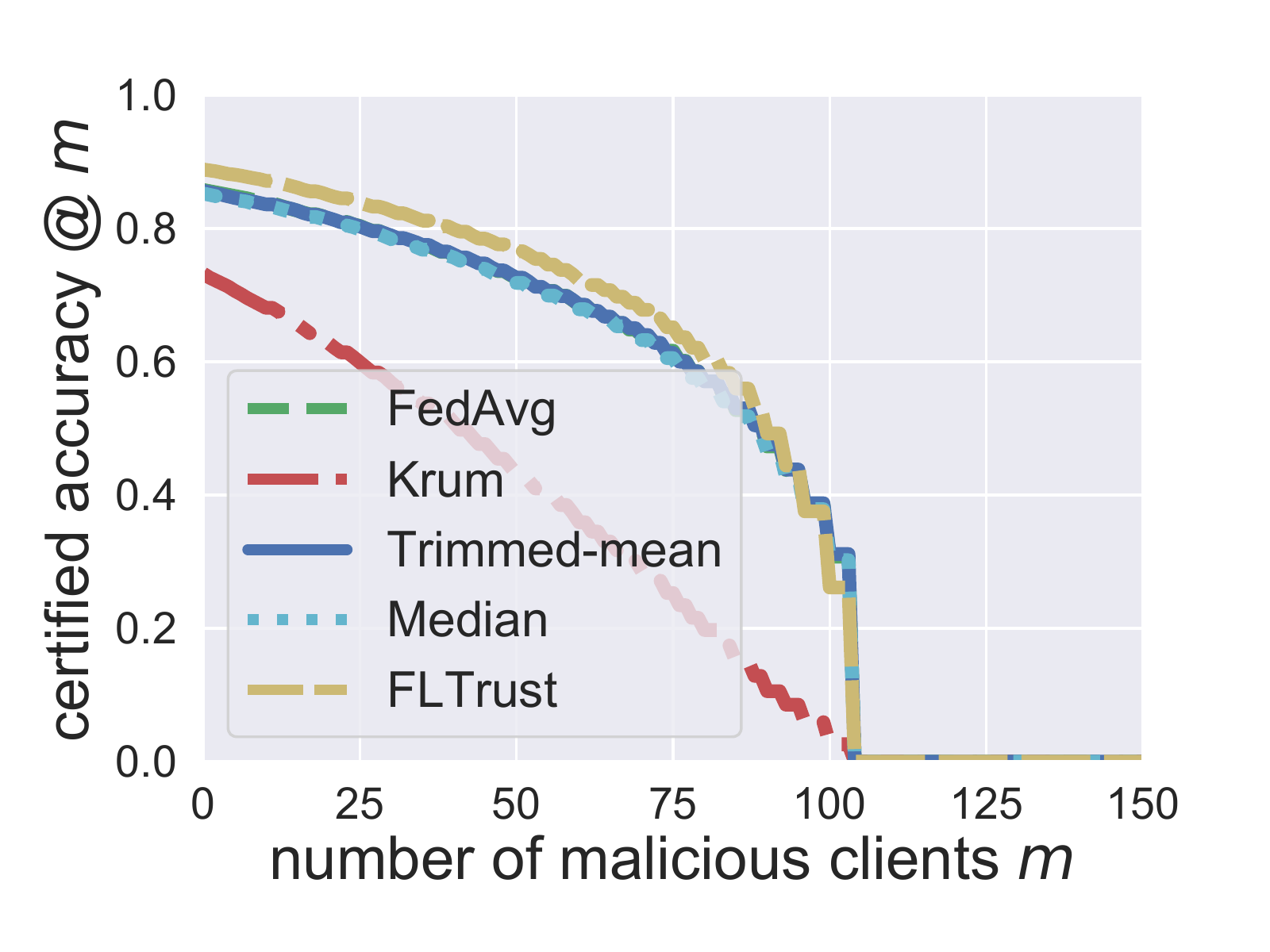}\label{fig:mnist0.5_base5p}}
    \subfloat[FLCert-D]{\includegraphics[width=0.24\textwidth]{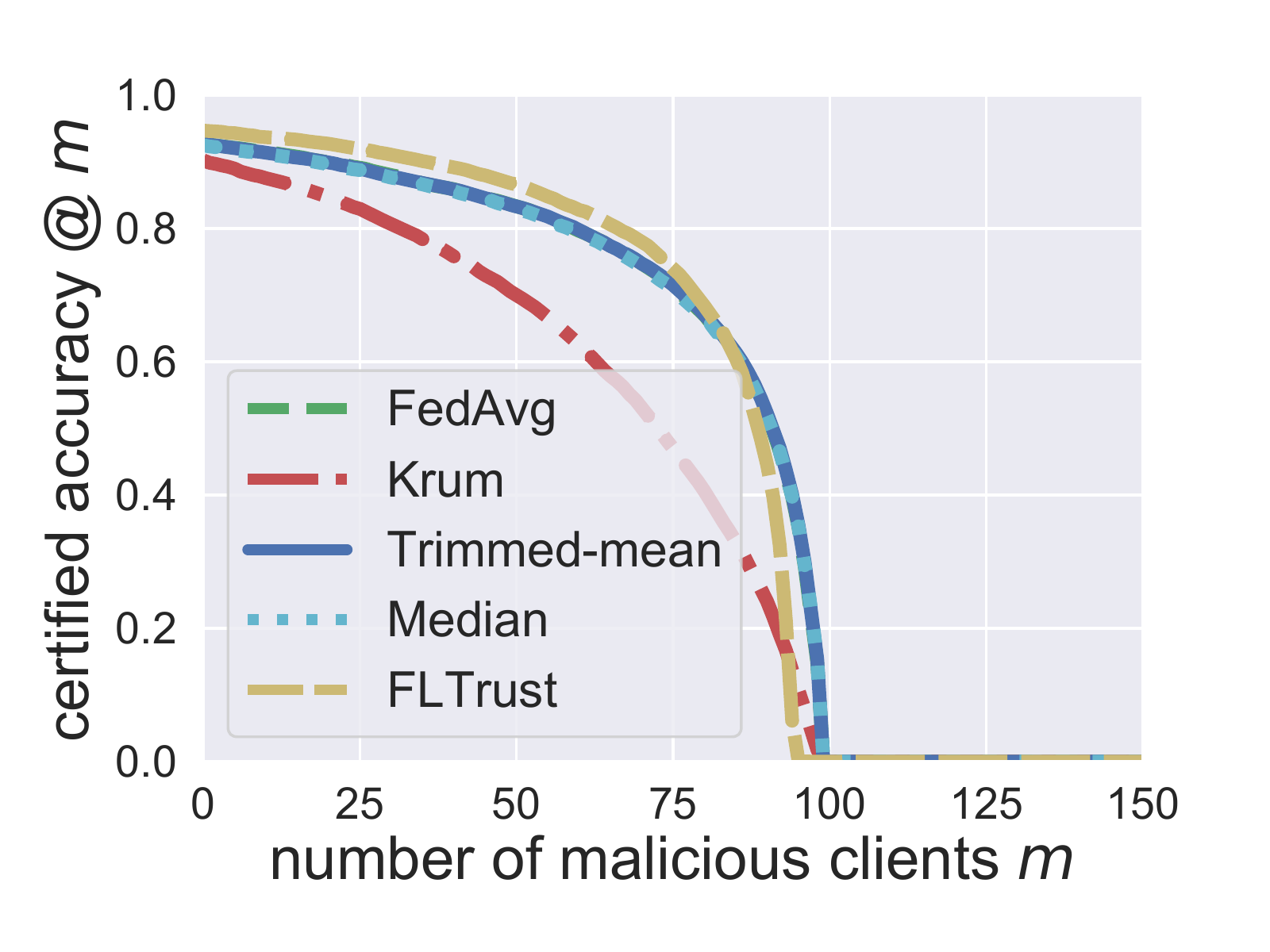}\label{fig:mnist0.5_base5d}}
    \caption{FLCert with different base FL algorithms.}
    \label{fig:mnist0.5-base}
\end{figure}

\myparatight{FLCert with different base FL algorithms} We explore FLCert with different base FL algorithms, where the dataset is MNIST-0.5 and $N=200$. We do not use the default $N=500$ because when $N=500$, the (expected) number of clients in each group is 2, for which Krum, Trimmed-mean, and Median are not defined. Figure \ref{fig:mnist0.5-base} shows the results.  
We notice that the certified accuracy of FLCert is similar when FedAvg, Trimmed-mean, or Median is used as the base FL algorithm, and is better than the certified accuracy when the base FL algorithm is Krum.
This is because Krum selects a single local model as the new global model while the other base FL algorithms consider all the received local models to update the global model. As a result, the global models learnt by Krum are less accurate than the global models learnt by the other base FL algorithms. Therefore, the ensemble global model of our FLCert has a lower certified accuracy when the base FL algorithm is Krum. Moreover, we observe that FLTrust achieves the highest certified accuracy when the number of malicious clients $m$ is small. This is because the server is assumed to hold a small clean dataset in FLTrust, which helps learn more accurate global models with small groups of clients.

\begin{figure}[!t]
    \center
    {
    \subfloat[]{\includegraphics[width=0.24\textwidth]{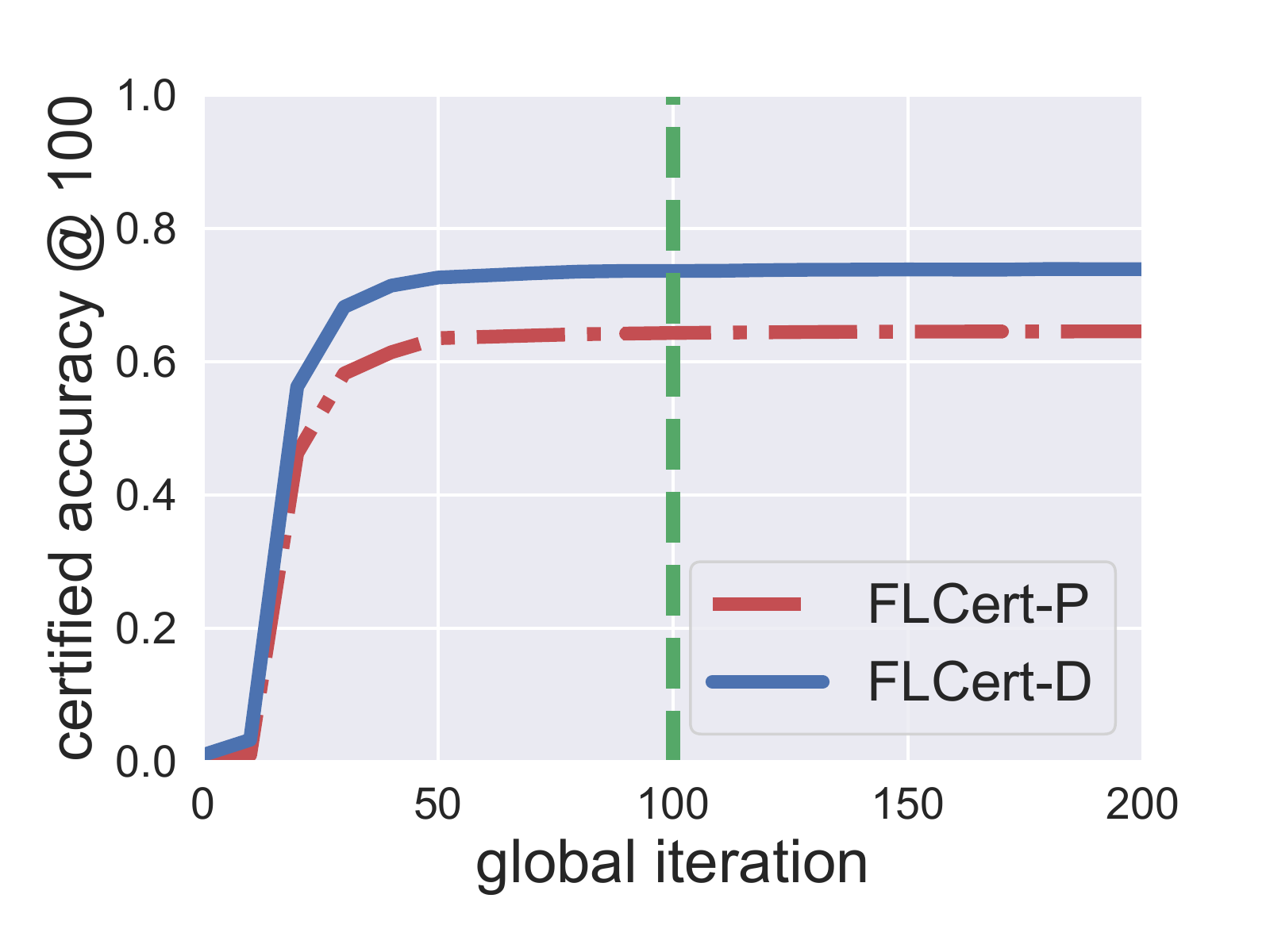}\label{fig:ca_com}}
    \subfloat[]{\includegraphics[width=0.24\textwidth]{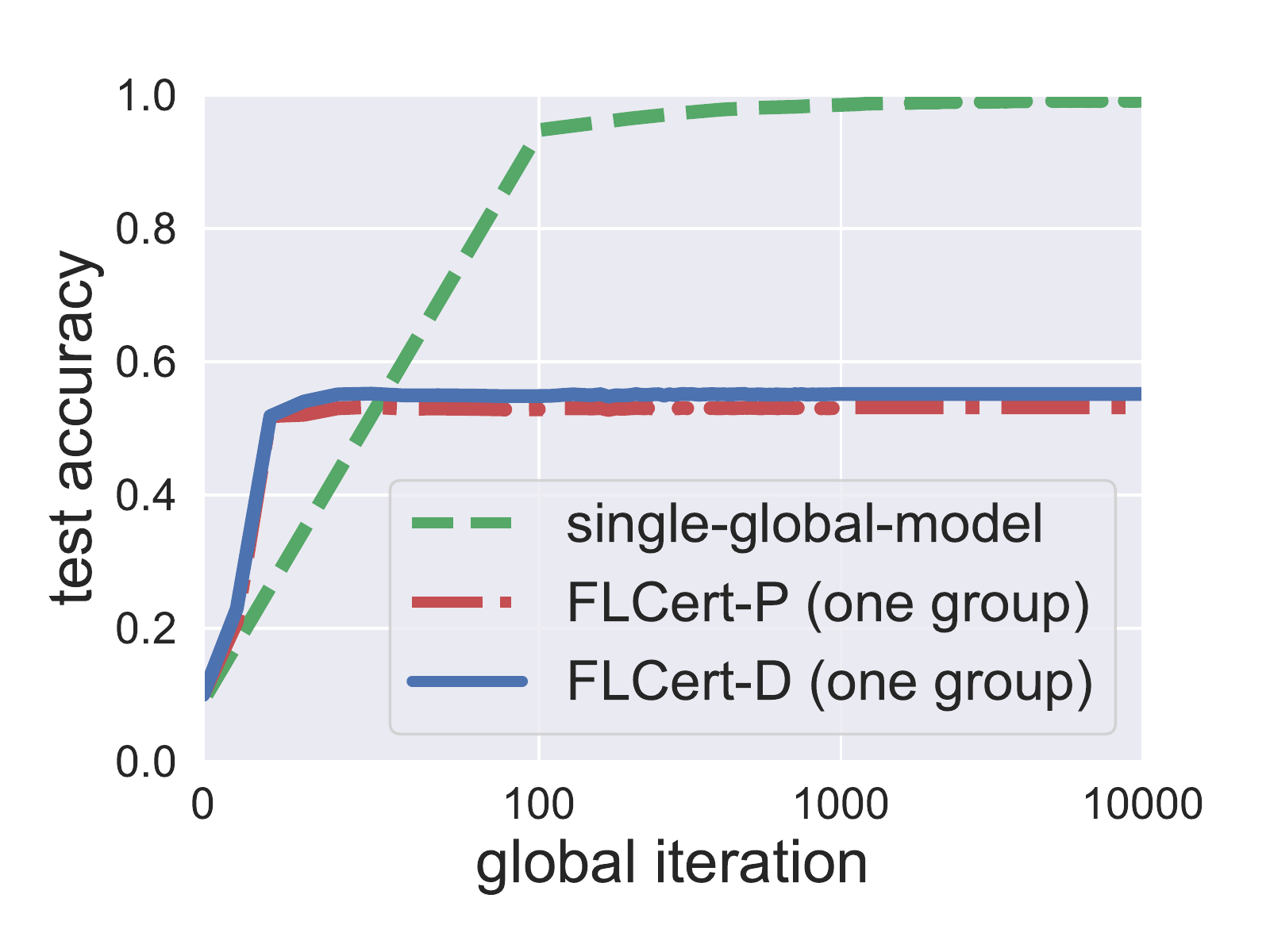}\label{fig:convergence}}}
    
    \caption{(a) The communication/computation-security trade-off of FLCert. The vertical line shows the setting of the number of global iterations, where our FLCert has the same communication/computation cost for the clients as the single-global-model setting. (b) Convergence of learning the global model in the single-global-model setting and learning a global model for a certain group in FLCert. The dataset is MNIST-0.5 and base FL algorithm is FedAvg.}
    \label{fig:trade-off}
\end{figure}

\subsubsection{Test Accuracy and Attack Success Rate}

{Figure \xc{\ref{fig:empirical}}} shows the test accuracy against \xc{existing} untargeted  attacks and attack success rate against \xc{existing} backdoor attacks for the conventional single-global-model settings and FLCert when different base FL algorithms are used, for MNIST-0.5 dataset.  
We use $N=200$ as Krum, Trimmed-mean, and Median are not defined when $N=500$. The numbers for the untargeted poisoning attacks represent  test accuracy and a larger test accuracy indicates a better FL method. The numbers for the backdoor attacks represent  attack success rate and a smaller number indicates a better FL method.  

We notice that in general, the single-global-model setting achieves comparable or higher test accuracy and lower attack success rate than FLCert when there are no malicious clients or the number of malicious clients is small. However, our FLCert becomes better when the number of malicious clients is larger. An exception is when FLTrust is the aggregation rule, where both single-global-model and FLCert are robust against different attacks. 
Our results indicate that  FLCert can tolerate more malicious clients against empirical attacks than single-global-model.

\subsubsection{Communication/Computation-Security Trade-off}
\label{sec:empirical_cost}

Suppose the base FL algorithm uses $T_e$ global iterations when learning each global model. Figure \ref{fig:ca_com} shows the certified accuracy of FLCert for MNIST-0.5 as $T_e$ increases, where FedAvg is the base FL algorithm, $\beta_e=1$,  $N=500$, and $m=100$. Our certified accuracy increases as $T_e$ increases and converges after $T_e$ is large enough. Suppose  the single-global-model setting uses the default parameter setting $T=1,000$ and $\beta=0.1$. 
The vertical line in Figure \ref{fig:ca_com} corresponds to the setting of $T_e$ with which our FLCert has the same communication/computation cost for a client as the single-global-model setting, i.e., $T_e=\beta T/\beta_e=0.1\times 1000/1=100$. We notice that with such setting of $T_e$, our FLCert-P and FLCert-D can already achieve a high certified accuracy. For instance, FLCert-D achieves a certified accuracy of 0.73, which is 99\% of 0.74,  the largest certified accuracy FLCert-D can achieve by using a large $T_e$. Our results show that, compared to the single-global-model setting, our FLCert can achieve provable security against a bounded number of malicious clients without additional communication and computation cost for the clients.

One reason is that learning a global model for a group in FLCert converges faster than learning a global model for all clients in the single-global-model setting. 
Figure \ref{fig:convergence} shows the test accuracy under no attacks in the single-global-model setting and a global model for a certain group in FLCert-P and FLCert-D  as the number of global iterations increases. We observe that the global model in the single-global-model setting converges with roughly 1,000 global iterations, while a global model for a group in our FLCert converges with less than 100 global iterations. The reason is that the global model in the single-global-model setting aims to fit the local training data on all clients, while a global model in our FLCert aims to fit the local training data on only a group of clients. 

\xc{We note that FLCert incurs some cost at inference time. Specifically, we need to store the parameters of $N$ global models. Moreover, we need to query the $N$ global models to make a prediction for a given input. However, we note that such cost is affordable. First, we have shown that a moderate $N$ (e.g., $N$=500) is enough to achieve a high certified accuracy. Second, we can leverage model compression techniques to further reduce the model size. Third, the $N$ global models can make predictions in parallel.}

\section{Conclusion, \xc{Limitations, and Future Work}}
In this work, we propose FLCert, an ensemble FL framework that provides provable security guarantees against poisoning attacks from malicious clients. We propose two variants FLCert-P and FLCert-D based on how the clients are grouped and derive their theoretical security guarantees. Moreover, we design a randomized algorithm to compute the certified security level for FLCert-P in practice. Our empirical results on multiple datasets show that our FLCert can effectively defend against poisoning attacks with provable security guarantees.  

\xc{One limitation of our work is that we do not leverage any prior knowledge on the learning task or the base FL algorithm when deriving our certified security levels. It is an interesting future work to involve such prior knowledge when deriving certified security guarantees.}

\section*{Acknowledgements}

We thank the anonymous reviewers and AE for constructive comments. This work was supported by National Science Foundation grant No. 2131859, 2125977, 2112562, and 1937786 as well as Army Research Office grant No. W911NF2110182.

\bibliographystyle{plain}
\bibliography{refs}

\section*{Proof of Theorem~\ref{certified_radius_p}}
\label{proof_of_certified_radius_p}

\begin{figure}[!t]
    \center
    {\includegraphics[width=0.3\textwidth]{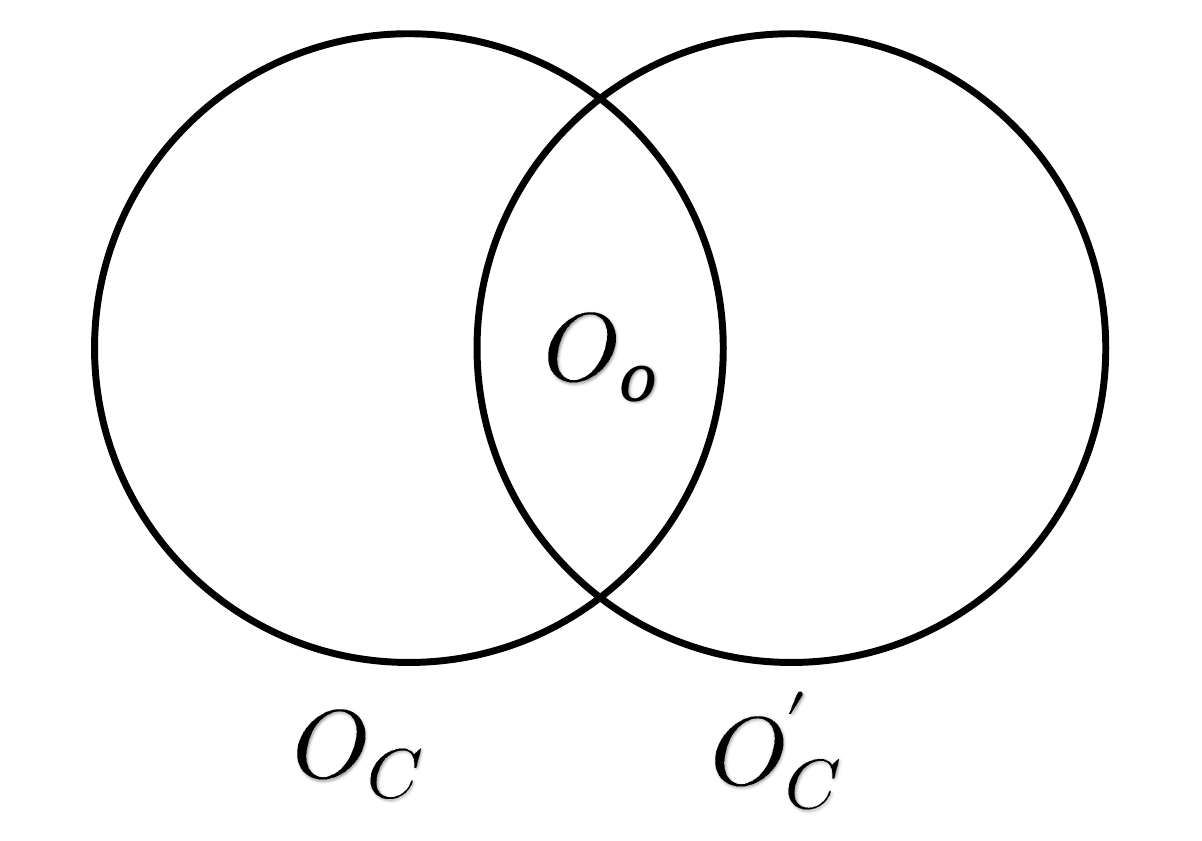}}
    \caption{Illustration of $\mathbf{O}_C,\mathbf{O}_C'$, and $\mathbf{O}_o$.}
    \label{fig:proof1}
\end{figure}

We first define a \emph{subsample} as a set of $k$ clients sampled from the $n$ clients uniformly at random without replacement. We further define the space of all subsamples of $k$ clients from $\mathbf{C}$ as $\mathbf{O}_C=\{\mathcal{S}(\mathbf{C}, k)\}$ and the space of all possible subsamples from $\mathbf{C}'$ as $\mathbf{O}_C'=\{\mathcal{S}(\mathbf{C}', k)\}$. Let $\mathbf{O}_o=\{\mathcal{S}(\mathbf{C}\cap\mathbf{C}', k)\}=\mathbf{O}_C\cap\mathbf{O}_C'$ denote the space of all possible subsamples from the set of normal clients $\mathbf{C}\cap\mathbf{C}'$, and $\mathbf{O} = \{\mathcal{S}(\mathbf{C}\cup\mathbf{C}', k)\} = \mathbf{O}_C\cup\mathbf{O}_C'$ denote the space of all possible subsamples from either $\mathbf{C}$ or $\mathbf{C}'$. Figure \ref{fig:proof1} illustrates $\mathbf{O}_C, \mathbf{O}_C'$, and $\mathbf{O}_o$. We use a random variable $\mathbf{X}$ to denote a subsample $\mathcal{S}(\mathbf{C}, k)$ and $\mathbf{Y}$ to denote a subsample $\mathcal{S}(\mathbf{C}', k)$ in $\mathbf{O}$. We know that $\mathbf{X}$ and $\mathbf{Y}$ have the following probability distributions:
\begin{align}
	&\text{Pr}(\mathbf{X}=a) = 
		\begin{cases}
 			\frac{1}{{n \choose k}}, &\text{ if } a \in \mathbf{O}_C\\
			0, 					   &\text{ otherwise},
		\end{cases}\\
	&\text{Pr}(\mathbf{Y}=a) = 
		\begin{cases}
 			\frac{1}{{n \choose k}}, &\text{ if } a \in \mathbf{O}_C'\\
			0, 					   &\text{ otherwise}.
		\end{cases}
\end{align}
Assume that a federated learning algorithm $f$ takes a subsample $a$ and a test input $\mathbf{x}$ as its input and outputs a label $f(a,\mathbf{x})$. 
We have the following equations:
\begin{align}
	p_y &= \text{Pr}(f(\mathbf{X},\mathbf{x})=y)\\
		&= \text{Pr}(f(\mathbf{X},\mathbf{x})=y|\mathbf{X}\in\mathbf{O}_o) \cdot \text{Pr}(\mathbf{X}\in\mathbf{O}_o) \nonumber\\\label{eq:subtract1}
		&\;+  \text{Pr}(f(\mathbf{X},\mathbf{x})=y|\mathbf{X}\in(\mathbf{O}_C-\mathbf{O}_o)) \nonumber\\
		&\quad\cdot \text{Pr}(\mathbf{X}\in(\mathbf{O}_C-\mathbf{O}_o)),\\
	p_y' &= \text{Pr}(f(\mathbf{Y},\mathbf{x})=y)\\\label{eq:subtract2}
		&= \text{Pr}(f(\mathbf{Y},\mathbf{x})=y|\mathbf{Y}\in\mathbf{O}_o) \cdot \text{Pr}(\mathbf{Y}\in\mathbf{O}_o) \nonumber\\
		&\;+  \text{Pr}(f(\mathbf{Y},\mathbf{x})=y|\mathbf{Y}\in(\mathbf{O}_C'-\mathbf{O}_o)) \nonumber\\ &\quad\cdot\text{Pr}(\mathbf{Y}\in(\mathbf{O}_C'-\mathbf{O}_o)).
\end{align}
Note that we have:
{\small
\begin{align}
	&\text{Pr}(f(\mathbf{X},\mathbf{x})=y|\mathbf{X}\in\mathbf{O}_o) = \text{Pr}(f(\mathbf{Y},\mathbf{x})=y|\mathbf{Y}\in\mathbf{O}_o),\\
	&\text{Pr}(\mathbf{X}\in\mathbf{O}_o) = \text{Pr}(\mathbf{Y}\in\mathbf{O}_o) = \frac{{n-m \choose k}}{{n \choose k}}, 
\end{align}
}
where $m$ is the number of malicious clients. Therefore, we know:
\begin{align}
	&\text{Pr}(f(\mathbf{X},\mathbf{x})=y|\mathbf{X}\in\mathbf{O}_o) \cdot \text{Pr}(\mathbf{X}\in\mathbf{O}_o)\nonumber\\
	= &\text{Pr}(f(\mathbf{Y},\mathbf{x})=y|\mathbf{Y}\in\mathbf{O}_o) \cdot \text{Pr}(\mathbf{Y}\in\mathbf{O}_o).
\end{align}
By subtracting (\ref{eq:subtract1}) from (\ref{eq:subtract2}), we obtain:
{\small
\begin{align}
	&p_y' - p_y \nonumber\\
	=& \text{Pr}(f(\mathbf{Y},\mathbf{x})=y|\mathbf{Y}\in(\mathbf{O}_C'-\mathbf{O}_o)) \cdot \text{Pr}(\mathbf{Y}\in(\mathbf{O}_C'-\mathbf{O}_o))\nonumber\\
			-& \text{Pr}(f(\mathbf{X},\mathbf{x})=y|\mathbf{X}\in(\mathbf{O}_C-\mathbf{O}_o)) \cdot \text{Pr}(\mathbf{X}\in(\mathbf{O}_C-\mathbf{O}_o)).
\end{align}
}
Similarly, we have the following equation for any $i\neq y$:
{\small
\begin{align}
	&p_i' - p_i \nonumber\\
	=  &\text{Pr}(f(\mathbf{Y},\mathbf{x})=i|\mathbf{Y}\in(\mathbf{O}_C'-\mathbf{O}_o)) \cdot \text{Pr}(\mathbf{Y}\in(\mathbf{O}_C'-\mathbf{O}_o))\nonumber\\
			-& \text{Pr}(f(\mathbf{X},\mathbf{x})=i|\mathbf{X}\in(\mathbf{O}_C-\mathbf{O}_o)) \cdot \text{Pr}(\mathbf{X}\in(\mathbf{O}_C-\mathbf{O}_o)).
\end{align}
}
Therefore, we can show:
{\small
\begin{align}
	&p_y'- p_i' \nonumber\\
	= &p_y - p_i + (p_y' - p_y) - (p_i' - p_i) \\
			= &p_y - p_i \nonumber\\
				+ & \left[\text{Pr}(f(\mathbf{Y},\mathbf{x})=y|\mathbf{Y}\in(\mathbf{O}_C'-\mathbf{O}_o)) \right.\nonumber\\
				&- \left.\text{Pr}(f(\mathbf{Y},\mathbf{x})=i|\mathbf{Y}\in(\mathbf{O}_C'-\mathbf{O}_o))\right] \cdot \text{Pr}(\mathbf{Y}\in(\mathbf{O}_C'-\mathbf{O}_o))\nonumber\\
				-& \left[\text{Pr}(f(\mathbf{X},\mathbf{x})=y|\mathbf{X}\in(\mathbf{O}_C-\mathbf{O}_o)) \right.\nonumber\\
				&- \left.\text{Pr}(f(\mathbf{X},\mathbf{x})=i|\mathbf{X}\in(\mathbf{O}_C-\mathbf{O}_o))\right] \cdot \text{Pr}(\mathbf{X}\in(\mathbf{O}_C-\mathbf{O}_o)).\label{eq:difference}
\end{align}
}
Note that we have:
\begin{align}
	&\text{Pr}(f(\mathbf{Y},\mathbf{x})=y|\mathbf{Y}\in(\mathbf{O}_C'-\mathbf{O}_o)) \nonumber\\&- \text{Pr}(f(\mathbf{Y},\mathbf{x})=i|\mathbf{Y}\in(\mathbf{O}_C'-\mathbf{O}_o)) \ge -1, 
\end{align}
\begin{align}
	&\text{Pr}(f(\mathbf{X},\mathbf{x})=y|\mathbf{X}\in(\mathbf{O}_C-\mathbf{O}_o)) \nonumber\\
	&- \text{Pr}(f(\mathbf{X},\mathbf{x})=i|\mathbf{X}\in(\mathbf{O}_C-\mathbf{O}_o)) \le 1,
\end{align}
\begin{align}
	&\text{Pr}(\mathbf{Y}\in(\mathbf{O}_C'-\mathbf{O}_o)) \nonumber\\=& \text{Pr}(\mathbf{X}\in(\mathbf{O}_C-\mathbf{O}_o)) \nonumber\\
	= &1 - \frac{{n-m \choose k}}{{n \choose k}}. 
\end{align}
 Therefore, (\ref{eq:difference}) gives:
 {\small
\begin{align}
	p_y'- p_i' &\ge p_y - p_i + (-1)\cdot \left[1 - \frac{{n-m \choose k}}{{n \choose k}}\right] - \left[1 - \frac{{n-m \choose k}}{{n \choose k}}\right]\\
			&= p_y - p_i - \left[2 - 2\cdot\frac{{n-m \choose k}}{{n \choose k}}\right]\\
			&= \frac{\left\lceil p_y \cdot {n \choose k}\right\rceil}{{n \choose k}} - \frac{\left\lfloor p_z \cdot {n \choose k}\right\rfloor}{{n\choose k}} - 2\left[1 - \frac{{n-m \choose k}}{{n \choose k}}\right] \\
			&\ge \frac{\left\lceil\underline{p_y} \cdot {n \choose k}\right\rceil}{{n \choose k}} - \frac{\left\lfloor\overline{p}_z \cdot {n \choose k}\right\rfloor}{{n\choose k}} - 2\left[1 - \frac{{n-m^* \choose k}}{{n \choose k}}\right]\\
			&> 0,
\end{align}
}
which indicates $F(\mathbf{C'}, \mathbf{x}) = y$. 

\section*{Proof of Theorem~\ref{tightness_theorem}}
\label{proof_of_tightness}
We prove Theorem~\ref{tightness_theorem} by constructing a federated learning algorithm $f^*$ 
such that $F(\mathbf{C'}, \mathbf{x}) \neq y$ or there exist ties. 

\begin{figure}[!t]
    \center
    \subfloat[Case 1]{\includegraphics[width=0.25\textwidth]{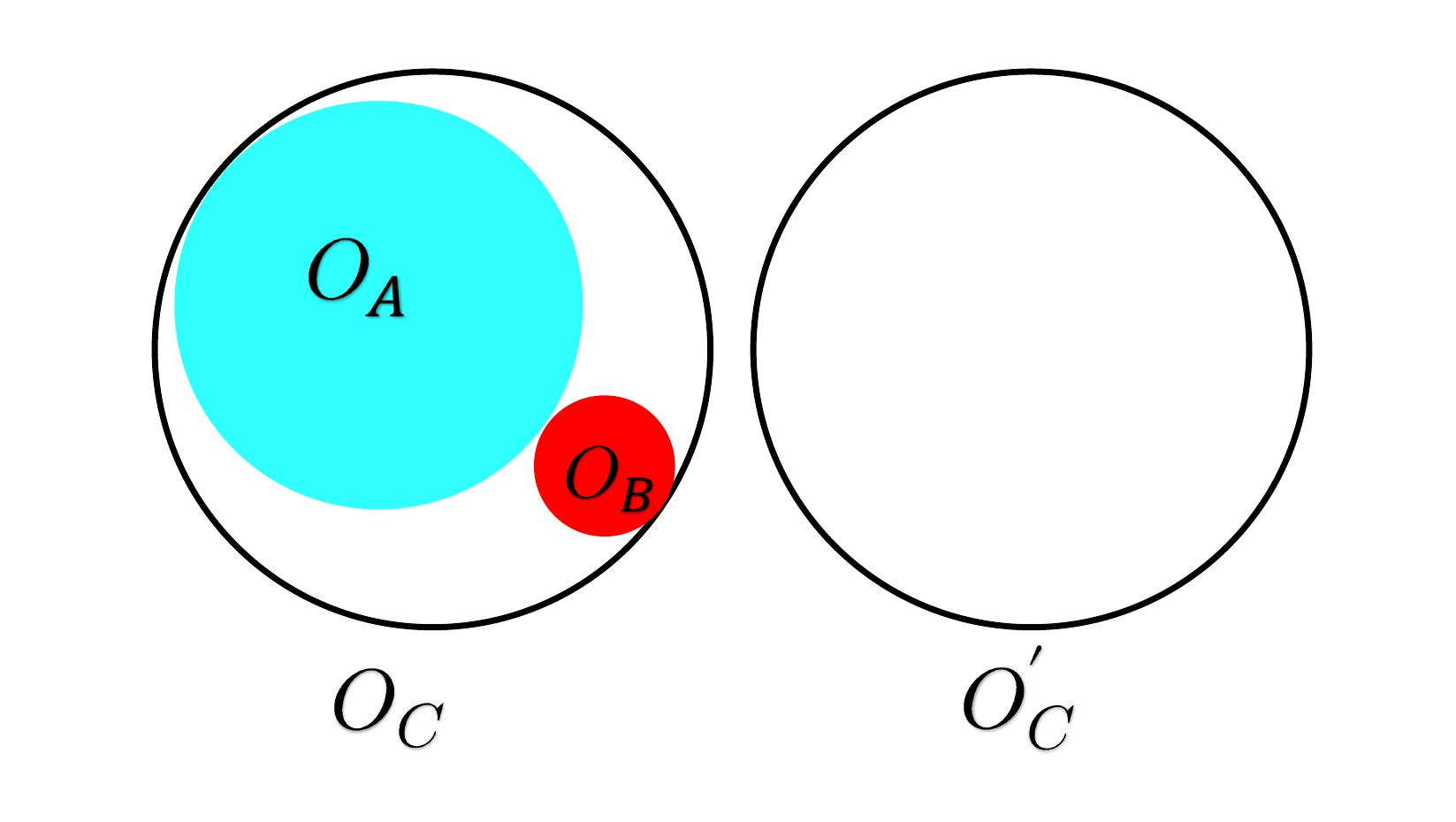}\label{fig:case1}}
    \subfloat[Case 2]{\includegraphics[width=0.25\textwidth]{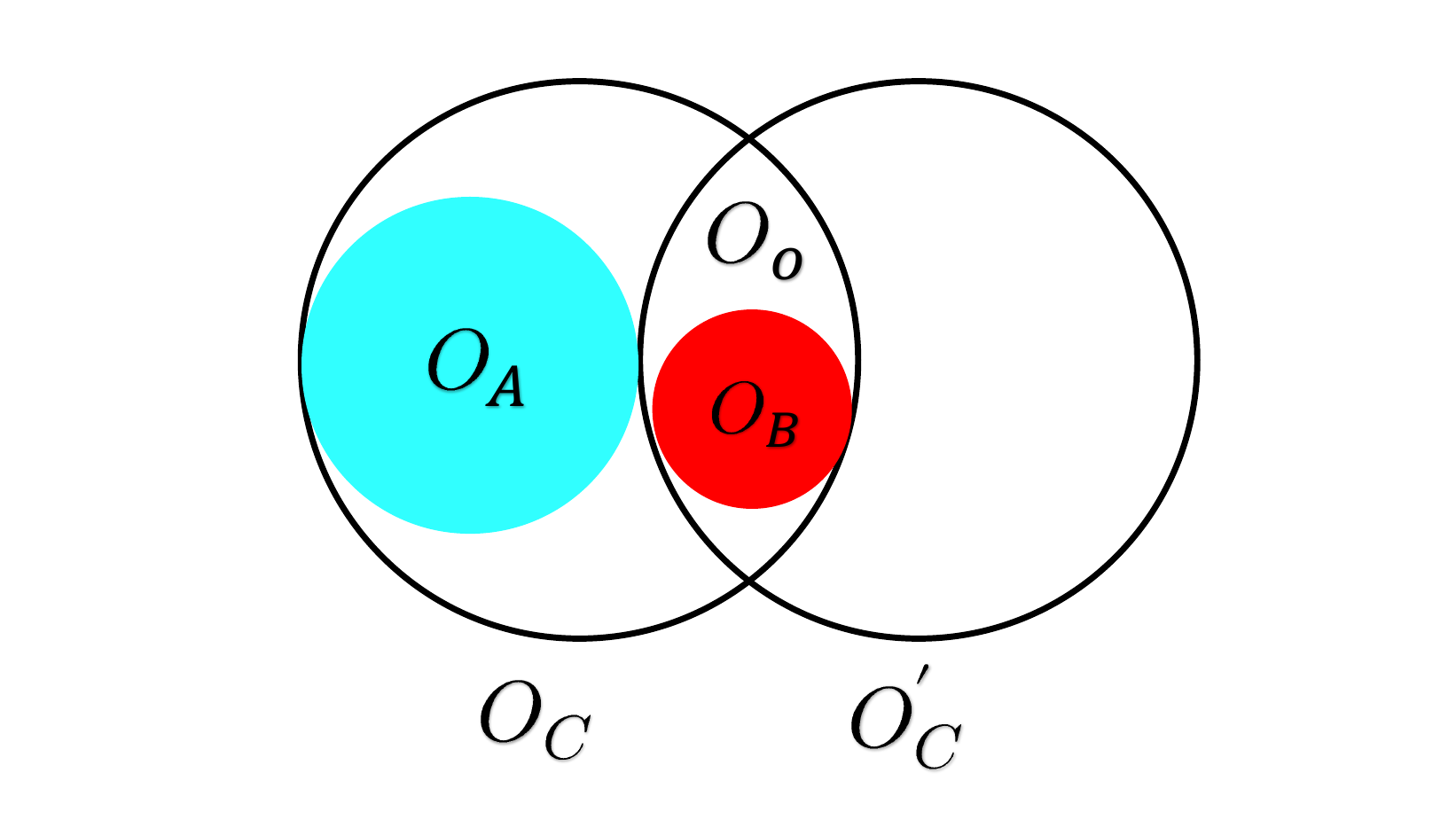}\label{fig:case2}}\\
    \subfloat[Case 3]{\includegraphics[width=0.25\textwidth]{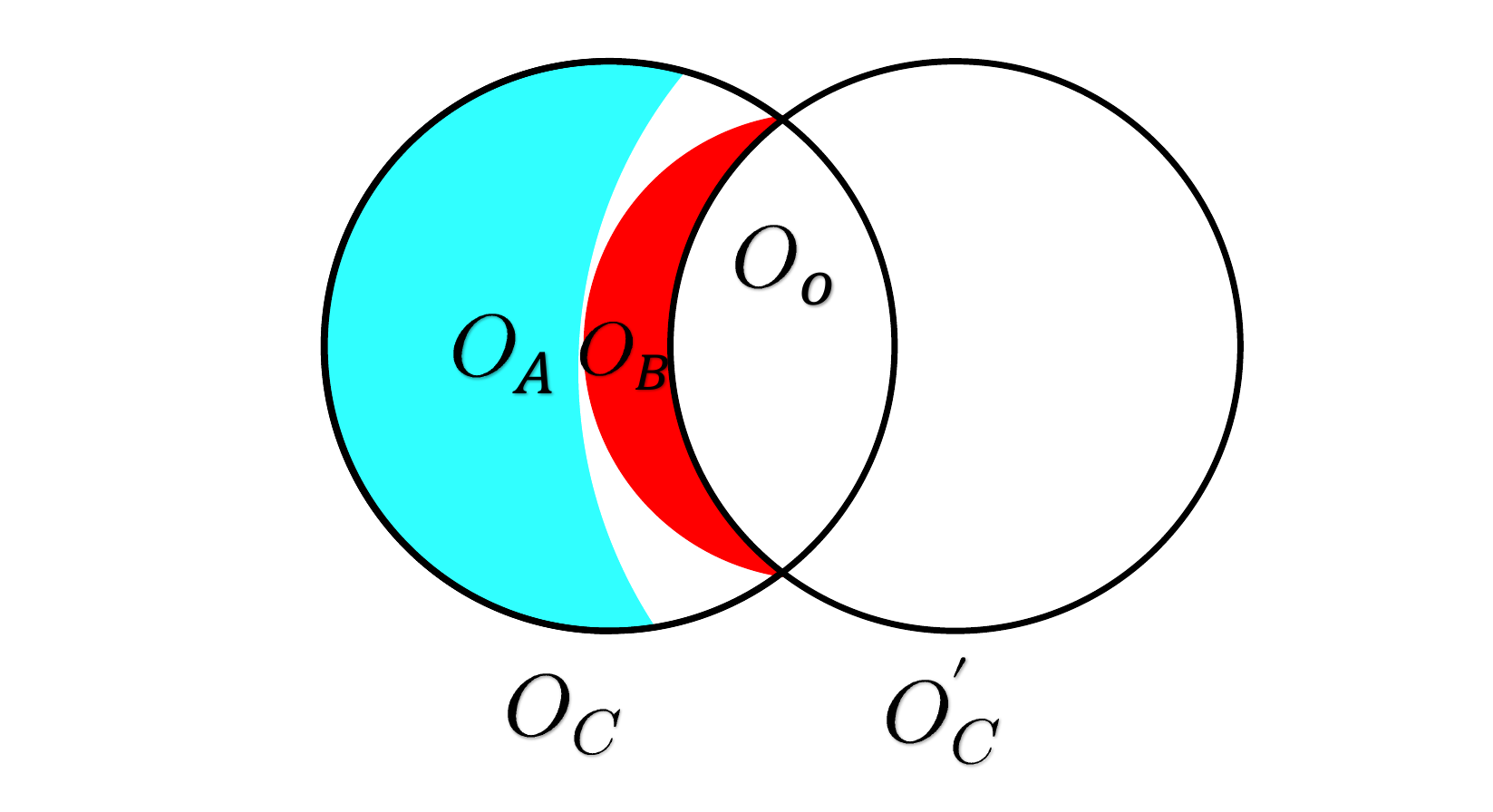}\label{fig:case3}}
    \subfloat[Case 4]{\includegraphics[width=0.25\textwidth]{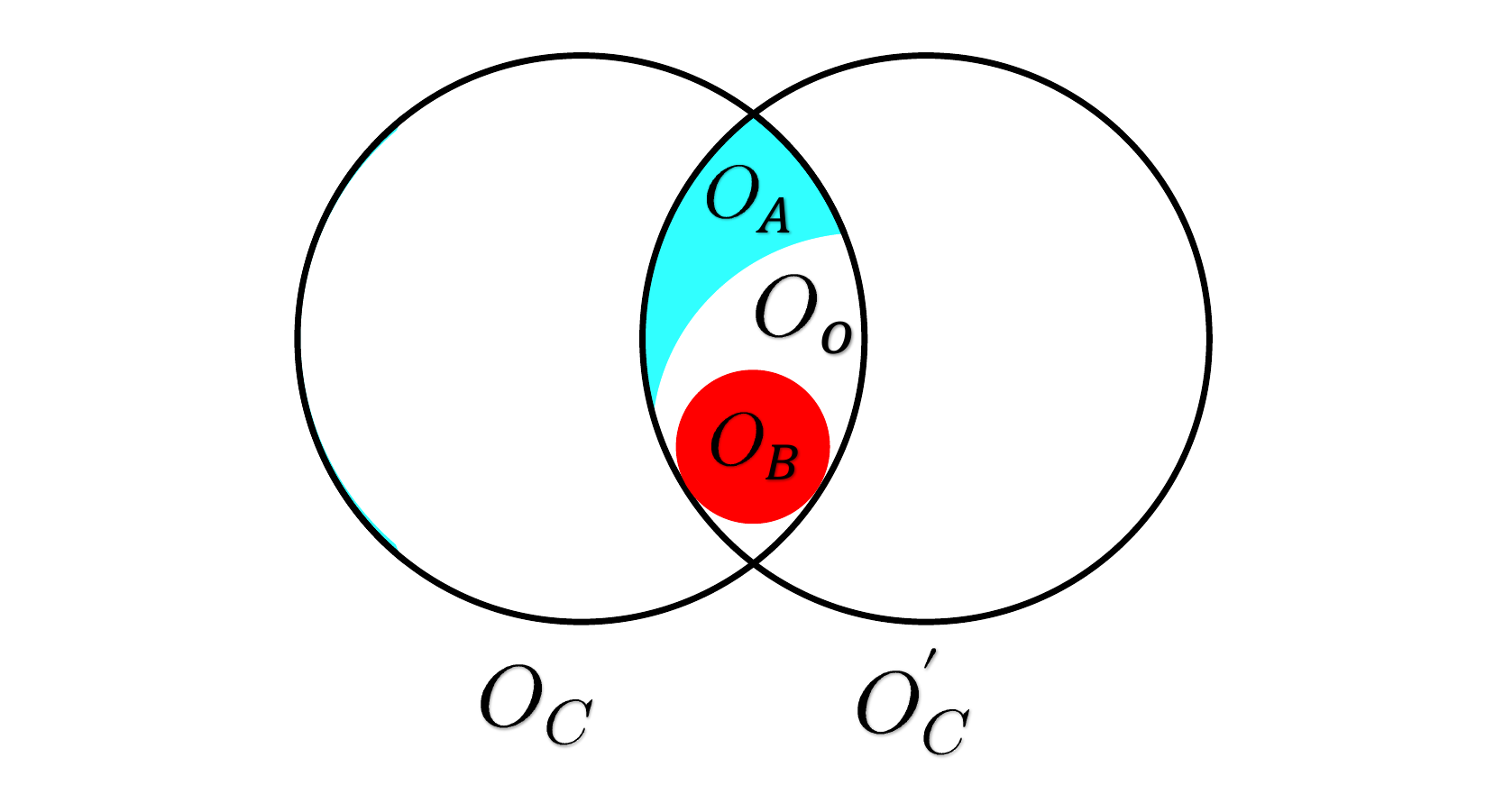}\label{fig:case4}}
    \caption{Illustration of $\mathbf{O}_C, \mathbf{O}_C', \mathbf{O}_o, \mathbf{O}_A,$ and $\mathbf{O}_B$ in the four cases.} 
    \label{fig:four_cases}
\end{figure}
We follow the definitions of $\mathbf{O}, \mathbf{O}_C, \mathbf{O}_C', \mathbf{O}_o, \mathbf{X}$, and $\mathbf{Y}$ in the previous proof.
Next, we consider four cases (Figure \ref{fig:four_cases} illustrates them).

\begin{enumerate}[label={\bfseries Case \arabic*:}, wide=0pt]

\item $m\ge n-k$.

In this case, we know $\mathbf{O}_o=\emptyset$. 
Let $\mathbf{O}_A \subseteq \mathbf{O}_C$ and $\mathbf{O}_B \subseteq \mathbf{O}_C$ such that $|\mathbf{O}_A|=\left\lceil\underline{p_y} \cdot {n \choose k}\right\rceil$, $|\mathbf{O}_B|=\left\lfloor\overline{p}_z \cdot {n \choose k}\right\rfloor$, and $\mathbf{O}_A\cap\mathbf{O}_B=\emptyset$. Since $\underline{p_y} + \overline{p}_z \leq 1$, we have:
\begin{align}
	|\mathbf{O}_A| + |\mathbf{O}_B| &= \left\lceil\underline{p_y} \cdot {n \choose k}\right\rceil + \left\lfloor\overline{p}_z \cdot {n \choose k}\right\rfloor \\
									&\le \left\lceil\underline{p_y} \cdot {n \choose k}\right\rceil + \left\lfloor(1-\underline{p_y})\cdot {n \choose k}\right\rfloor \\
									&= \left\lceil\underline{p_y} \cdot {n \choose k}\right\rceil +  {n \choose k} - \left\lceil\underline{p_y} \cdot {n \choose k}\right\rceil \\
									&= {n \choose k}=|\mathbf{O}_C|.
\end{align}
Therefore, we can always find such a pair of disjoint sets $(\mathbf{O}_A, \mathbf{O}_B)$. Figure \ref{fig:case1} illustrates $\mathbf{O}_A, \mathbf{O}_B,\mathbf{O}_C$, and $\mathbf{O}_C'$.
We can construct $f^*$ as follows:
\begin{align}
		f^*(a,\mathbf{x})=
		\begin{cases}
 			y, &\text{ if } a\in \mathbf{O}_A\\
			z, &\text{ if } a\in \mathbf{O}_B\cup\mathbf{O}_C'\\
 			i, i\neq y \text{ and } i\neq z, &\text{ otherwise}.
		\end{cases}
\end{align}
We can show that such $f^*$ satisfies the following probability properties:
\begin{align}
	p_y &= \text{Pr}(f^*(\mathbf{X},\mathbf{x})=y) = \frac{|\mathbf{O}_A|}{|\mathbf{O}_C|} = \frac{\left\lceil\underline{p_y} \cdot {n \choose k}\right\rceil}{{n \choose k}}\ge \underline{p_y},\\
	p_z &= \text{Pr}(f^*(\mathbf{X},\mathbf{x})=z) = \frac{|\mathbf{O}_B|}{|\mathbf{O}_C|} = \frac{\left\lfloor\overline{p}_z \cdot {n \choose k}\right\rfloor}{{n \choose k}} \le\overline{p}_z.
\end{align}
Therefore, $f^*$ satisfies the probability condition (\ref{eq:prob_condition}). However, we have:
\begin{align}
	p_z' = \text{Pr}(f^*(\mathbf{Y},\mathbf{x})=z) = 1,
\end{align}
which indicates $F(\mathbf{C'},  \mathbf{x}) = z \neq y$.

 \item $m^* < m < n-k$, $0 \le \underline{p_y} \le 1-\frac{{n-m \choose k}}{{n \choose k}}$, and $0\le \overline{p}_z \le \frac{{n-m \choose k}}{{n \choose k}}$.

Let $\mathbf{O}_A\subseteq \mathbf{O}_C - \mathbf{O}_o$ such that $|\mathbf{O}_A|=\lceil\underline{p_y} \cdot {n \choose k}\rceil$. 
Let $\mathbf{O}_B\subseteq \mathbf{O}_o$ such that $|\mathbf{O}_B|=\lfloor\overline{p}_z \cdot {n \choose k}\rfloor$. 
Figure \ref{fig:case2} illustrates $\mathbf{O}_A, \mathbf{O}_B,\mathbf{O}_C, \mathbf{O}_C'$, and $\mathbf{O}_o$. We can construct a federated 
learning algorithm $f^*$ as follows:
	\begin{align}
		f^*(a,\mathbf{x})=
		\begin{cases}
 			y, &\text{ if } a\in \mathbf{O}_A\\
			z, &\text{ if } a\in \mathbf{O}_B\cup(\mathbf{O}_C'-\mathbf{O}_o)\\
 			i, i\neq y \text{ and } i\neq z, &\text{ otherwise}.
		\end{cases}
	\end{align}
	We can show that such $f^*$ satisfies the following probability conditions:
	\begin{align}
		p_y &= \text{Pr}(f^*(\mathbf{X},\mathbf{x})=y)= \frac{|\mathbf{O}_A|}{|\mathbf{O}_C|} = \frac{\left\lceil\underline{p_y} \cdot {n \choose k}\right\rceil}{{n \choose k}}\ge \underline{p_y},\\
		p_z &= \text{Pr}(f^*(\mathbf{X},\mathbf{x})=z) = \frac{|\mathbf{O}_B|}{|\mathbf{O}_C|} = \frac{\left\lfloor\overline{p}_z \cdot {n \choose k}\right\rfloor}{{n \choose k}} \le\overline{p}_z,
	\end{align}
	which indicates $f^*$ satisfies (\ref{eq:prob_condition}). However, we have:
	\begin{align}
		p_y' - p_z' &= \text{Pr}(f^*(\mathbf{Y},\mathbf{x})=y) - \text{Pr}(f^*(\mathbf{Y},\mathbf{x})=z)\\
					&= 0 - \frac{|\mathbf{O}_B|+|\mathbf{O}_C'-\mathbf{O}_o|}{|\mathbf{O}_C'|}\\
					&= -\frac{\left\lfloor\overline{p}_z \cdot {n \choose k}\right\rfloor}{{n\choose k}} - 1 + \frac{{n-m \choose k}}{{n \choose k}}\\
					&< 0, 
	\end{align}
	which implies $F(\mathbf{C'},  \mathbf{x}) \neq y$.

  \item $m^* < m < n-k$, $0\le \underline{p_y} \le 1-\frac{{n-m \choose k}}{{n \choose k}}$, and $\frac{{n-m \choose k}}{{n \choose k}}\le\overline{p}_z \le 1-\underline{p_y}$.

Let $\mathbf{O}_A\subseteq \mathbf{O}_C-\mathbf{O}_o$  and $\mathbf{O}_B\subseteq \mathbf{O}_C-\mathbf{O}_o$ such that $|\mathbf{O}_A|=\lceil\underline{p_y} \cdot {n \choose k}\rceil$, $|\mathbf{O}_B|=\lfloor\overline{p}_z \cdot {n \choose k}\rfloor - {n-m \choose k}$, and $\mathbf{O}_A \cap \mathbf{O}_B=\emptyset$. Note that $|\mathbf{O}_C-\mathbf{O}_o| = {n \choose k}-{n-m \choose k}$, and we have:
{\small
\begin{align}
&|\mathbf{O}_A|+|\mathbf{O}_B| \nonumber\\=& \left\lceil\underline{p_y} \cdot {n \choose k}\right\rceil + \left\lfloor\overline{p}_z \cdot {n \choose k}\right\rfloor - {n-m \choose k} \\
							\le& \left\lceil\underline{p_y} \cdot {n \choose k}\right\rceil + \left\lfloor(1-\underline{p_y}) \cdot {n \choose k}\right\rfloor - {n-m \choose k} \\
							=& \left\lceil\underline{p_y} \cdot {n \choose k}\right\rceil + \left[{n \choose k} - \left\lceil\underline{p_y} \cdot {n \choose k}\right\rceil\right] -{n-m \choose k} \\
							=& {n \choose k} - {n-m \choose k}.
\end{align}
}
Therefore, we can always find a pair of such disjoint sets $(\mathbf{O}_A, \mathbf{O}_B)$. Figure \ref{fig:case3} illustrates $\mathbf{O}_A, \mathbf{O}_B,\mathbf{O}_C, \mathbf{O}_C'$, and $\mathbf{O}_o$. We can  construct an algorithm $f^*$ as follows:
	\begin{align}
		f^*(a,\mathbf{x})=
		\begin{cases}
 			y, &\text{ if } a\in \mathbf{O}_A\\
			z, &\text{ if } a\in \mathbf{O}_B\cup\mathbf{O}_C'\\
 			i, i\neq y \text{ and } i\neq z, &\text{ otherwise}.
		\end{cases}
	\end{align}
	We can show that such $f^*$ satisfies the following probability conditions:
	\begin{align}
		p_y &= \text{Pr}(f^*(\mathbf{X},\mathbf{x})=y) = \frac{|\mathbf{O}_A|}{|\mathbf{O}_C|} = \frac{\left\lceil\underline{p_y} \cdot {n \choose k}\right\rceil}{{n \choose k}}\ge \underline{p_y},\\
		p_z &= \text{Pr}(f^*(\mathbf{X},\mathbf{x})=z) = \frac{|\mathbf{O}_B|+|\mathbf{O}_o|}{|\mathbf{O}_C|} = \frac{\left\lfloor\overline{p}_z \cdot {n \choose k}\right\rfloor}{{n \choose k}} \nonumber\\&\le\overline{p}_z,
	\end{align}
	which are consistent with the probability conditions (\ref{eq:prob_condition}). However, we can show the following:
	\begin{align}
		 p_z' =  \text{Pr}(f^*(\mathbf{Y},\mathbf{x})=z)=1,
	\end{align}
	which gives $F(\mathbf{C'},  \mathbf{x}) = z \neq y$.

  \item $m^* < m < n-k$, $1-\frac{{n-m \choose k}}{{n \choose k}}< \underline{p_y} \le 1$, and $0\leq \overline{p}_z \le 1 - \underline{p_y}<\frac{{n-m \choose k}}{{n \choose k}}$.

Let $\mathbf{O}_A\subseteq \mathbf{O}_o$ and $\mathbf{O}_B\subseteq \mathbf{C}_o$ such that $|\mathbf{O}_A|=\left\lceil\underline{p_y} \cdot {n \choose k}\right\rceil + {n-m \choose k} - {n \choose k}$, $|\mathbf{O}_B|=\left\lfloor\overline{p}_z \cdot {n \choose k}\right\rfloor$, and $\mathbf{O}_A \cap \mathbf{O}_B=\emptyset$. Note that $|\mathbf{O}_o| = {n-m \choose k}$, and we have:
{\small
\begin{align}
&|\mathbf{O}_A|+|\mathbf{O}_B| \nonumber\\
=& \left\lceil\underline{p_y} \cdot {n \choose k}\right\rceil + {n-m \choose k} - {n \choose k} + \left\lfloor\overline{p}_z \cdot {n \choose k}\right\rfloor \\
							\le& \left\lceil\underline{p_y} \cdot {n \choose k}\right\rceil + {n-m \choose k} - {n \choose k} + \left\lfloor(1-\underline{p_y}) \cdot {n \choose k}\right\rfloor \\
							=& \left\lceil\underline{p_y} \cdot {n \choose k}\right\rceil + {n-m \choose k} - {n \choose k} + \left[{n \choose k} - \left\lceil\underline{p_y} \cdot {n \choose k}\right\rceil\right]  \\
							=& {n-m \choose k}.
\end{align}
}
Therefore, we can always find such a pair of disjoint sets $(\mathbf{O}_A$, $\mathbf{O}_B)$. Figure \ref{fig:case4} illustrates $\mathbf{O}_A, \mathbf{O}_B,\mathbf{O}_C, \mathbf{O}_C'$, and $\mathbf{O}_o$. Next, we can construct an algorithm $f^*$ as follows:
	\begin{align}
		f^*(a,\mathbf{x})=
		\begin{cases}
 			y, &\text{ if } a\in \mathbf{O}_A\cup(\mathbf{O}_C - \mathbf{O}_o)\\
			z, &\text{ if } a\in \mathbf{O}_B\cup(\mathbf{O}_C' - \mathbf{O}_o)\\
 			i, i\neq y \text{ and } i\neq z, &\text{ otherwise}.
		\end{cases}
	\end{align}
	We can show that $f^*$ has the following properties:
	\begin{align}
		p_y &= \text{Pr}(f^*(\mathbf{X},\mathbf{x})=y) = \frac{|\mathbf{O}_A|+|\mathbf{O}_C-\mathbf{O}_o|}{|\mathbf{O}_C|} = \frac{\left\lceil\underline{p_y} \cdot {n \choose k}\right\rceil}{{n \choose k}}\nonumber\\ 
		&\ge \underline{p_y},\\
		p_z &= \text{Pr}(f^*(\mathbf{X},\mathbf{x})=z) = \frac{|\mathbf{O}_B|}{|\mathbf{O}_C|} = \frac{\left\lfloor\overline{p}_z \cdot {n \choose k}\right\rfloor}{{n \choose k}} \le\overline{p}_z,
	\end{align}
	which implies $f^*$ satisfies the probability condition (\ref{eq:prob_condition}). However, we  also have:
	{\small
	\begin{align}
		&p_y' - p_z' \nonumber\\
		=& \text{Pr}(f^*(\mathbf{Y},\mathbf{x})=y) -  \text{Pr}(f^*(\mathbf{Y},\mathbf{x})=z)\\
					=& \frac{|\mathbf{O}_A|}{|\mathbf{O}_C'|} - \frac{|\mathbf{O}_B|+|\mathbf{O}_C'-\mathbf{O}_o|}{|\mathbf{O}_C'|}\\
					=& \frac{\left\lceil\underline{p_y} \cdot {n \choose k}\right\rceil + {n-m \choose k} - {n \choose k}}{{n \choose k}} - \frac{\left\lfloor\overline{p}_z \cdot {n \choose k}\right\rfloor - {n-m \choose k} + {n \choose k}}{{n \choose k}}\\
					=& \frac{\left\lceil\underline{p_y} \cdot {n \choose k}\right\rceil}{{n \choose k}} - \frac{\left\lfloor\overline{p}_z \cdot {n \choose k}\right\rfloor}{{n\choose k}} - \left[2 - 2\cdot \frac{{n-m \choose k}}{{n \choose k}}\right].
	\end{align}
	}
	 Since $m>m^*$, we have:
	\begin{align}
	\frac{\left\lceil\underline{p_y} \cdot {n \choose k}\right\rceil}{{n \choose k}} - \frac{\left\lfloor\overline{p}_z \cdot {n \choose k}\right\rfloor}{{n\choose k}}  \le \left[2 - 2\cdot \frac{{n-m \choose k}}{{n \choose k}}\right].
	\end{align}
	Therefore, $p_y' - p_z' \le 0$,
	which indicates $F(\mathbf{C'},  \mathbf{x}) \neq y$ or there exist ties.

\end{enumerate}
To summarize, we have proven that in any possible cases, Theorem \ref{tightness_theorem} holds, indicating that our derived certified Byzantine size is tight.

\section*{Proof of Theorem~\ref{probability_of_certify}}
\label{proof_of_probability}
Based on the Clopper-Pearson method, we have:  

\begin{align}
    &\text{Pr}(\underline{p_{y_t}} \leq \text{Pr}(f(\mathcal{S}(\mathbf{C},k), \mathbf{x}_t)=y_t) \land \overline{p}_{z_t}\nonumber\\
    \geq& \text{Pr}(f(\mathcal{S}(\mathbf{C},k), \mathbf{x}_t)=i), \forall i \neq y_t) \nonumber\\
    \geq& 1 - \frac{\alpha}{d}.
\end{align}
Therefore, for a test input $\mathbf{x}_t$, if our algorithm does not abstain for $\mathbf{x}_t$, the probability that it returns an incorrect certified Byzantine size is at most $\frac{\alpha}{d}$: 

\begin{align}
    &\text{Pr}((\exists \mathbf{C}'\in \mathbf{\Omega}(\mathbf{C},{\hat{m}_t^*}), h(\mathbf{C}', k ,\mathbf{x}_t)\neq\hat{y}_t)|\hat{y}_t \neq \text{ABSTAIN} ) \nonumber\\\leq &\frac{\alpha}{d}. 
\end{align}

Then, we have the following: 
\begin{align}
&\quad\text{Pr}(\cap_{\mathbf{x}_t \in \mathcal{D}} ((\forall \mathbf{C}'\in \mathbf{\Omega}(\mathbf{C},{\hat{m}_t^*}), h(\mathbf{C}', k,\mathbf{x}_t)=\hat{y}_t)|\nonumber\\&\quad\quad\quad\hat{y}_t\neq \text{ABSTAIN})) \nonumber\\
\label{theorem_4_apply_oooles_inequality_1}
&= 1 - \text{Pr}(\cup_{\mathbf{x}_t \in \mathcal{D}} ((\exists \mathbf{C}'\in \mathbf{\Omega}(\mathbf{C},{\hat{m}_t^*}), h(\mathbf{C}', k,\mathbf{x}_t)\neq\hat{y}_t )|\nonumber\\
&\quad\quad\quad\hat{y}_t\neq \text{ABSTAIN})) \\
\label{theorem_4_apply_oooles_inequality_2}
& \geq  1 - \sum_{\mathbf{x}_t \in \mathcal{D}}\text{Pr}((\exists \mathbf{C}'\in \mathbf{\Omega}(\mathbf{C},{\hat{m}_t^*}), h(\mathbf{C}', k,\mathbf{x}_t)\neq\hat{y}_t )|\nonumber\\
&\quad\quad\quad\hat{y}_t\neq \text{ABSTAIN}) \\
& \geq  1- d \cdot \frac{\alpha}{d} \\
& = 1 -\alpha .
\end{align}
We have (\ref{theorem_4_apply_oooles_inequality_2}) from (\ref{theorem_4_apply_oooles_inequality_1}) based on the Boole's inequality. 

\begin{IEEEbiography}[{\includegraphics[width=1in,height=1.25in,clip,keepaspectratio]{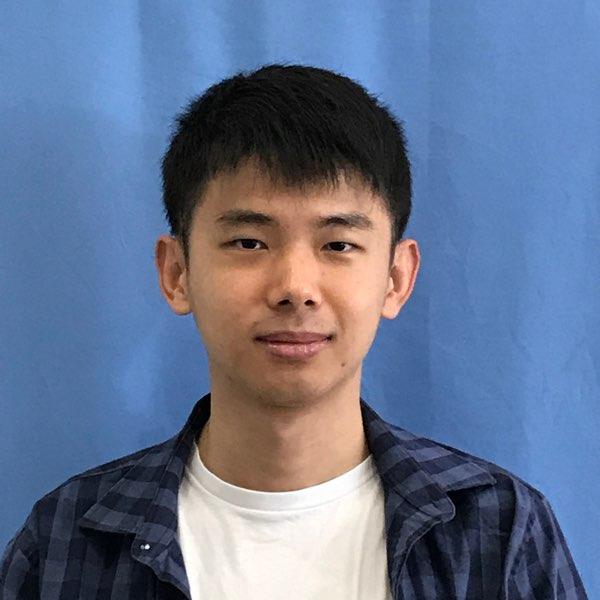}}]
{Xiaoyu Cao} received an B.Eng. degree from the department of gifted young, University of Science and Technology of China (USTC), M.Eng. degree from Iowa State University, and Ph.D degree from Duke University in 2016, 2019, and 2022, respectively. 
He is currently a research scientist at Meta Platforms. His research interests are in the area of machine learning security and privacy, with special interest in federated learning security. 
\end{IEEEbiography}

\begin{IEEEbiography}[{\includegraphics[width=1in,height=1.25in,clip,keepaspectratio]{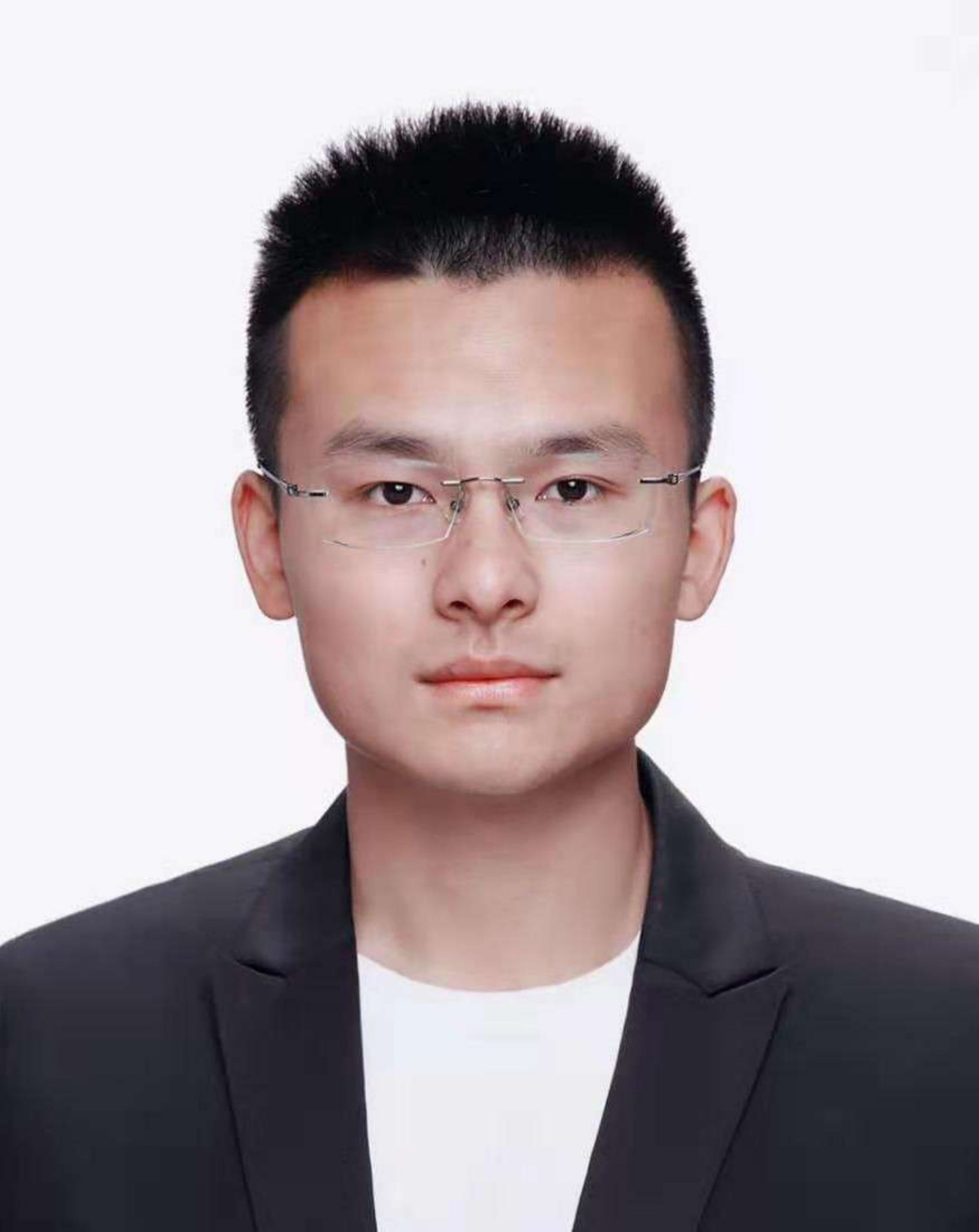}}]
{Zaixi Zhang} received an BS degree from the department of gifted young, University of Science and Technology of China (USTC), in 2019. He is currently working toward the Ph.D. degree in the School of Computer Science and Technology at USTC. His main research interests include data mining, machine learning security \& privacy, and graph representation learning. He has published papers in referred conference proceedings, such as IJCAI, NeurIPS, KDD, and AAAI.
\end{IEEEbiography}

\begin{IEEEbiography}[{\includegraphics[width=1in,height=1.25in,clip,keepaspectratio]{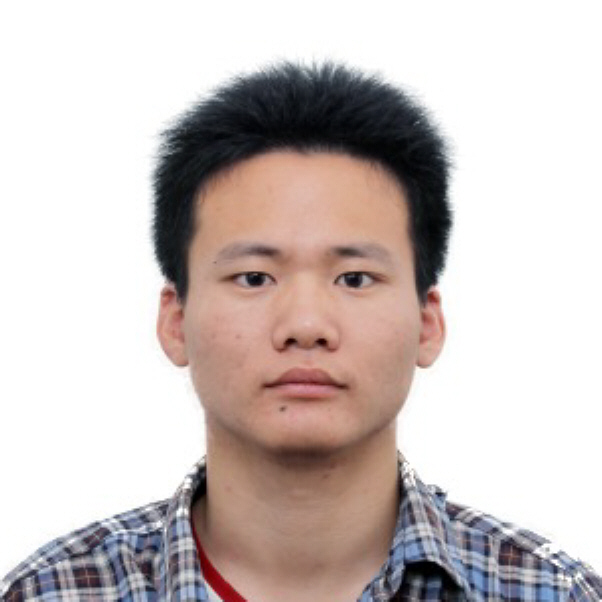}}]
{Jinyuan Jia} received an B.Eng. degree from University of Science and Technology of China (USTC), M.Eng. degree from Iowa State University, and Ph.D degree from Duke University in 2016, 2019, and 2022, respectively. 
He is a postdoc at the University of Illinois Urbana-Champaign and will be an Assistant Professor at The Pennsylvania State University starting in July 2023. His research involves security, privacy, and machine learning, with a recent focus on the intersection among them. 
\end{IEEEbiography}

\begin{IEEEbiography}[{\includegraphics[width=1in,height=1.25in,clip,keepaspectratio]{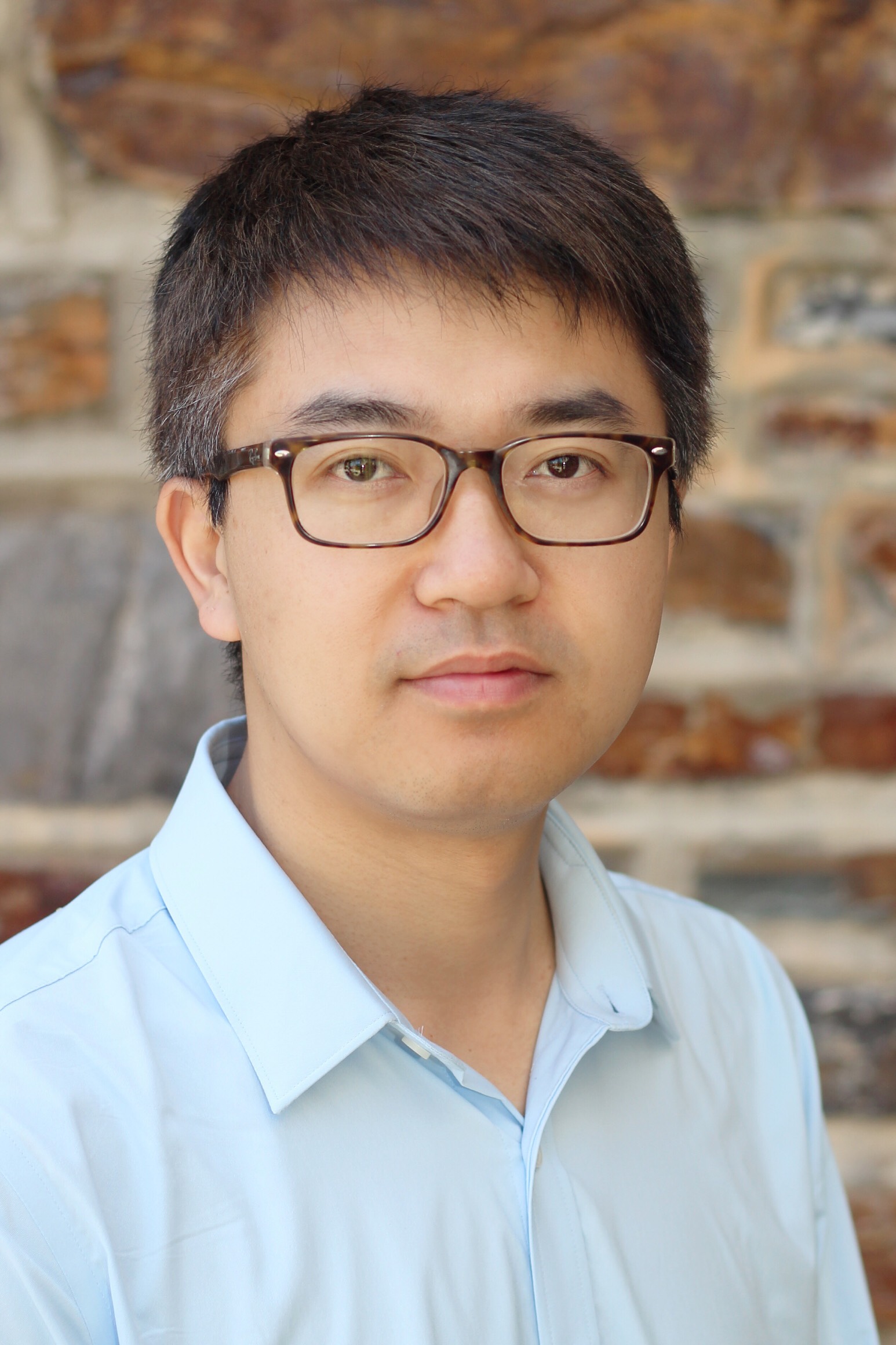}}]
{Neil Zhenqiang Gong} received an B.Eng. degree from University of Science and Technology of China (USTC) in 2010 and Ph.D. in Computer Science from University of California Berkeley in 2015. He is currently an Assistant Professor at Duke University. His research interests are cybersecurity, privacy, machine learning security, and social networks security. He has received multiple prestigious awards such as Army Research Office Young Investigator Program Award and NSF CAREER Award. 

\end{IEEEbiography}
\end{document}